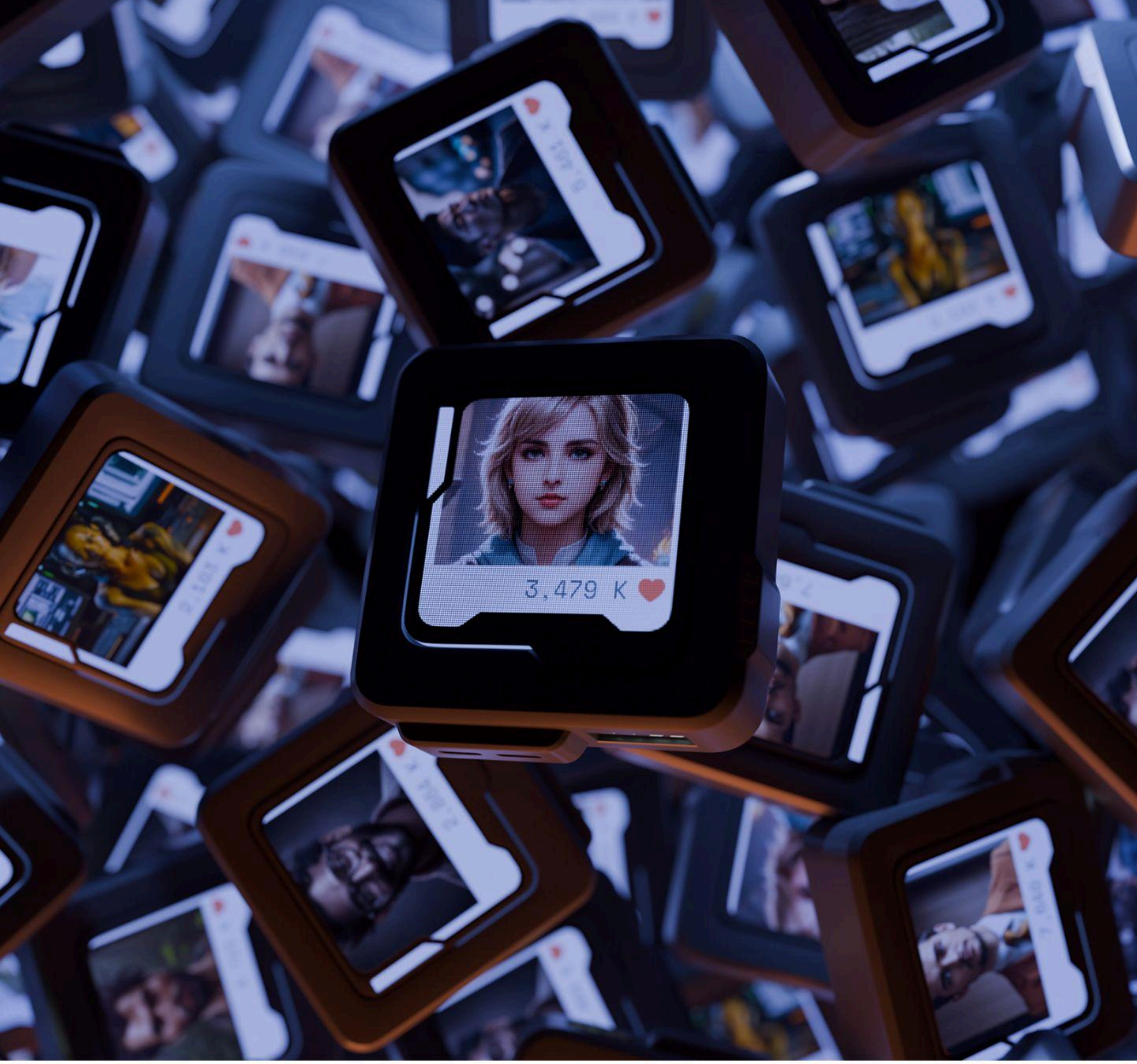

# AI-GENERATED ALGORITHMIC VIRALITY

How synthetic AI imagery and agentic AI accounts try to game TikTok and Instagram.

AI FORENSICS

# Credits

Authors: Natalia Stanusch, Martin Degeling, Salvatore Romano, Raziye Buse Çetin, Miazia Schüler, Silvia Semenzin.

The contribution from AI Forensics is funded by core grants from Open Society Foundations, Luminate, Omidyar Network, and Limelight Foundation.

Graphic & Brand Design: Denis Constant / Ittai Studio https://ittai.co/
Special Thanks to Marcus Bösch for constructive feedback.

Email: info@aiforensics.org

# Disclaimer:

This report presents statistics, analyses, and conclusions based on our specific research and publicly available data as of June 2025.

Any statistics or content numbers we refer to are valid only for the specific research we have carried out. Different statistics can be found by conducting data searches on the same platforms using different criteria.

Due to the vast amount of content on Instagram and TikTok, we have chosen to analyse data from a very short period (a few days) for three European Union countries, based on searches using a few selected keywords (hashtags). This report is by no means exhaustive and only provides an overview for a given period. We selected our field of research with the aim of being as representative as possible, and we explained and detailed our specific methodology and limitations in the relevant sections.

Sporadic mistakes may have arisen from manual annotation. Datasets and analysis codes supporting our research conclusions are attached.

# Executive summary

There is a growing discussion about social media feeds being increasingly filled with AI-generated content.[1] Due to its visual plausibility, low cost, and fast production speed, AI-generated content is said to be highly effective in "gaming the algorithm" and going viral. Popularly referred to as "AI slop," this phenomenon arguably leads to the presence of sloppy and potentially deceptive content at a scale unseen before.

This investigation offers a systematic analysis of AI-generated content and its labelling in TikTok's and Instagram's search results across 13 hashtags (see Appendix) in three European countries (Spain, Germany, and Poland) over the course of June 2025. We manually annotated and analyzed the 30 top search results on political (#trump, #zelensky, #pope) and broader topics (e.g.,#health, #history) to understand the relation between synthetic (content that is partially or entirely made using generative AI) and non-synthetic content across languages and countries. We then explored the emerging phenomenon of accounts producing generative AI content at scale by analyzing 153 accounts and proposing a new categorization schema of what we termed Agentic AI Accounts.

Our main findings are:

- On TikTok, one video out of four shows AI-generated content. Roughly 25% of the top 30 search results on TikTok for hashtags like #health, #history, or #trump contain synthetic AI imagery. On Instagram, search results show significantly less AI content.
- The majority of AI content on platforms is posted by what we call "Agentic AI Accounts." These accounts, which specialize in automated production and dissemination of (solely or primarily) AI content using generative AI tools, produced over 80% of the content containing synthetic AI imagery in search results on TikTok and over 15% on Instagram, marking a new era of generative AI integration with social media.
- Platforms seem not to sufficiently label AI content. Only around half of all content containing synthetic AI imagery on TikTok is labelled as AI content. While we found significantly fewer cases of synthetic AI imagery on Instagram (only 13 posts among the top results), an even lower percentage (23%) was labelled as AI content.
- Over 80% of AI content is photorealistic, increasing its deceptive potential.
- Even when present, AI labels have limited visibility. On TikTok, users' disclosures are often hidden in a long list of hashtags and descriptions, and are not immediately visible on the app's interface.

[1] See, e.g., 404 Media's reporting on "AI Slop Is a Brute Force Attack on the Algorithms That Control Reality"

- AI labels are non-existent for desktop users on Instagram. The official AI label does not appear at all on Instagram's web version, and is often hidden behind a second click on the app.

This study highlights a pressing need for a stronger enforcement and standardization of AI content labeling across platforms, particularly in light of their own commitments under the Digital Services Act (DSA). While TikTok and Instagram acknowledge AI content as a systemic risk, their current labeling practices remain inconsistent and largely ineffective. As generative AI tools become more accessible and Agentic AI Accounts continue to flood social media with plausible and often misleading content, platforms must not only ensure visible and accurate AI labels across all versions of their services but also reevaluate how the new Agentic AI Accounts can game their algorithms to increase their reach.

# Introduction

With the recent improvements and increased availability of generative AI tools, users on social media platforms reported seeing generative AI content shoved into their feeds. This generative AI content, which consists of synthetic AI imagery (videos, still images) and audio, has been used to sell questionable health supplements, visualize Bible-related content for Christians, and reanimate murdered children to narrate the stories of their demise as true crime content. Donald Trump reposted an "AI slop" reimagination of Gaza alongside other synthetic AI imagery of himself. Political parties turned to synthetic images as campaign materials. The production of synthetic AI imagery is cheap, easy, and fast, and whether meant to scam, harm, or amuse, it seems to crack the code for lucrative algorithmic vitality.

Whether users seek it out or not, it appears that content made of synthetic AI imagery takes a position on their feeds. 404 Media (an independent media focused on technology) refers to this phenomenon as AI slop "brute force attacking the algorithms" of social media platforms. "Brute forcing" is an approach to solving a problem by trying a lot of random solutions until one that works is found (i.e., trying all the combinations of a lock). In this context, it means a content creator can use generative AI tools to mass produce content, hoping that a few pieces of content will become engaging enough to gain algorithmic virality.

This practice can thus become profitable through content monetization. The 404 Media article claims that this strategy is, in fact, both enabled and allowed by the platforms as a part of attention-monetizing business models. Indeed, many examples of synthetic AI content have gained millions of views. In May 2025, four of the top ten YouTube channels with the most subscribers featured AI-generated material in every video, though some channels showed patterns of artificially inflated growth and view metrics. Some platforms, such as YouTube, began to recognize this systemic loophole and announce a "crackdown" on AI Slop. In its recent announcement, YouTube states, "*on July 15, 2025, YouTube is updating our guidelines to better identify mass-produced and repetitious content. This update better reflects what 'inauthentic' content looks like today.*" It remains to be seen how effective such actions are in relation to AI slop and other AI content, especially given the ongoing integration of generative AI tools that the platforms offer to content creators and advertisers.

# Synthetic imagery: deepfakes, generative AI, and AI slop

This investigation focuses on synthetic AI imagery in the context of social media platforms, meaning images and videos that include content made entirely or partially with generative AI tools, as well as deepfakes. As this investigation accounts for a spectrum of synthetic visual content - moving images, still images, and deepfakes - audio deepfakes and AI-generated voice overs are largely outside the scope of this report.

'Deepfake' is a term that originated before the development and popularity of generative AI tools, building primarily off the technical possibilities offered by deep learning architecture called GANs (generative adversarial networks). Deepfakes are fabricated moving-image content where a face or a body of a person is either superimposed over existing footage (e.g., a video where one's face is 'swapped' for another) or manipulated to fit a different audio. Deepfakes have been used for different purposes, ranging from political satire, humor, and nonconsensual, sexually explicit content.

Generative AI tools produce a still image or footage clip, making the content synthetically generated from start to finish. On a technical level, while deepfakes usually are operated by 'swapping' faces one for another, generative AI tools produce synthetically constructed content via algorithmic processes based on a prompt (e.g., written instructions given to a model). One of the leading uses for synthetic AI outputs is AI-made nonconsensual sexual imagery of adults and children, public and non-public figures, and fictive explicit content.[2] Another popular use case of generative AI outputs has been dubbed 'AI slop,' meaning mass-produced, low-quality AI-made content shared across the web. On social media platforms, AI slop includes seemingly uncanny, senseless content (such as 'Shrimp Jesus') that has gained visibility throughout 2024 and 2025, utilizing photorealistic formal qualities with unrealistic scenes and imagery.

# Agentic Slop: Automating research, creation, and distribution of AI slop

Recent advancements have also seen the introduction of AI agents, capable of carrying out complex tasks and interacting with environments independently, thus presenting a new set of implications for social media. An AI agent is an autonomous software program that performs specific tasks. A set of AI agents can compose an agentic AI system that can coordinate several specialized AI agents to achieve complex goals, displaying higher autonomy with the ability to manage multi-step, complex tasks. Agentic AI can autonomously research trends, draft and schedule personalized posts across social media platforms, repurpose long-form content into social media snippets, auto-generate accompanying images, and surface audience intent signals to optimize engagement and identify warm leads at scale. Moreover, they can perform all these tasks in series, starting from a single prompt. Potentially, AI agents can be used to fully automate the creation of content, at a speed that would be unsustainable for a human, creating a new market for AI Slop Guru. The potential for agentic AI to generate, disseminate, and even adapt content autonomously raises concerns about the scale and sophistication of future AI slop and manipulative campaigns, further challenging detection and moderation efforts.

[2] For some relevant investigations see WIRED's An AI Image Generator's Exposed Database Reveals What People Really Used It For, 404 Media's Chinese AI Video Generators Unleash a Flood of New Nonconsensual Porn, and Bellingcat's Faking It: Deepfake Porn Site's Link to Tech Companies.

## Scope of our investigation

To contribute to the ongoing discussion on the implications of generative AI on social media, this investigation offers a systematic analysis of TikTok's and Instagram's search results in relation to synthetic AI content. We focused on the search results instead of the algorithmically curated feeds as the latter heavily depend on behavioural information: if the user engages with synthetic AI content, they will encounter more of it on their feeds, whereas if the user avoids it, such type of content will likely show up less. To be able to understand the prevalence of synthetic AI content with respect to certain topics and to understand to what extent the platform's algorithms push such content in the search results, we analyzed whether content containing synthetic AI imagery appears to be recommended more often without searching for it specifically in the search results.

We further monitored the engagement, views, and positioning patterns of selected search results pages across countries: Spain, Germany, and Poland; and languages: English, Spanish, German, and Polish. This investigation focuses on examining three countries in the EU, one from the southern, western, and eastern parts of Europe. The annotators who qualitatively analyzed the content were either native or fluent speakers of the languages spoken in the respective countries. We subsequently derived a taxonomy of what we term Agentic AI Accounts that post content containing synthetic AI imagery. We also assessed the platforms' success rate of labelling AI content accordingly.

# Legal framework of labelling AI content on social media platforms

In this investigation, we assess whether content containing synthetic AI imagery is correctly labelled as AI-made. This is a particularly relevant concern in light of the European regulatory context. The AI Act is the first EU-wide regulation for AI systems. Entered into force in July 2024, it is being enforced gradually. The AI Act places transparency requirements on providers and deployers of AI systems capable of generating synthetic content (Art. 50). These requirements put the onus on technical solutions to make AI systems' outputs detectable and do not apply to platforms that only disseminate AI-generated content. Concerning this matter, Recital 136 of the AI Act refers to the risk assessment and mitigation obligations in the Digital Services Act (DSA).

The DSA is an earlier regulatory framework that evolved from the EU's 2000 E-Commerce Directive. It applies to online intermediaries. The DSA requires very large online platforms (VLOPs) and very large online search engines (VLOSEs) such as TikTok and Instagram to identify and mitigate systemic risks that might arise from the distribution of content, including AI-generated content. For instance, Article 35.1(k)[3] of the DSA clearly states that synthetic media that might be mistaken for existing persons, objects, places, entities, or events should be distinguishable through prominent markings, and users should be able to indicate this easily.

In March 2024, the European Commission published guidelines specific to elections to mitigate systemic risks online. The guidelines recommended specific mitigation measures for AI-generated content, such as clearly labeling deepfakes, providing users with a standard and easy-to-use interface for disclosing AI-generated content, testing and adopting efficient labels (Art 40(b)). Furthermore, the Code of Conduct on Disinformation, now part of the DSA, also includes commitments for transparency of AI systems. Under Commitment 15, relevant signatories are tasked with "proactively detecting" and "warning their users" against malicious use of AI-generated content that is outlined in the AI Act. The Code of Conduct acts as a benchmark for compliance with the DSA.

In addition to European-level regulations, Spain announced it will impose fines on companies for not labelling AI content accordingly. Earlier this year, the Spanish government introduced a bill aiming to fine non-compliance with proper labelling of AI-generated content as a "serious offence" that can lead to fines of up to 35 million euros, in line with the AI Act.

[3] "(k) ensuring that an item of information, whether it constitutes a generated or manipulated image, audio or video that appreciably resembles existing persons, objects, places or other entities or events and falsely appears to a person to be authentic or truthful is distinguishable through prominent markings when presented on their online interfaces, and, in addition, providing an easy to use functionality which enables recipients of the service to indicate such information."

In line with the regulatory framework, the platforms imposed guidelines that introduce an obligation to disclose synthetic AI content (on the users' side), and proclaim to have the technical capacity to detect and label such content themselves (see Table 1).

| | TikTok | Meta |
|---|---|---|
| Label name | "Creator labeled as AI-generated"<br>"AI generated" | "AI info" |
| Content defined as synthetic | "Images, video and/or audio that is generated or modified by artificial intelligence, such as artificial visuals, videos, or sounds that may portray realistic human likenesses or depictions created in a particular artistic style (e.g. painting, cartoons, and anime)." | "Video, audio and image content when we detect industry-standard AI image indicators or when people disclose that they're uploading AI-generated content." |
| Obligation to label on the users' side | Yes, to label "realistic AIGC, and prohibit content that can harmfully mislead or impersonate others." | Yes, to label content which has "photorealistic video or realistic-sounding audio that was digitally created, modified or altered, including with AI." |
| Possibility to label on the platform's side | Yes, TikTok "may automatically apply the 'AI-generated' label to content we identify as completely generated or significantly edited with AI. This may happen when a creator uses TikTok AI effects or uploads AI-generated content that has Content Credentials attached." | Yes, if a piece of content, like an image, "has a signal that shows it was edited using AI. Meta's systems read these signals to determine if the content needs a label." |
| Possibility for the user to label content using a method other than the label itself | Yes, "a text, a hashtag sticker, or context in the post's description is also listed as valid disclosure." | No. |

Table 1. TikTok's and Meta's guidelines for labelling synthetic AI content on organic (non-ads) content.

In May 2024, TikTok announced that it will label content "created using artificial intelligence." TikTok's definition of "AI-generated content (AIGC)" seems to us rather robust compared to Meta's. TikTok defines AI content that includes "*images, video and/or audio,*" *both* "*generated or modified by artificial intelligence*," and accounts for photorealistic and non-photorealistic AI content (e.g., painting, cartoons, and anime). TikTok "asks" users to disclose AI content as part of its Community Guidelines on Integrity and Authenticity. TikTok requires "*people to label realistic AIGC, and prohibit content that can harmfully mislead or impersonate others.*" Similarly to Instagram, on TikTok, users "can" disclose AI content by using a "creator label," however, "text, a hashtag sticker, or context in the post's description" is also listed as valid disclosure. In its "Integrity and Authenticity" guidelines, TikTok also explicitly mentions not only impersonations but also "realistic-appearing" content as requiring an AI-disclosure; "*we require you to label AIGC or edited media that shows realistic-appearing scenes or people. This can be done using the AIGC label, or by adding a clear caption, watermark, or sticker of your own.*"

TikTok states that it "may" automatically label content made with TikTok effects if they use AI, otherwise, users are asked to label content themselves. However, when content has Content Credentials attached, which TikTok defines as "*a technology from the Coalition for Content Provenance and Authenticity (C2PA). Content Credentials attach metadata to content that we can use to recognize and label AIGC instantly*," then the "AI-generated" "Auto label" may be applied by the platform.

<table>
<tr><th colspan="4">TikTok labels for synthetic AI content</th></tr>
<tr><th colspan="2">Labeled by TikTok as "AI generated"</th><th colspan="2">Labeled by TikTok as "Creator labeled as AI-generated"</th></tr>
<tr><td>mobile</td><td>browser</td><td>mobile</td><td>browser</td></tr>
<tr>
<td>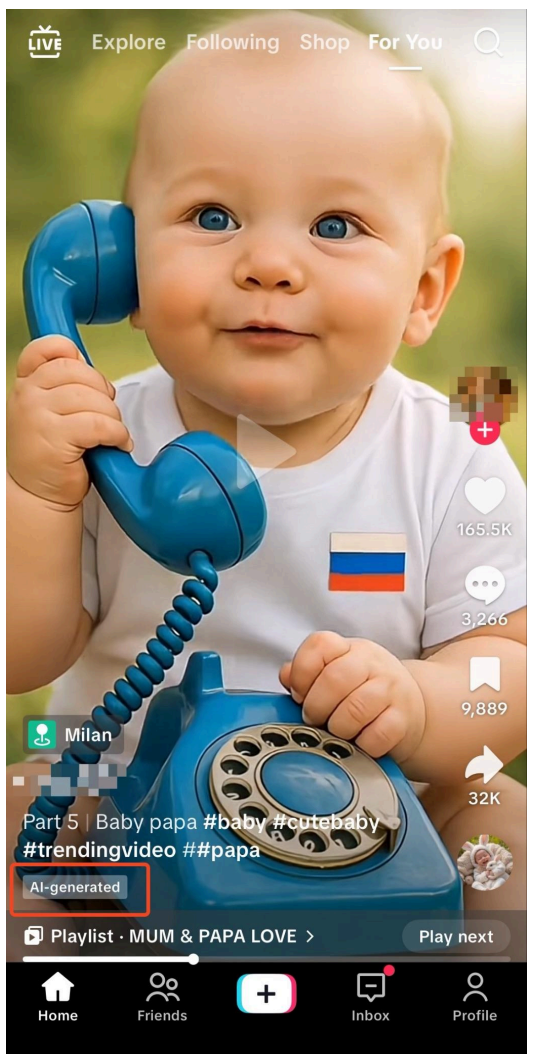
</td>
<td>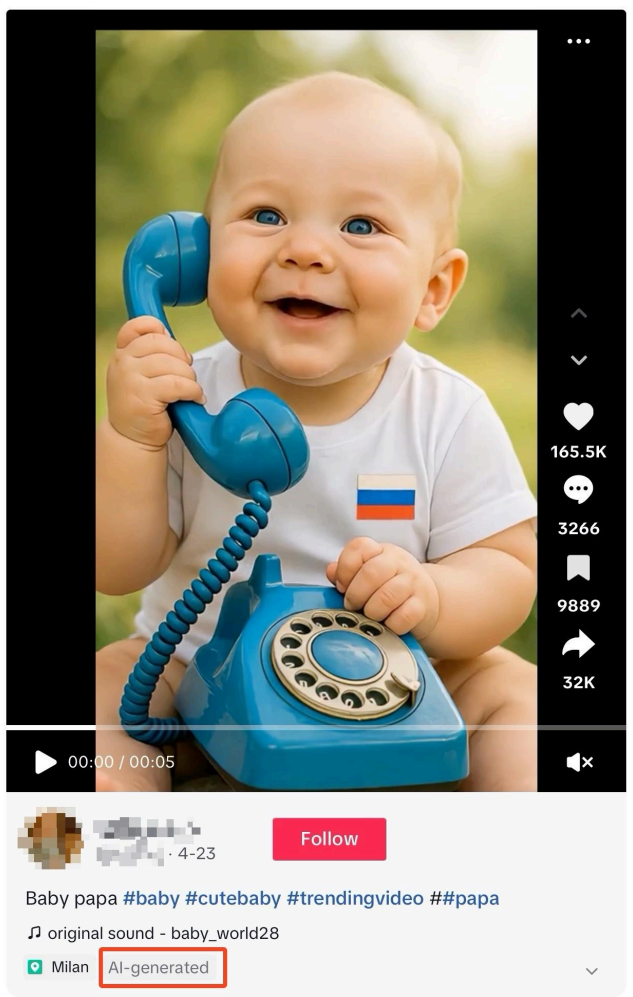
</td>
<td>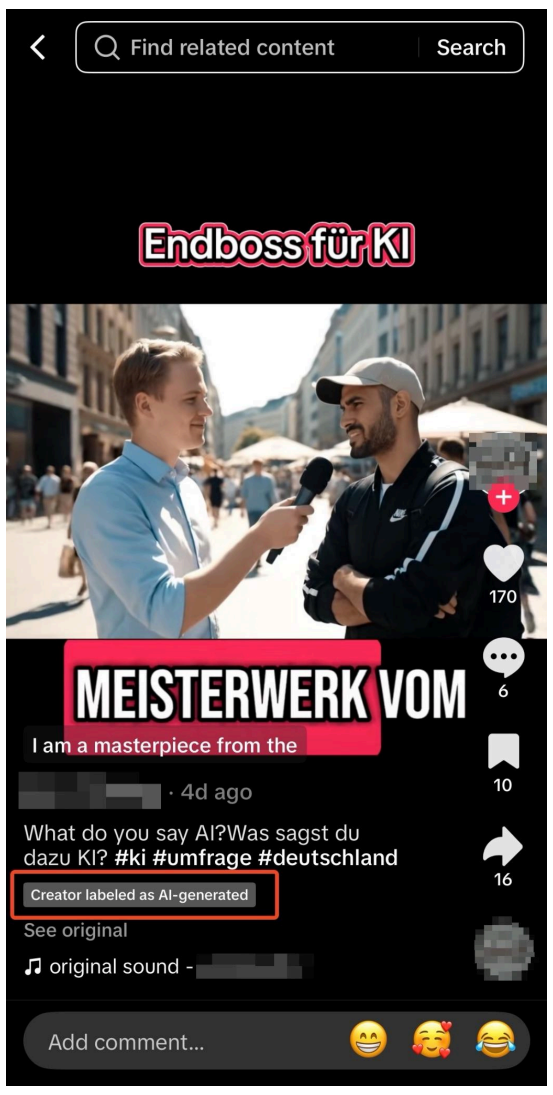
</td>
<td>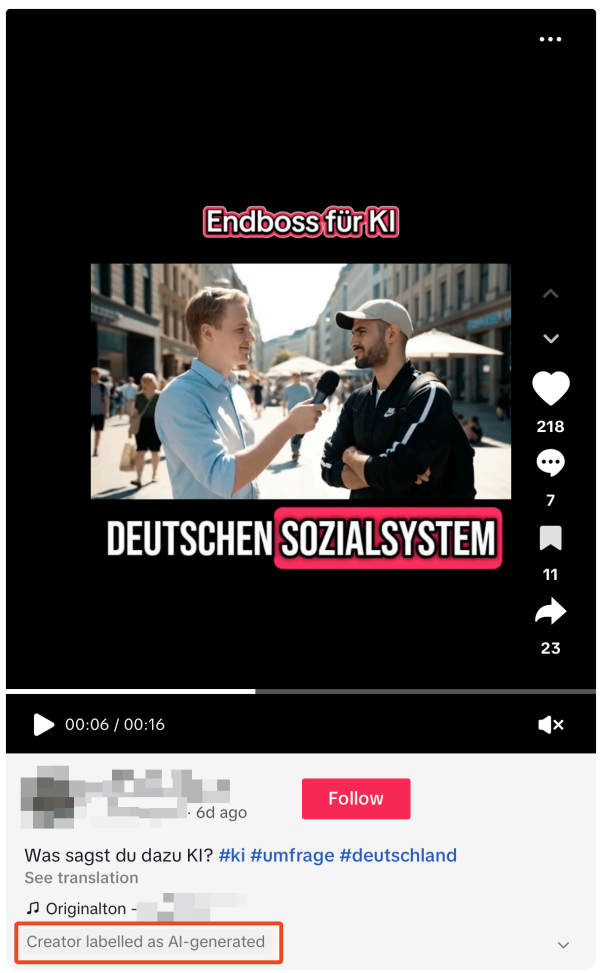
</td>
</tr>
<tr><td>Figure 1A [link]</td><td>Figure 1B</td><td>Figure 1C [link]</td><td>Figure 1D</td></tr>
</table>

Table 2. Examples of the visibility of TikTok's AI labels on the platform across mobile and web versions. The rectangle highlighted in red outlines the positioning of TikTok's official labels

## Visibility of labels on TikTok

TikTok's labels for content containing synthetic AI imagery have a consistent design and are visible across mobile and browser versions of the platform (Table 2, figures 1A-1D). We consider this to be a good example of how such labels should be implemented to ensure that users see the relevant AI information the moment they view the content. However, TikTok's policy of allowing content creators to use hashtags and text/context information in the description as a valid disclosure for synthetic AI imagery does not appear to be a systematic or safe enough measure. Indeed, as Tables 3 and 4 show (figures 2A-2C; 3A-3B), if content creators include an AI disclosure "deeper" in the description or list of hashtags, users have to click to unfold the full description of the post to see the disclosure. This puts further burden on the users to fact-check the content they are viewing, rather than content creators who are uploading the content or the platform itself that is moderating such content.

<table>
<tr><td colspan="3">TikTok's alternative disclaimer for labelling synthetic AI content on the app</td></tr>
<tr><td colspan="3">An example of alternative labelling (hashtag//text/context in the description)</td></tr>
<tr><td>mobile (invisible)</td><td>mobile (invisible)</td><td>mobile (visible)</td></tr>
<tr><td>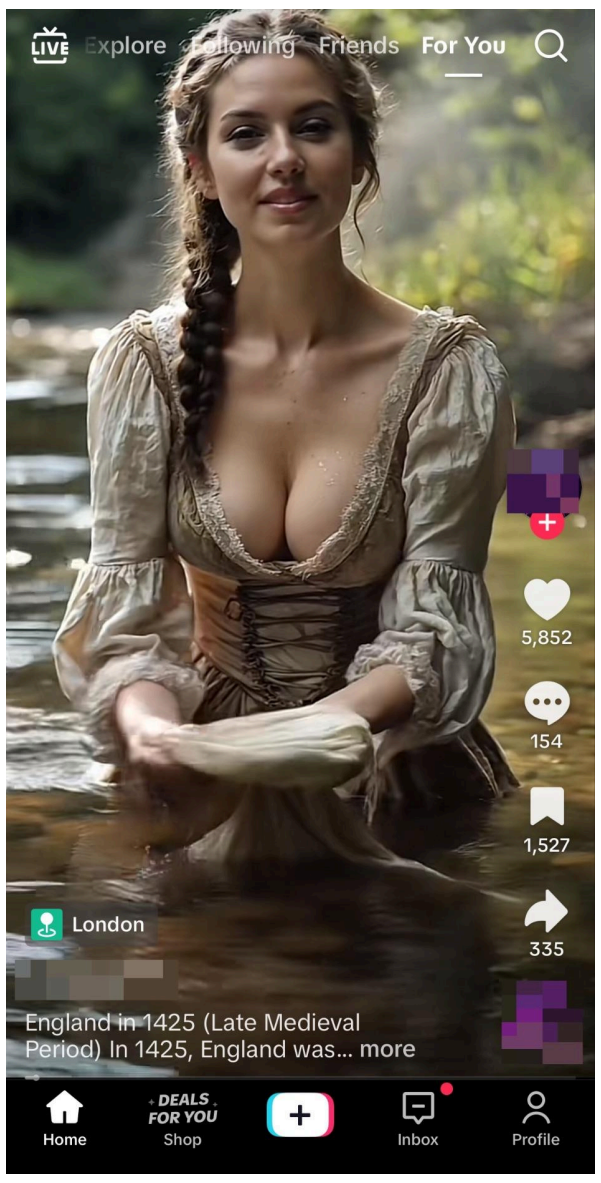
</td><td>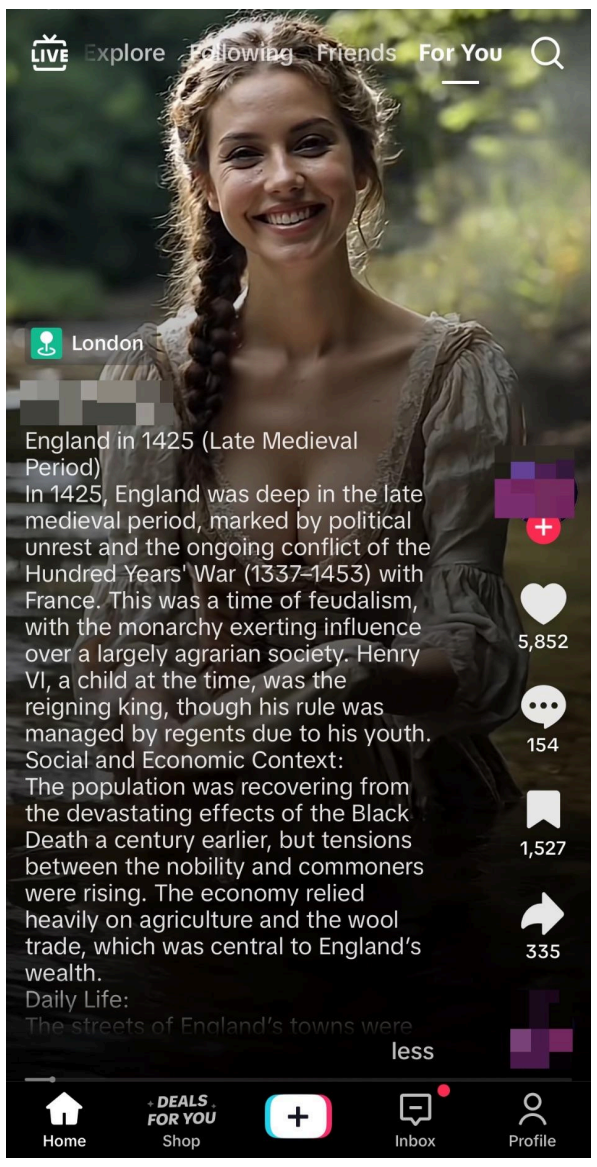
</td><td>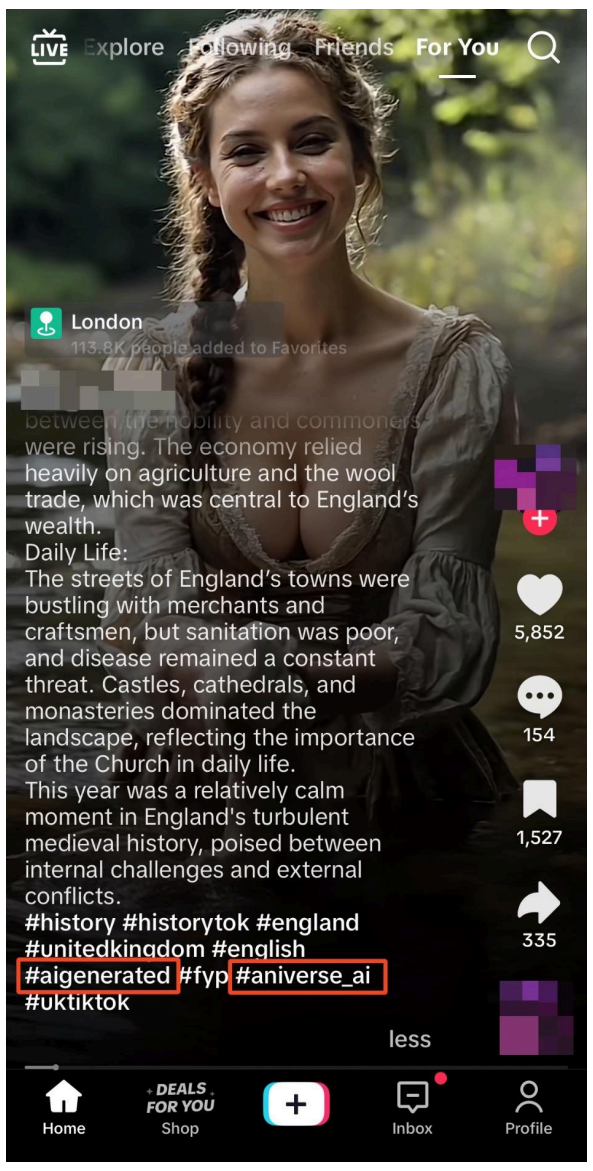
</td></tr>
<tr><td>Figure 2A [link]</td><td>Figure 2B</td><td>Figure 2C</td></tr>
</table>

Table 3. Examples of the visibility of TikTok's alternative users' disclaimer of AI content on the platform's app version. The rectangle highlighted in red outlines the positioning of the user's AI disclosure, which, in this case, is a hashtag.

<table>
<tr><td colspan="2">TikTok's alternative disclaimer for labelling synthetic AI content on the browser</td></tr>
<tr><td colspan="2">An example of alternative labelling (hashtag//text/context in the description)</td></tr>
<tr><td>browser (invisible)</td><td>browser (visible)</td></tr>
<tr><td>
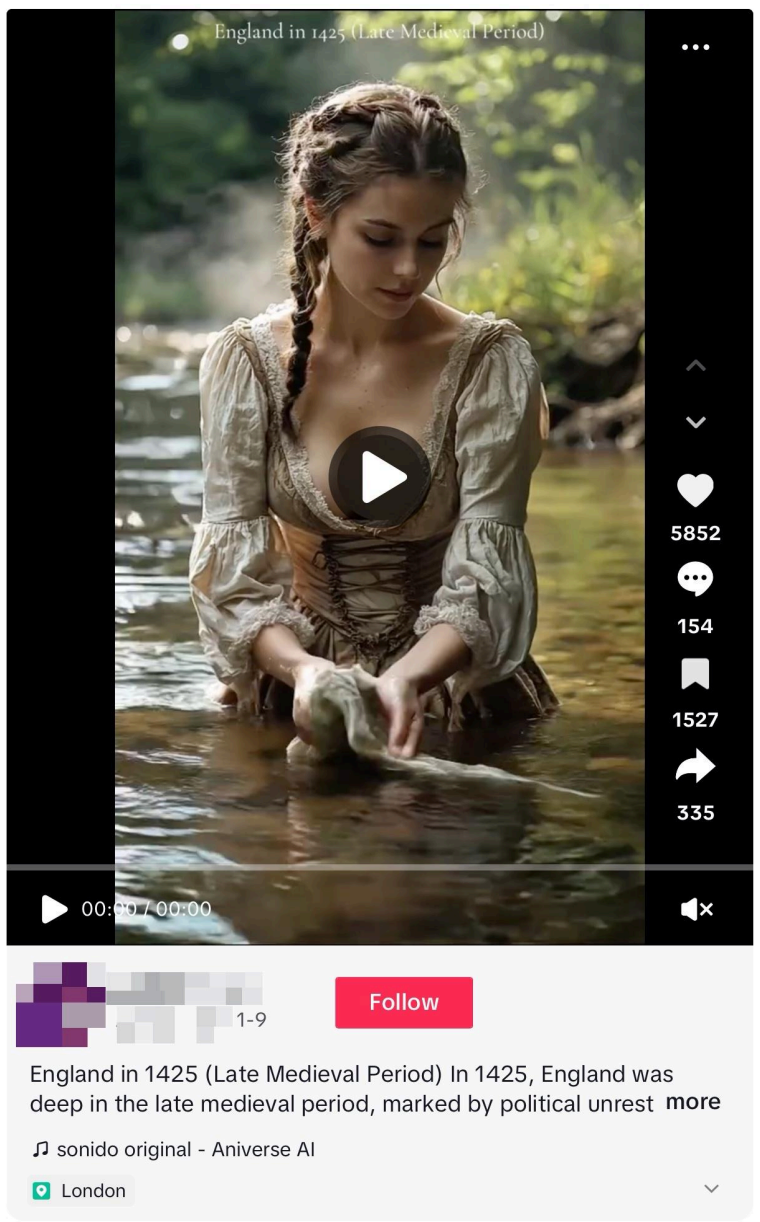

</td><td>
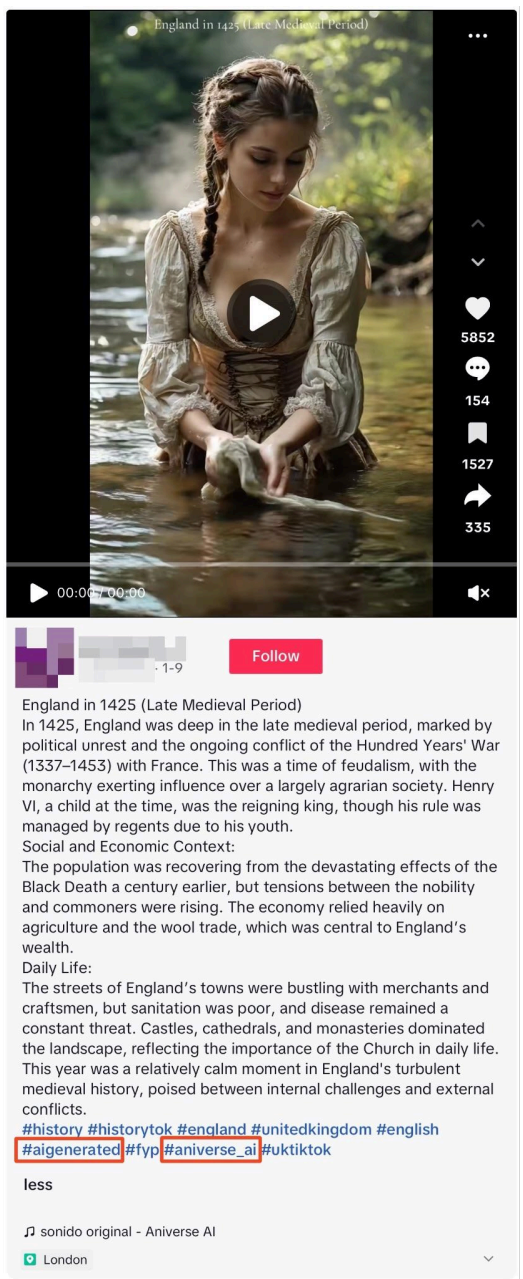

</td></tr>
<tr><td>Figure 3A [link]</td><td>Figure 3B</td></tr>
</table>

Table 4. Examples of the visibility of TikTok's alternative users' disclaimer of AI content on the platform's web version. The rectangle highlighted in red outlines the positioning of the user's AI disclosure, which, in this case, is a hashtag.

# Visibility of labels on Instagram

For Instagram, as outlined in "Our Approach to Labeling AI-Generated Content and Manipulated Media," Meta claims that it "*display[s] the 'AI info' label for content we detect was generated by an AI tool and share whether the content is labeled because of industry-shared signals or because someone self-disclosed.*" The label, upon clicking, displays more information on the post and its labelling reason.

In their follow-up on AI labels from "Our Approach to Labeling AI-Generated Content and Manipulated Media (April 5, 2024)," Meta claims that it will distinguish between AI "made" content and content that "was only modified or edited by AI tools," by containing the "AI Info" for the latter one to the post's menu. In the guidelines "When the AI info label is required," Meta states that it "*requires an AI label when content has photorealistic video or realistic-sounding audio that was digitally created, modified or altered, including with AI.*" However, Meta states that static images "that have been created or modified with AI" do not need to be labelled by the users. At the same time, they "may" receive the label assigned by Meta if AI generation or modification of the content is detected. [Meta includes ](#)the following types of content that are required to be labeled: “realistic” videos, realistic songs, and AI-generated voiceovers. Users [“may” face](#) penalties if they fail to comply. Videos with no humans in sight or in a non-photorealistic style (e.g., “resembling a cartoon”) do not require an AI label.

<table>
<tr><td colspan="3">Instagram’s label for synthetic AI content</td></tr>
<tr><td colspan="3">still image (image-based post, or a carousel post)</td></tr>
<tr><td>mobile (invisible)</td><td>mobile (visible)</td><td>browser (invisible)</td></tr>
<tr>
<td>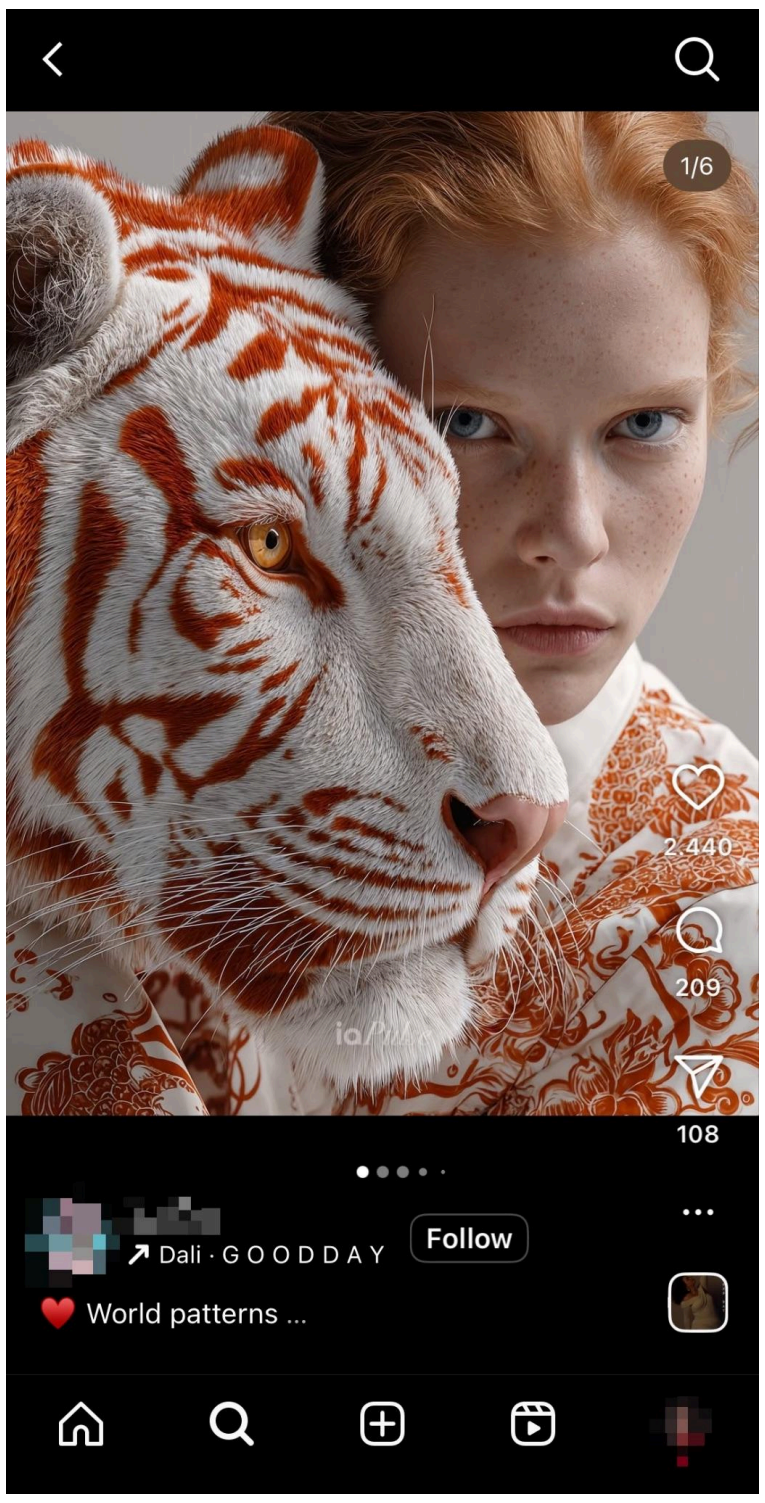
</td>
<td>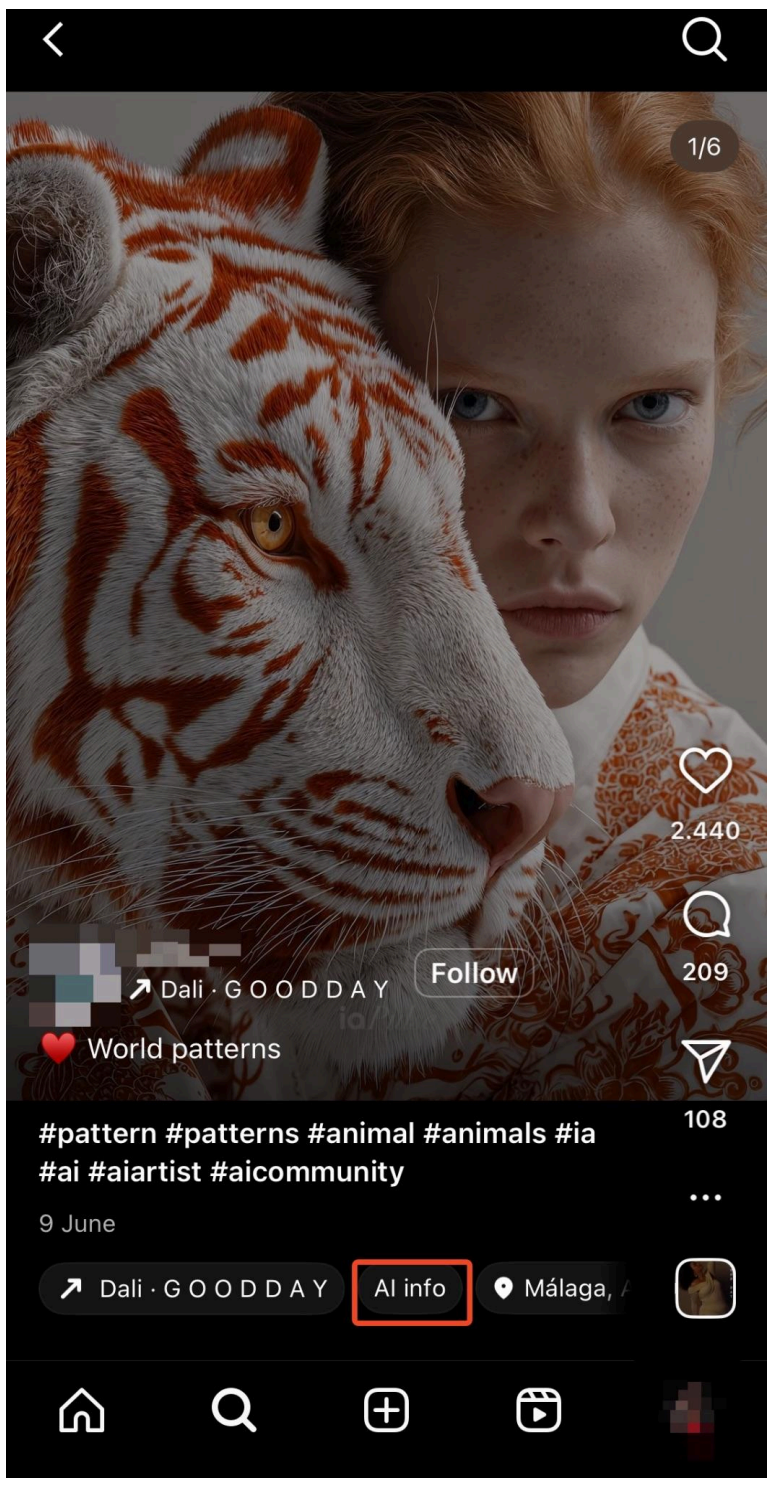
</td>
<td>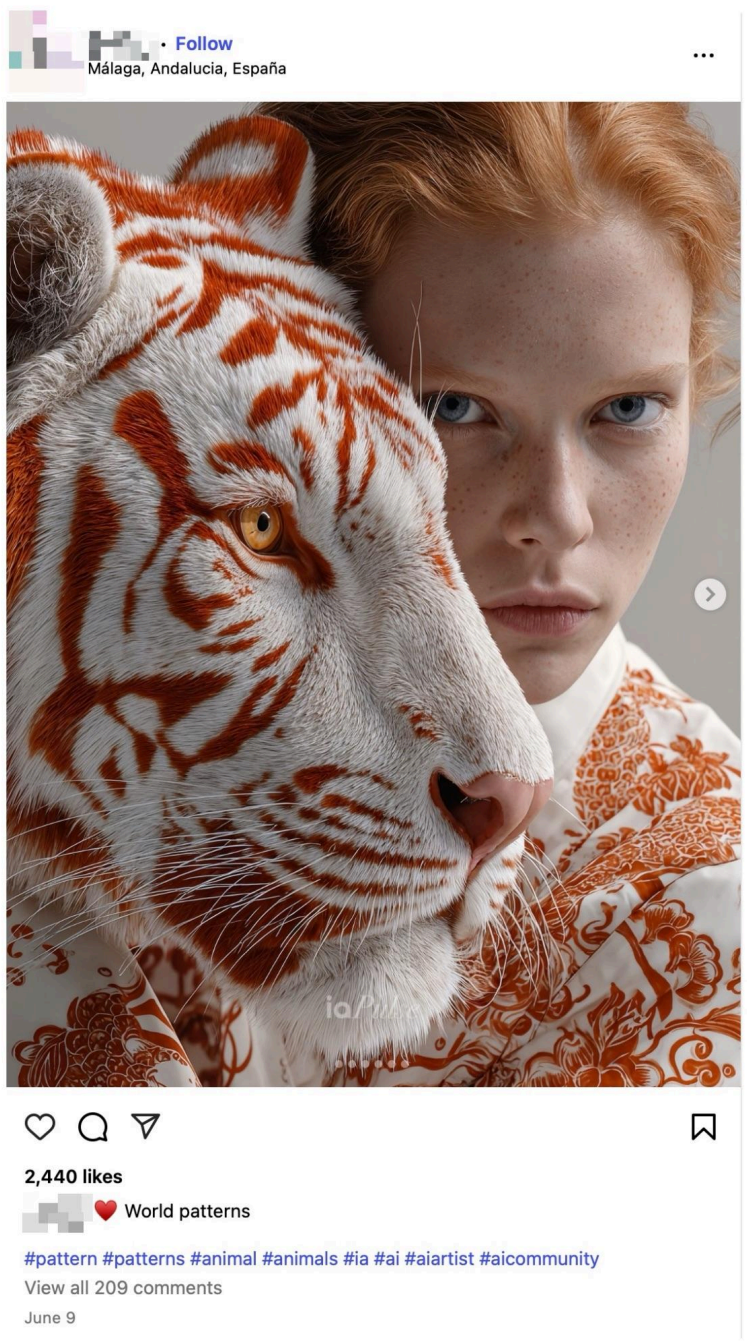
</td>
</tr>
<tr><td>Figure 4A [link]</td><td>Figure 4B</td><td>Figure 4C</td></tr>
</table>

Table 5. Instagram’s AI label positioning across the mobile and web versions of the platform. The rectangle highlighted in red outlines the positioning of Instagram’s AI label.

Unlike TikTok, Meta’s (Instagram’s) guidelines include only one labeling method, which is the official AI Label. However, we have observed the label is not visible by default in many posts, but rather hidden among other information, forcing the user to search for the AI label appearing in different parts of the screen depending on the post, and being hidden from the initial view of the content (Table 5, figures 4A-4B and Table 5, 5A-5B). Furthermore, Instagram’s AI Label was not displayed on the web version of the platform (figures 4C and 5C) during our research.

<table>
<tr><td colspan="3">Instagram's label for synthetic AI content</td></tr>
<tr><td colspan="3">moving image (Reel)</td></tr>
<tr><td>mobile (invisible)</td><td>mobile (visible)</td><td>browser (invisible)</td></tr>
<tr><td>
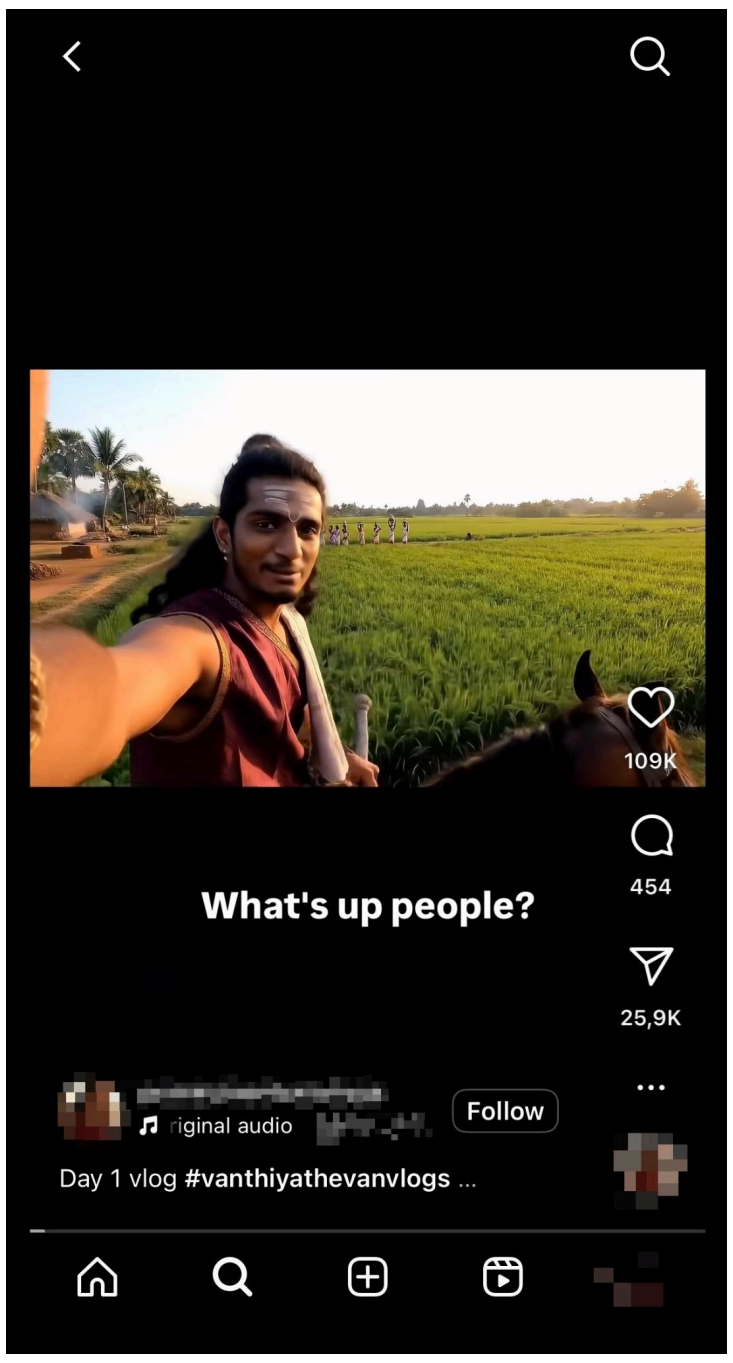

</td><td>
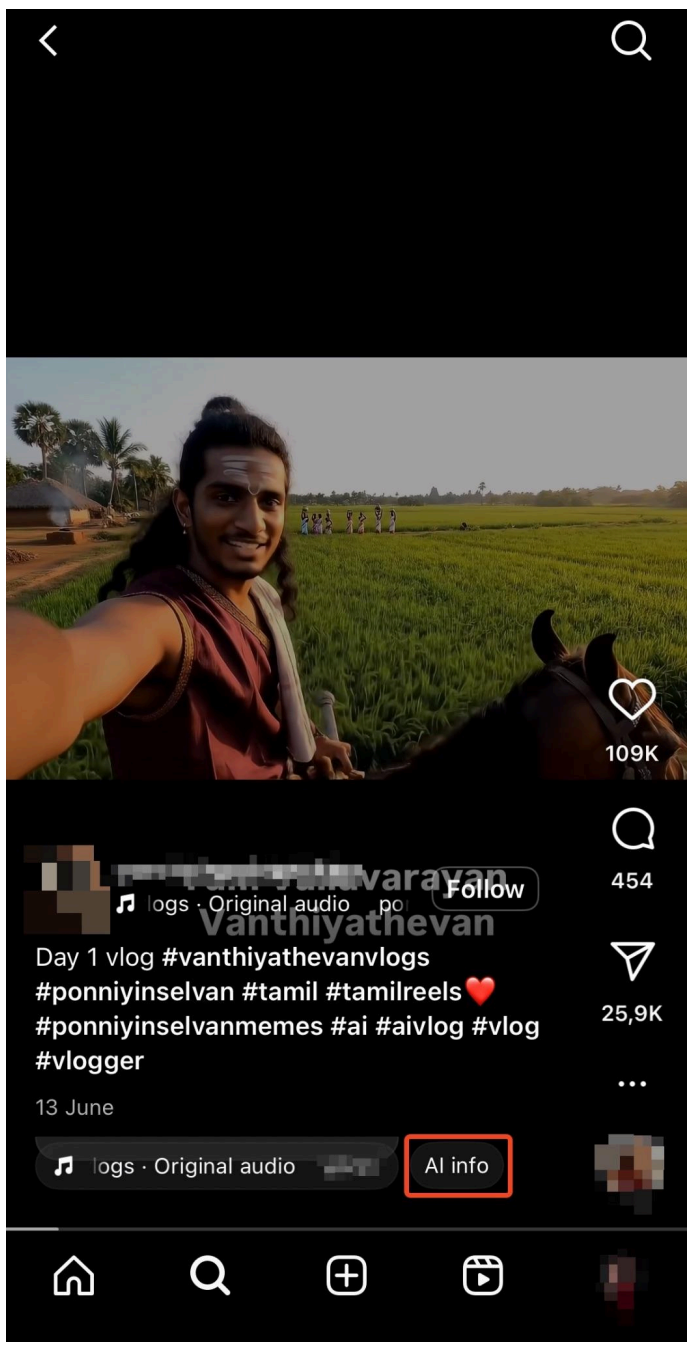

</td><td>
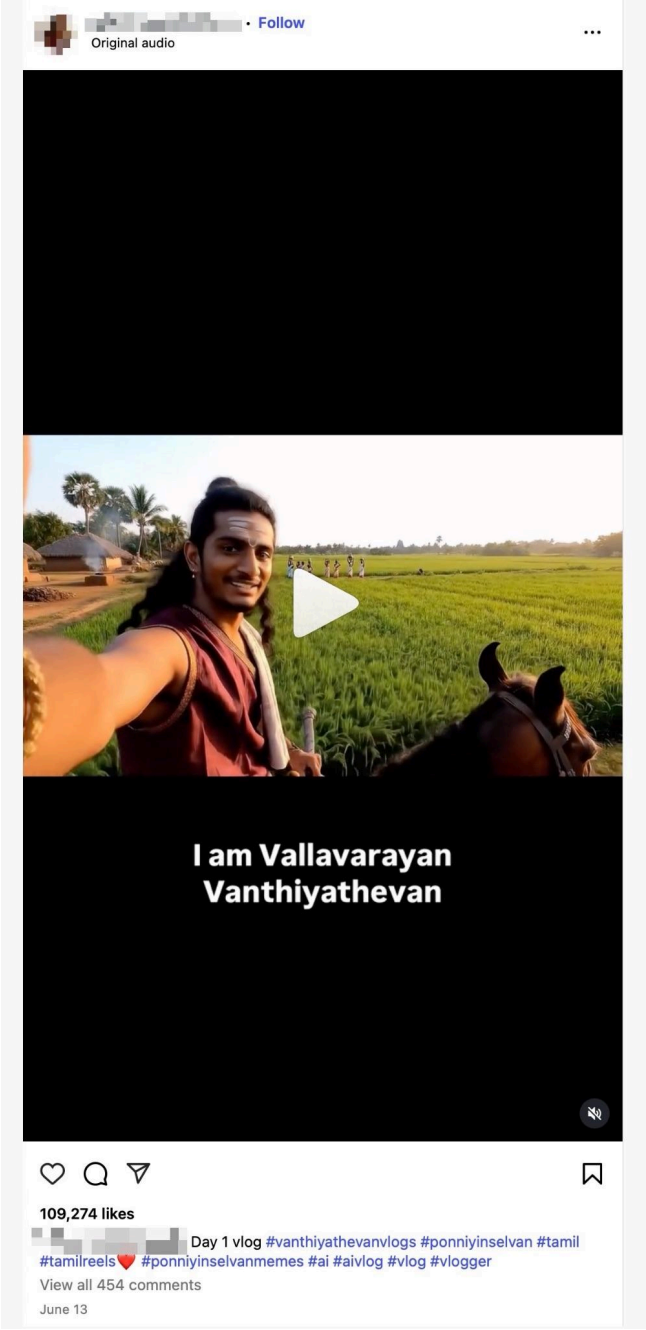

</td></tr>
<tr><td>Figure 5A [link]</td><td>Figure 5B</td><td>Figure 5C</td></tr>
</table>

Table 6. Instagram's AI label positioning across the mobile and web versions of the platform. The rectangle highlighted in red outlines the positioning of Instagram's AI label.

# Methodology

We performed a cross-platform (TikTok and Instagram) semi-automated data collection to measure the proportion of content containing synthetic AI imagery on the search results pages for various hashtags. We further compared the synthetic AI content to non-synthetic content and its engagement metrics. We assessed the presence of the AI labels across all the content encountered on both platforms. We further analyzed the accounts that shared content containing synthetic AI imagery based on their posting behaviour and the presence of synthetic AI content in their feeds.

Our approach utilized automated collection and qualitative analysis in a systematic way to objectively assess the consistency of TikTok's and Instagram's labelling mechanism of synthetic AI content, as well as the presence of synthetic AI content in the search results across hashtags on varied topics. We acknowledge that the selections of hashtags (and thus, topics) and the total number of posts we analysed (see Table 7) are by no means exhaustive, and further and more robust investigations of this kind should follow. While we investigated both the technical consistency and possible reasoning behind the presence of labeled and unlabeled content containing synthetic AI imagery across search results pages, we also performed qualitative content analysis of accounts to assess the behaviors and types of content shared by what we define as Agentic AI Accounts; however, this report does not draw any legal conclusions from this analysis, and is not intended to make any accusations against the platforms concerned.

| | Number of hashtags | Number of unique posts | Number of accounts | Number of qualitatively analyzed posts per query |
|---|---|---|---|---|
| TikTok first data collection | 13 | 1541 | 1315 | top 30 pieces of content as ordered on the search results page |
| TikTok second data collection | 13 | 1466 | 1214 | top 30 pieces of content as ordered on the search results page |
| Instagram | 13 | 1403 | 996 | top 30 pieces of content as ordered on the search results page |

Table 7. Search results across topics and hashtags for TikTok and Instagram

## Data collection

We semi-automatically collected the search results from TikTok's web version. We were logging-in by emulating users' physical residential IP addresses via a residential proxy to create new accounts. It was necessary to create accounts as TikTok only shows six search results per search term for non-logged-in users. We fetched the list of the first 100 search results for each search term and downloaded the original video files using a separate library to retain the files for annotation, to allow for annotation even if some videos were taken down between the data collection and the analysis. We collected data on TikTok on June 16 and June 26, 2025. To check the presence of TikTok's two types of AI labels and users' AI disclaimer, we automated the check for the presence of TikTok's official labels in the metadata as well as the presence of relevant keywords for users' AI disclaimer in the content of the posts' descriptions, hashtags, and stickers. We composed a list of keywords based on a qualitative exploratory reading of a sample of TikTok posts, accounting for punctuation properties to limit the number of false positives (see Table A in the Appendix). Given that we scraped results from the TikTok web version, we acknowledge the possible differences in the results displayed between the web and app versions of the platform.

On Instagram, in order to collect search results, we had to create accounts using residential proxies and retrieve search results using an unofficial API. While the API returns basic metadata on posts, it does not show whether a post is labeled as AI-generated. To collect this information we reviewed each post using an automated Android phone and the Instagram app and extracted the information from the context menu. We collected data on Instagram on June 18, 2025.

We collected search results for hashtags spanning various topic areas: politics/current affairs (#trump, #zelensky), entertainment (#history), wellbeing (#health), and culture/current affairs (#pope) (see table B in the Appendix). The search for a hashtag instead of a keyword was chosen to account for a primary platform affordance on both TikTok and Instagram that allows users to locate content related to a given topic. As the hashtags selected are not explicitly related to synthetic AI content (unlike, e.g., #genai; #brainrot), their selection allows us to see how much synthetic AI content appears in the search results when users do not search for it. The hashtag selection took place in May 2025, focusing on topics and relevant events covered in media worldwide rather than specific to only one of the three countries examined, allowing for a cross-country comparison of the search results. The choice of relevant hashtags was also driven by the popular cases in which synthetic AI imagery achieved virality and a status of a controversy, such as Donald Trump sharing synthetic images of himself and reimagined Gaza Strip, the Balenciaga Pope meme, and AI slop images framed as news footage from the Israel-Iran conflict. We therefore chose diverse topics to verify whether synthetic AI content is more prevalent in relation to specific topic areas. As the data collection included results from three countries (Spain, Germany, and Poland), the hashtags were queried in English and relevant translations. Due to the labelling and time constraints, we limited data collection to include a maximum of four hashtags per topic, including the hashtags' translations. We acknowledge that both the selection of topic areas and related hashtags, as well as the dates of data collections, might have impacted the data sample we analyzed; other hashtags and topics might give different results as to the percentages and types of synthetic AI content present in the search results on both platforms.

The following analysis for TikTok and Instagram focuses on total (=all) pieces of content, including the content duplicates; on average, we found 31% duplicates across hashtag queries and countries for TikTok and 37% duplicates for Instagram.

## Analysis to detect AI content

Given that there are no reliable tools to automatically detect the variety of synthetic AI content in the scope of this investigation, we turned to manual coding of content based on a developed codebook and a detection guidebook.[4] While some tools were found to perform relatively well in detecting synthetic AI imagery in high-quality images, much of the content shared on social media consists of blurry frames of moving images or pictures of low quality. It is also common to post a mix of synthetic AI imagery from different sources or with non-AI material. Synthetic AI content on both TikTok and Instagram was therefore coded independently by two coders, using the annotation environment Label Studio.

The choice of Label Studio as an annotation environment allowed researchers to annotate content based on the visual material itself (with no view of AI labels, hashtags, etc.) to avoid confirmation bias from seeing textual clues. Each disagreement between the two annotators was discussed on a case-by-case basis. The annotators first labelled a sample of content, upon which the disagreement rate was circa 10%. Once the annotators agreed upon the initial labelling sample, the following codings had a disagreement rate of circa 4% before reconciling the disagreements. While we took a conservative approach on coding unclear content, we acknowledge an unavoidable margin of error in the manual coding of the data.

The codebook (attached in the Appendix) served as a consistent coding scheme for detecting and labelling synthetic visual content. The codebook builds upon AI Forensics' previous report on the use of generative AI imagery in French political campaigns during the 2024 European Parliament and legislative elections across social media. The following categories were annotated:

1. GenAI - content consists of synthetic AI imagery, that is, synthetic visuals (moving and/or still images).
    1.1 photorealistic - content which is labelled as synthetic and consists of a representation of stylistic realism; it imitates the stylistic exactness of a photograph or film capture (unlike, e.g., a cartoon).
2. Partial GenAI - content includes both synthetic and non-synthetic imagery, for example, synthetic AI images intertwined with stock images.
    2.1 photorealistic - content which is labelled as Partial GenAI and consists of a representation of stylistic realism; it imitates the stylistic exactness of a photograph or film capture (unlike, e.g., a cartoon).
3. Unclear - no definitive conclusion on the nature of the content can be drawn from the information at hand.
4. Not GenAI - the content is definitely not made using generative AI tools.

[4] "AIF Guidebook: A Human Guide to Detecting Synthetic AI Imagery"

# Results of our research

Based on our research, synthetic AI content is present to a similar extent across Spain, Germany, and Poland in the search results on TikTok and Instagram (Table 8). We found considerably less synthetic AI content in the Instagram search results than on TikTok. On TikTok, the amount of synthetic AI imagery content in the top 30 results per country is similar, accounting for circa 25% of all search results; on Instagram, we found significantly less synthetic AI imagery, accounting for only circa 4,9% of all results. Over 80% of all synthetic AI imagery on TikTok and Instagram alludes to photorealistic formal qualities (figure 6). We found that the majority of synthetic AI content on TikTok consisted of posts made entirely using synthetic imagery; in the first TikTok data collection, 60 out of 183 synthetic posts contained combined non-AI imagery interwoven in the content, whereas in the second TikTok data collection, 41 out of all 161 posts contained combined non-AI and AI imagery.

| | Country | Number of all annotated results | Number of AI content items in the annotated results | Total percentage of AI content items in the annotated results | Total percentage of AI content items containing photorealistic imagery |
|---|---|---|---|---|---|
| TikTok 16.06.2025 | Spain | 240 | 55 | 26.31% | 90.91% |
| | Germany | 240 | 64 | 26.78% | 84.38% |
| | Poland | 240 | 64 | 26.89% | 85.94% |
| TikTok 26.06.2025 | Spain | 210[5] | 49 | 23.44% | 87.76% |
| | Germany | 240 | 50 | 20.83% | 80.00% |
| | Poland | 240 | 62 | 22.80% | 87.10% |
| Instagram 18.06.2025 | Spain | 240 | 5 | 2.09% | 92.31% |
| | Germany | 240 | 5 | 2.09% | 88.89% |
| | Poland | 210[6] | 9 | 2.75% | 87.50% |

Table 8. Average percentage of synthetic AI imagery and photorealistic AI imagery in the content on TikTok and Instagram, divided by country and data collection.

Given our conservative approach to coding synthetic AI imagery, we identified circa 3,65% of content on TikTok and 0.7% on Instagram as 'unsure,' meaning that there is a lack of definitive evidence to label this content as synthetic AI imagery or non-AI content. A notable number of 6 videos where annotators were unsure surfaced under the hashtag #zelensky due to the existence of several deepfakes, some of which the annotators managed to identify, while with other suspicious cases, the annotators remained conservatively wary if no official source was available to confirm that the content was not manipulated.

[5] Due to a misconfiguration in the second data collection, no data was collected on #trump in Spain on Jun 26.

[6] Similarly we did not collect data on #history in Poland on Instagram because of a technical failure.

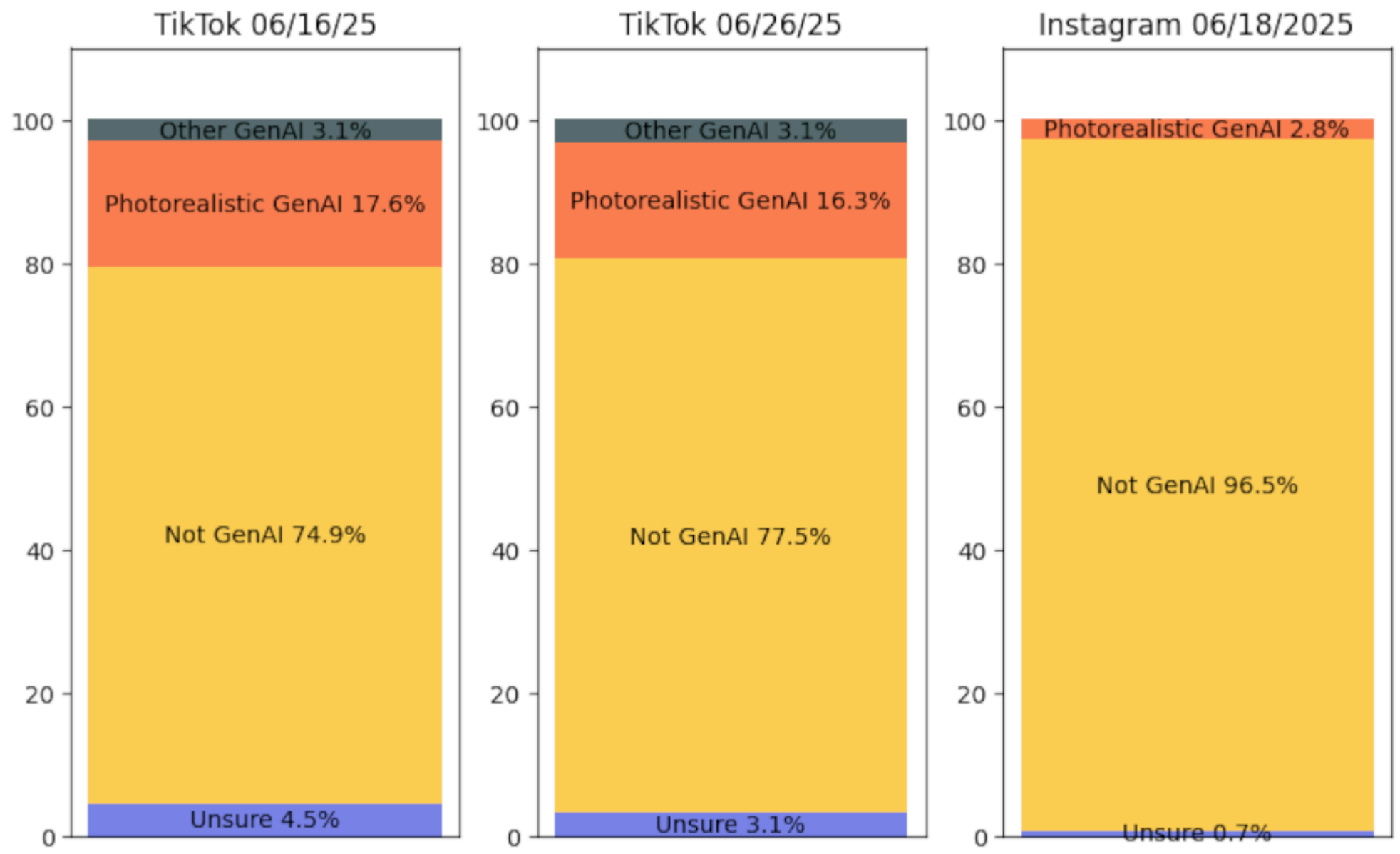

Figure 6. Comparison of percentages of non-AI content, content labeled as ‘unsure,’ and content containing photorealistic synthetic AI imagery across all data collections.

# Outlining AI Slop Topology: examples of synthetic AI content

In the following section, we briefly describe and categorize examples of synthetic AI content we encountered across TikTok’s and Instagram’s search results as a mapping (and, thus, by definition, an incomplete categorization) of AI Slop topology. This topology follows the order of topic areas we queried for, outlined by the relevant hashtags, with the examples chosen as the most notable cases by content, form, and possible implications for the users who encounter them. We note that many of the following examples of synthetic AI imagery are not hashtag or topic-specific but span across the platforms as certain conventions and subdivisions of AI content, hence the topological character of this classification. For a quantitative distribution of synthetic AI imagery content and its related account types, see the [Taxonomy of Agentic AI Accounts](#) section.

# AI slop as authority satire

**Infantilization** and **Nonrelistic Situations**

Most synthetic AI imagery in the content searched for with hashtags #trump and #zelensky exemplified AI slop's aesthetics and subject matter for the purposes we can broadly term as (political) satire aimed at mocking authority figures. Two types of such AI slop as authority satire include infantilization and nonrealistic situations. Infantilization describes content where political or otherwise authoritative figures are shown as toddlers or adults with hyperbolized childlike features (figures 7A and 7B). By nonrelistic situations, we describe photorealistic AI slop which retains some level of realistic rendering of human figures, but positions them in improbable and exaggerated situations (see figures 7C and 7D). Such posts can thus be considered examples of political satire and playful exploration of the tradition of internet memes and political cartoons. Most of the AI slop as authority satire content is photorealistic, yet its subject matter makes it highly implausible to appear as 'real.'

<table>
<tr>
<td>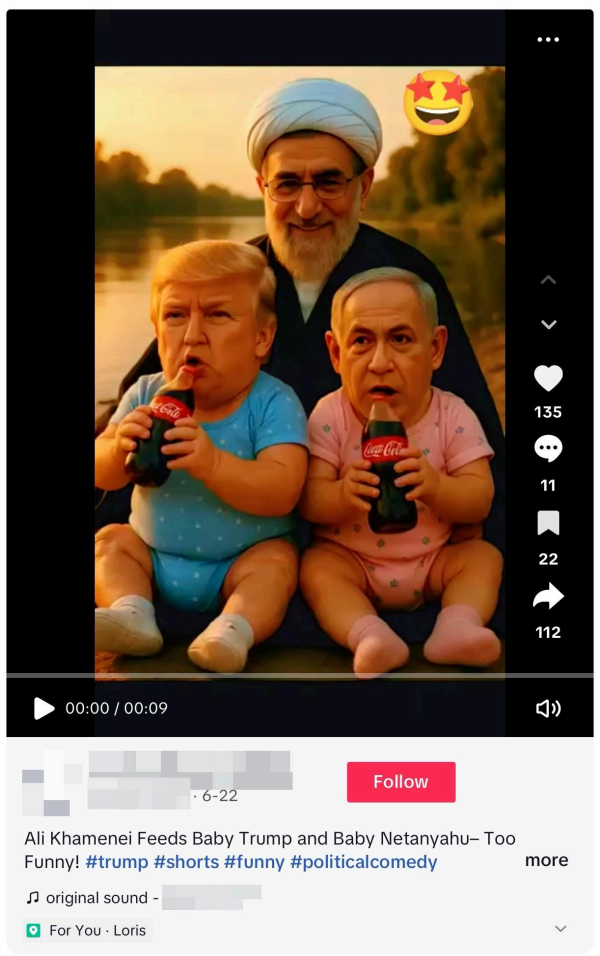
</td>
<td>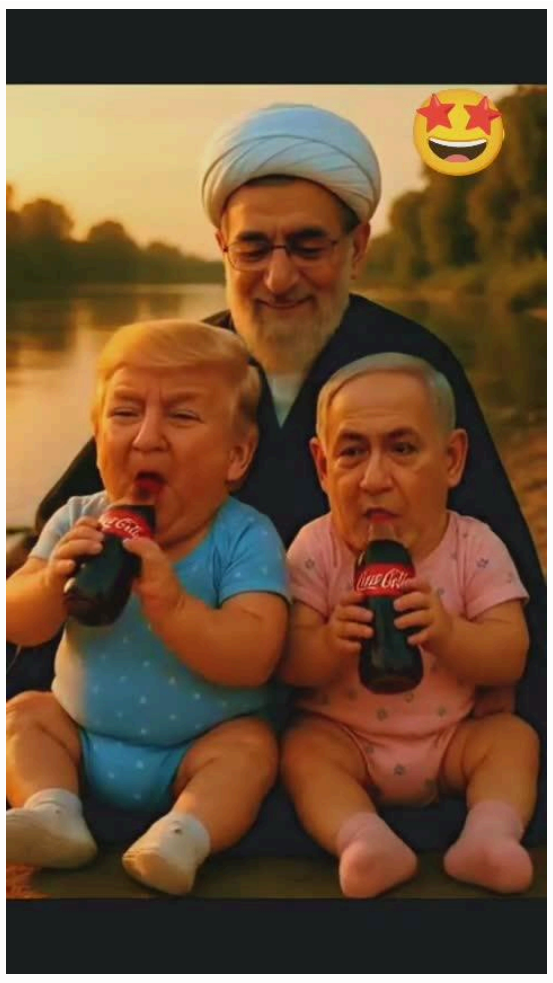</td>
<td>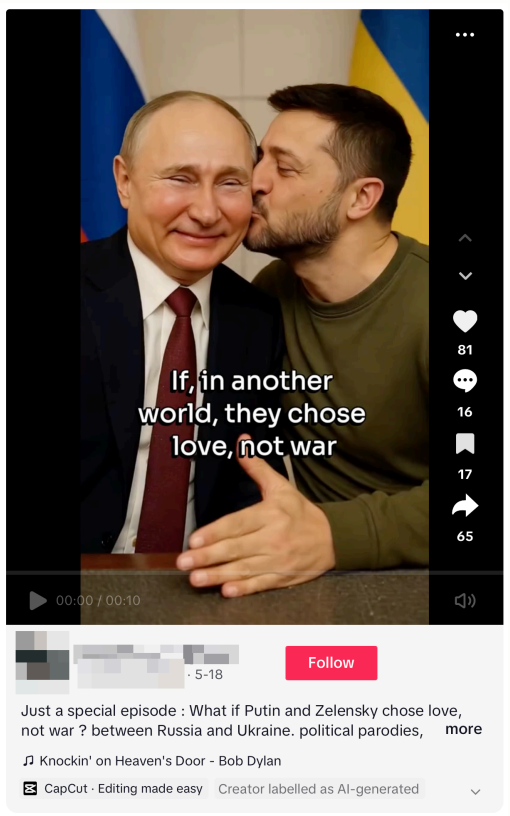
</td>
<td>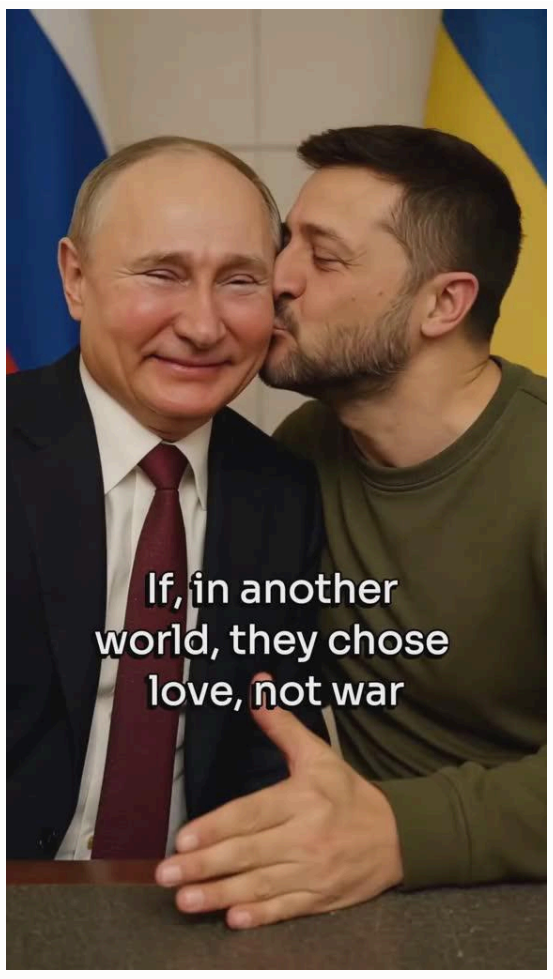
</td>
</tr>
<tr>
<td>fig. 7A [link]</td>
<td>fig. 7B</td>
<td>fig. 7C [link]</td>
<td>fig. 7D</td>
</tr>
</table>

Table 9. Examples of content in TikTok's search results for #trump ( 10A-10B) and #zelensky (7C-7D).

# Synthetic AI imagery as plausible (dis)information

**Synthetic Interviews** and **Synthetic Citizen Journalism**

We came across two potentially problematic examples of the undisclosed (fig. 8A) and disclosed (figs. 9A-9C) use of synthetic AI imagery as plausible (dis)information. These examples appeared within the top 30 results on TikTok's and Instagram's search results pages under #trump. Each of the two examples falls under one of the two broader types of AI slop: synthetic interviews and synthetic citizen journalism. By synthetic interviews, we describe a type of AI slop content which replicates the characteristics of interviews, especially the format of media interviews with ‘people on the streets.’ By synthetic citizen journalism, we define a type of AI slop that simulates events (such as explosions, natural disasters, or similar), emulating the visual language of amateur footage ‘on the ground,’ as if it were recorded with a smartphone by a witness of the scene. The TikTok video exhibiting the qualities of synthetic citizen journalism type (fig. 8A), is a misattribution of another synthetic clip showing what appears to be a nuclear-like explosion. The clip was initially shared as a YouTube Short on a YouTube channel specializing in synthetic AI content of explosions. Originally, the content creator disclosed the syntheticity of the clip by including the disclosure in the description and hashtags of the Short, alongside a label-like disclosure visible on YouTube (fig. 8B). The clip was later reshared across social media platforms including TikTok as real footage from Iran.

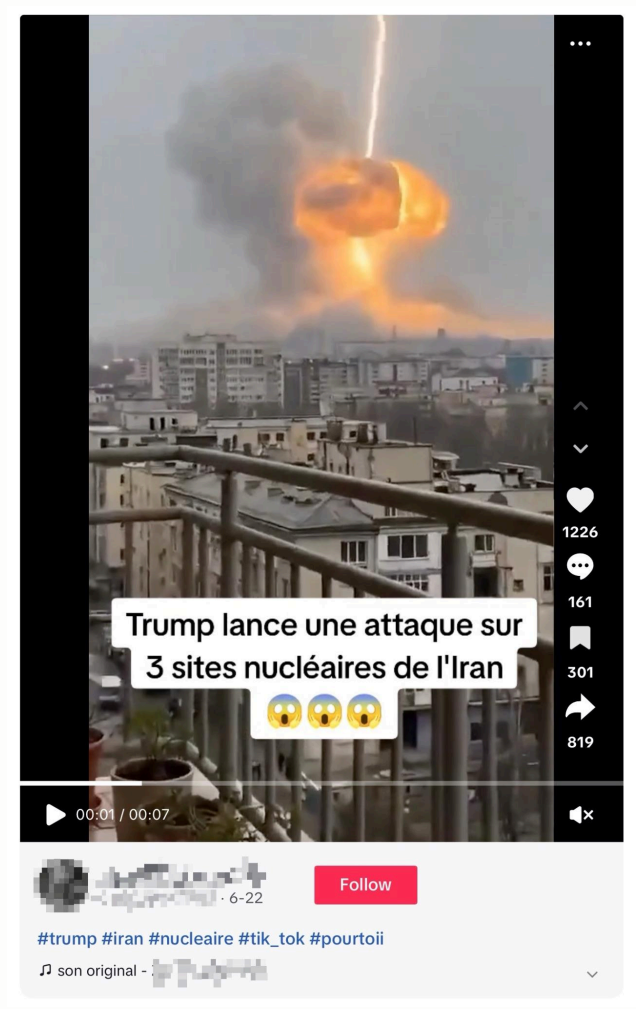

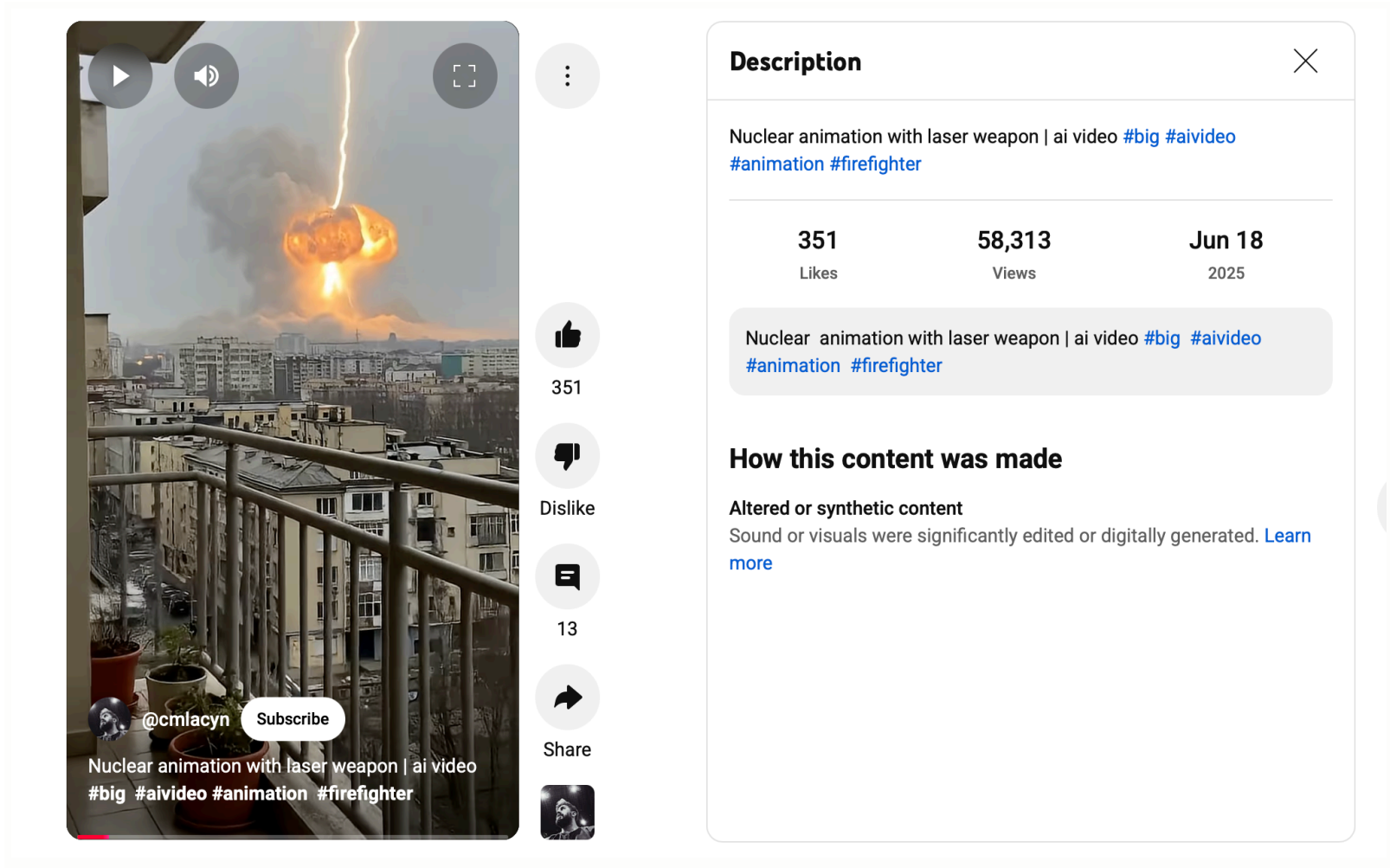

| | |
|---|---|
| fig. 8A [link] | fig. 8B [link] |

Table 10. A screenshot of the unlabelled and undisclosed reshared synthetic AI clip of an exposure on TikTok (figure 8A) and a screenshot of the original clip posted on YouTube as a YouTube Short with a correct AI disclosure (figure 8B).

The second problematic piece of content, an example of synthetic interviews type, comes from Instagram (figures 9A-9C). It is a Reel consisting solely of synthetic AI imagery in a form of phorealistic and plausible interview clips with citizens in Iran after an assumed fall of the Iranian regime. The content is framed in the description as satire and contains the hashtag “#ai,” but the AI label, which Meta requires in this case, is missing. The AI disclosure in the description of the post is hidden from the view unless the user ‘expands’ the description of the post to see the list of hashtags; instead, the post’s description starts with a possibly leading “EXCLUSIVE: Interviewing Iranians.”

<table>
<tr>
<td>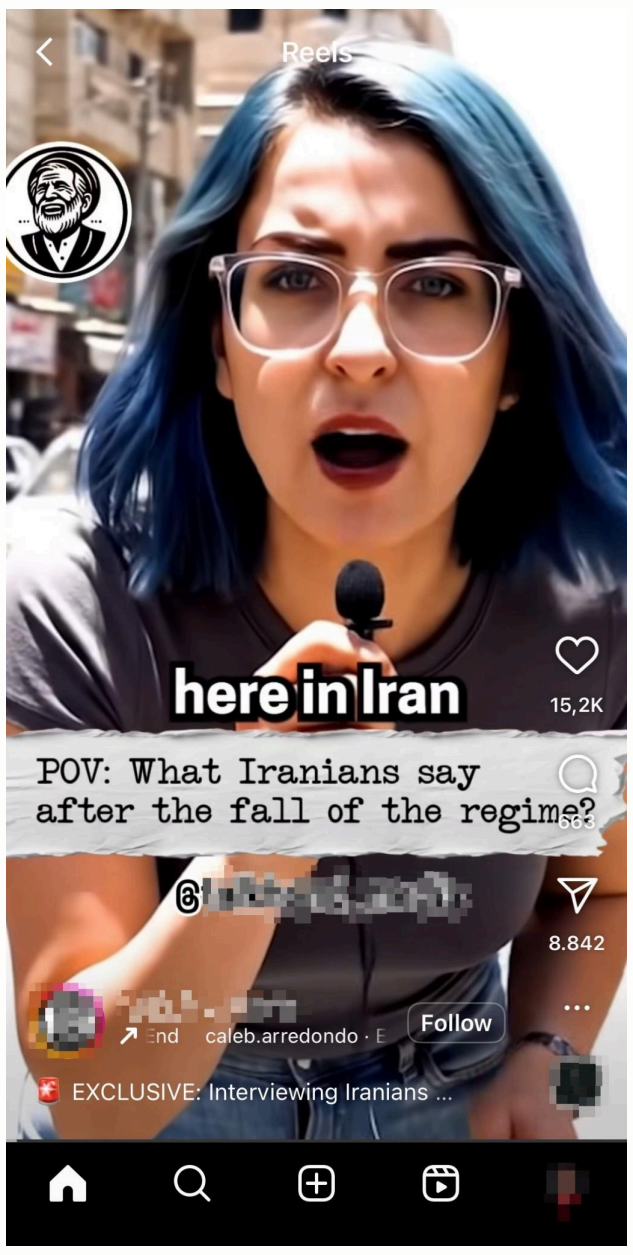
</td>
<td>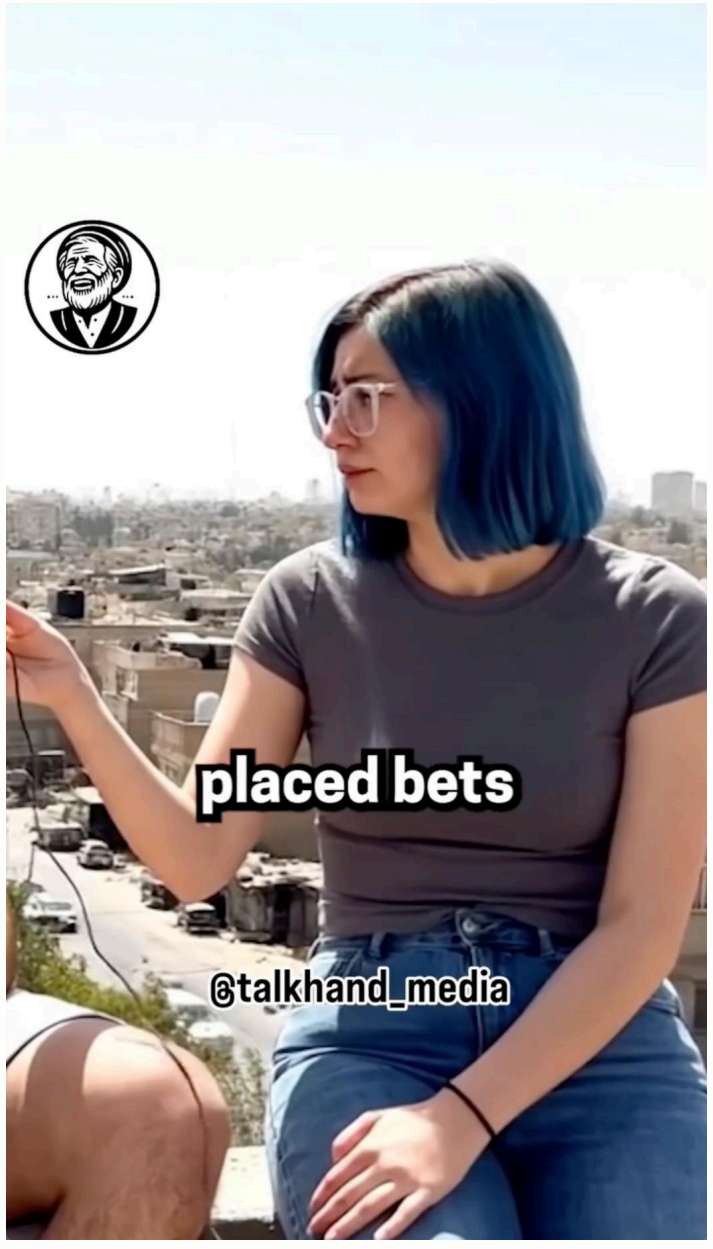
</td>
<td>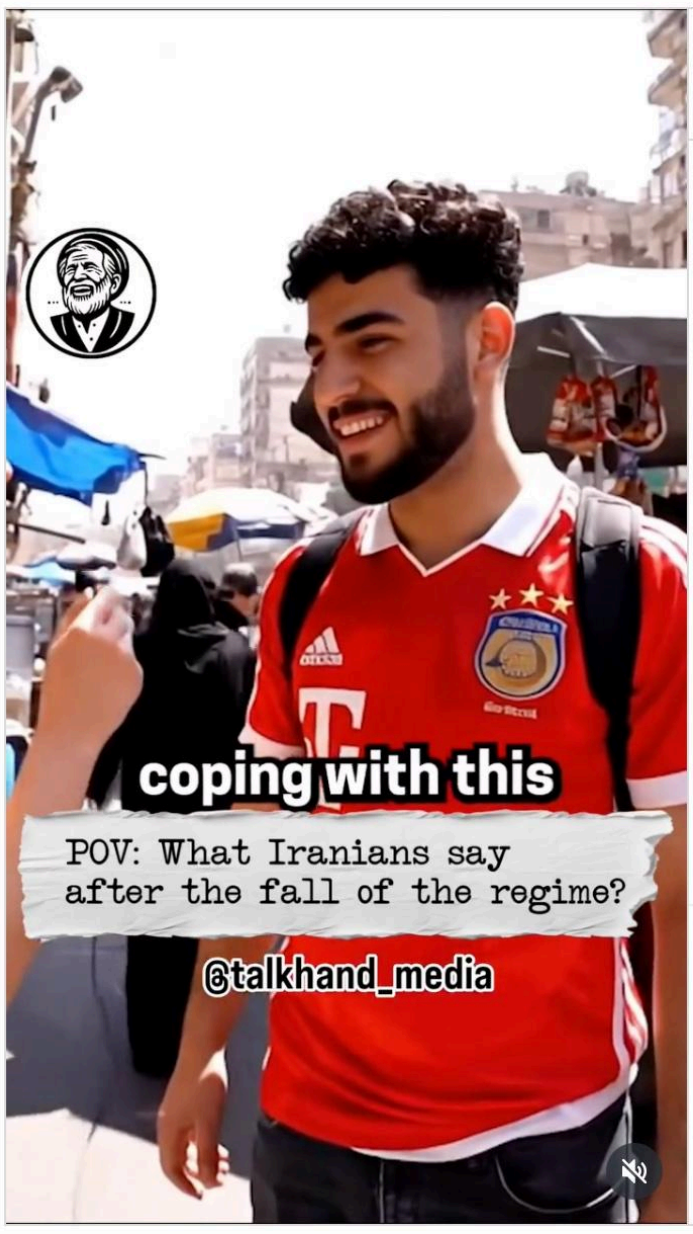
</td>
</tr>
<tr>
<td>fig. 9A [link]</td>
<td>fig. 9B</td>
<td>fig. 9C</td>
</tr>
</table>

Table 11. Screenshots (figures 9A-9C) depicting the Instagram Reel showing fake photorealistic interview clips 'from' Iran, made entirely using synthetic AI imagery and lacking proper labelling.

## Synthetic storytelling and illusive historicizing

**Impossible photography** and **synthetic Point of View (POVs) footage**

The synthetic content in the search results for #history and its relevant translations contained AI slop materials which belong to the category we term synthetic storytelling and illusive historicizing, as they attempt to visualize particular historical events or imagined situations, resulting in some historical distortions (figures 10A-10D). We distinguish between two types of this category of AI slop: impossible photography and synthetic Point of View (POVs) footage. By impossible photography, we refer to examples of synthetic AI imagery showing photorealistic renderings of scenes, historical characters, or locations to illustrate actual historical facts (figures 10A and 10B). The use of impossible photography is also used in a particular way in the context of AI slop that 'illustrates' various sensational stories with a mix of regular and synthetic content, a case described more in detail in the Taxonomy of Agentic AI Accounts section as 'hybrid' Agentic AI Accounts. Synthetic Point of View (POVs) footage exemplifies AI Slop images and videos that are organized around the 'POV' visual storytelling, often containing the "POV: ..." text in its description or as text on the visual content itself (figures 10C-10D). Such content is meant to visualize 'what it would be like to be' in a specific historical place or during a historical event, often emulating a first-person perspective.

| | | | |
|---|---|---|---|
| 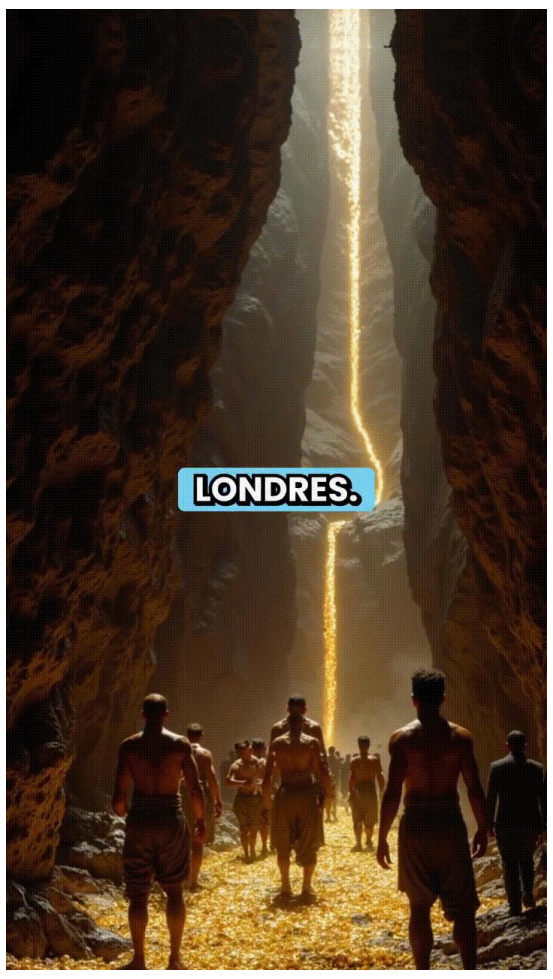 | 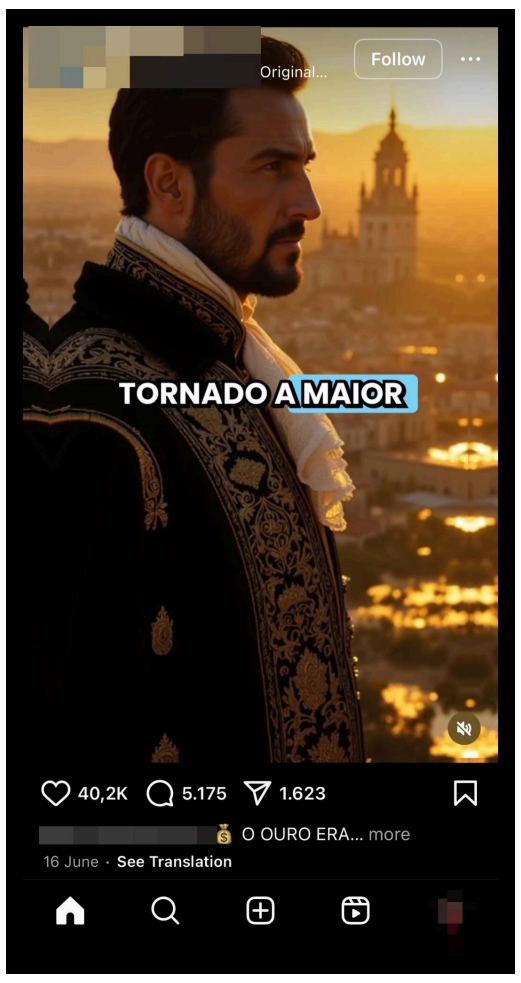 | 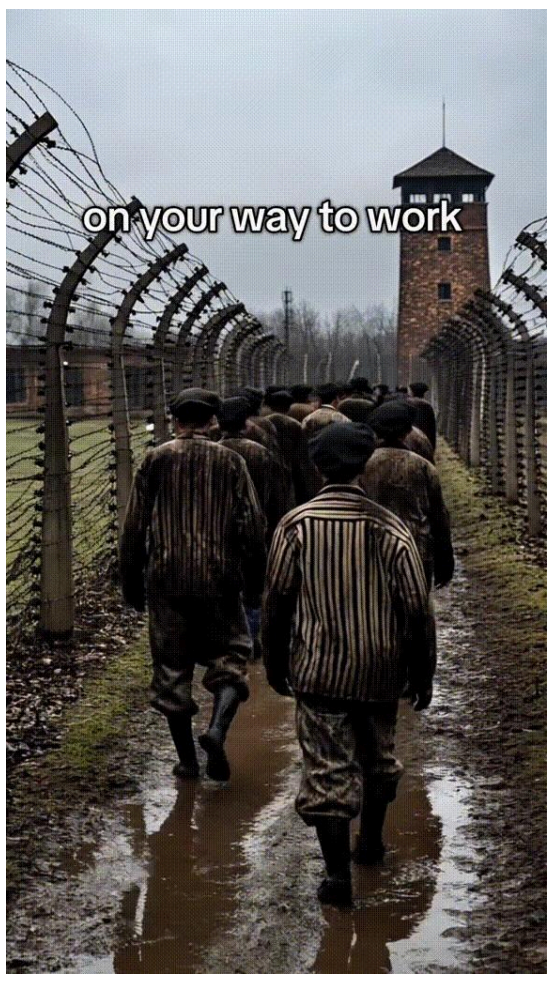 | 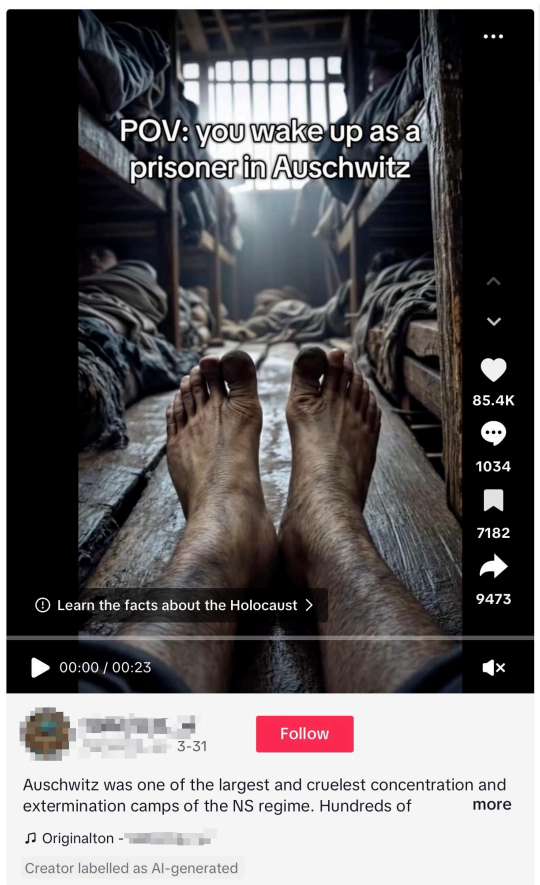 |
| fig. 10A [link] | fig. 10B | fig. 10C [link] | fig. 10D |

Table 12. Examples of synthetic AI imagery for #history on Instagram (13A-13B) and TikTok (13C-13D)

# (AI) Slop as Stock images

**Generating ideas as stock images** and **generating the invisible body**

In the search results for #health and its relevant translations, many examples of posts containing synthetic AI imagery included what we term AI Slop as Stock images. We differentiate between two approaches to using AI Slop as Stock images: generating ideas as stock images and generating the invisible body. By generating ideas as stock images, we refer to content that is meant as a visual aid in illustrating specific ideas, abstract concepts, and metaphors, often reproducing stock images' visual qualities and format (figures 11A-11B). Generating the invisible body uses both photorealistic and animated synthetic imagery (figures 11C-11D) to visualize processes that occur in the body or on a microscopic level.

| | | | |
|---|---|---|---|
| 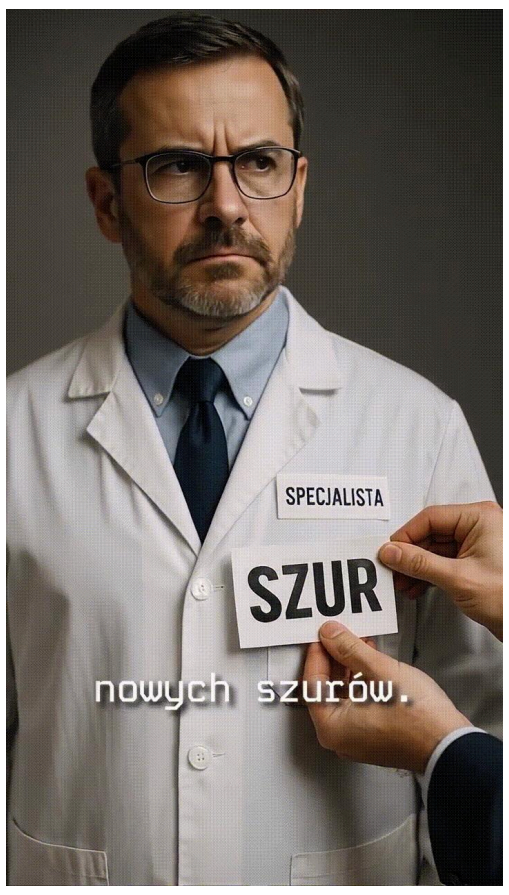 | 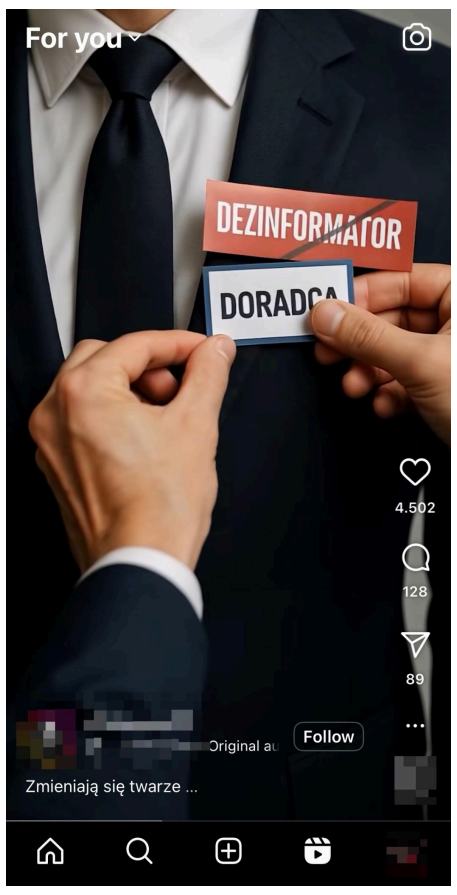 | 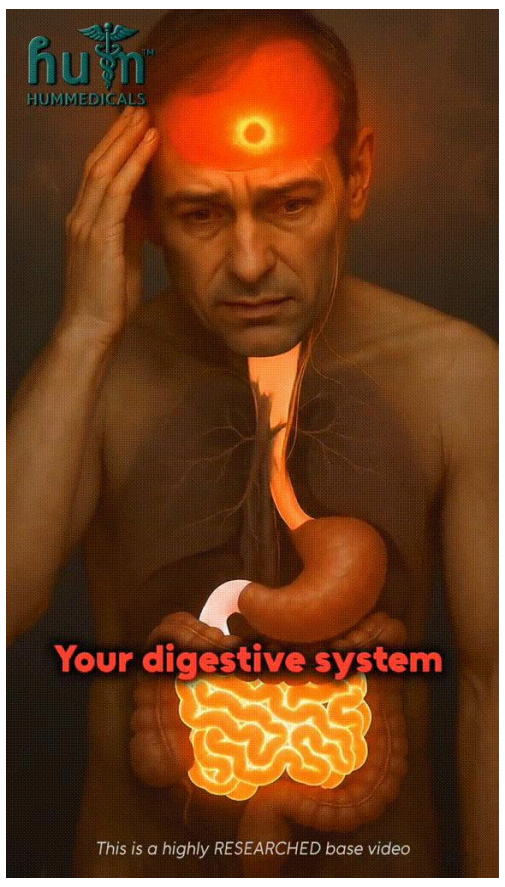 | 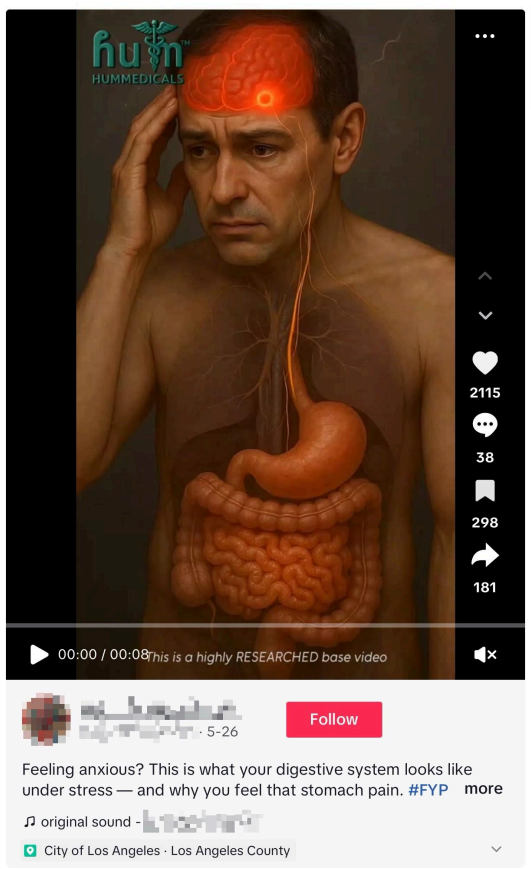 |
| fig. 11A [link] | fig. 11B | fig. 11C [link] | fig. 11D |

Table 13. Examples of synthetic AI imagery appearing under #health on Instagram (11A-11B) and TikTok (11C-11D)

# Rendering synthetic past-futures

**Reanimating surrogate clips** and **generating synthetic surrogate clips**

Most of the content we encountered in the search results for the hashtag #pope, which contained synthetic AI imagery, was of a similar category we refer to as rendering synthetic past-futures. This category is exemplified by two types of AI slop: reanimating surrogate clips and generating synthetic surrogate clips. By reanimating surrogate clips, we mean that such AI slop content is mixing regular and synthetic imagery by 'bringing to life' or animating real photos using generative AI tools (figures 12A-12B). Generating synthetic surrogate clips constitutes examples of synthetic AI slop content showing real figures in historically impossible settings or (often silly) actions (figures 12C-12D).

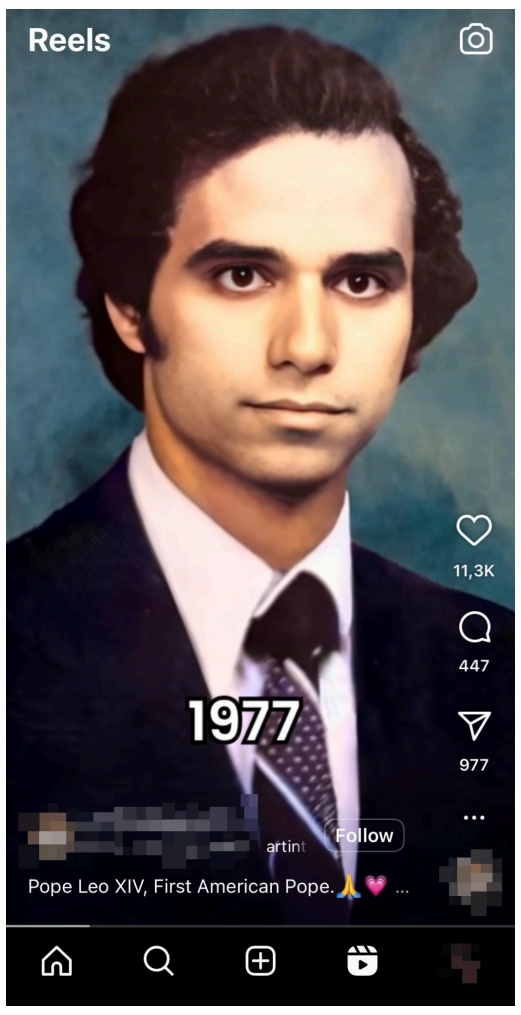

fig. 12A [link]

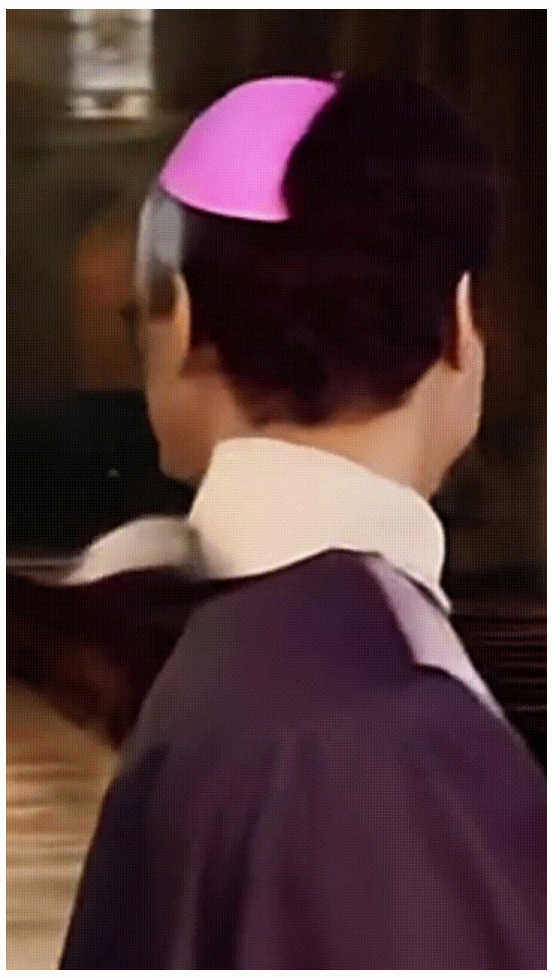

fig. 12B

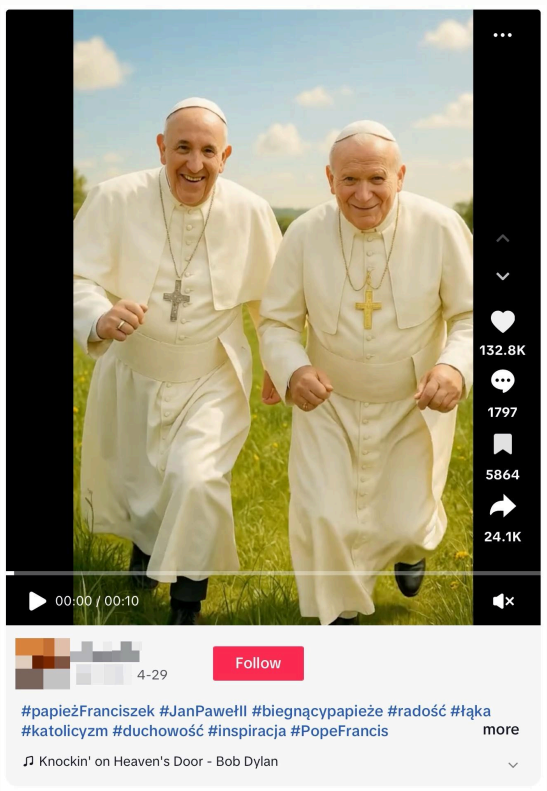

fig. 12C [link]

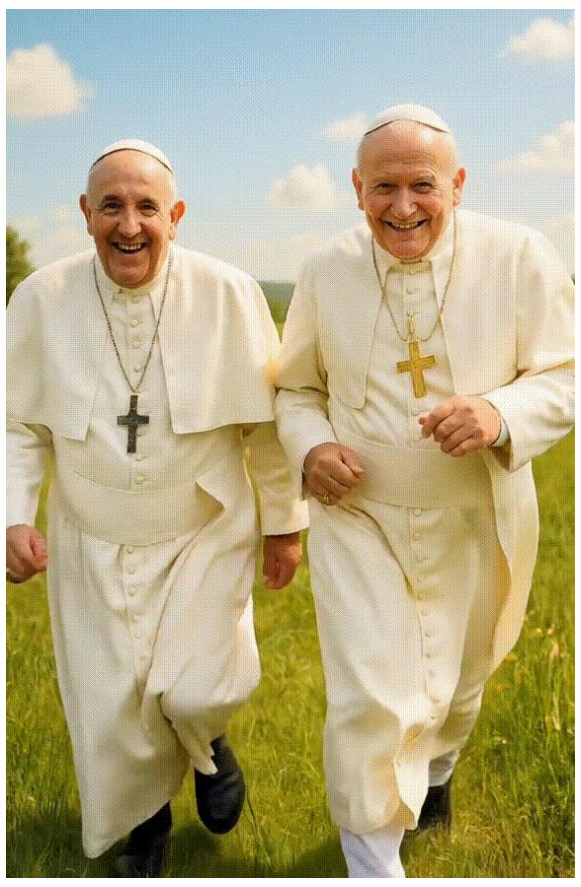

fig. 12D

Table 14. Examples of synthetic AI imagery appearing under #pope on Instagram (figures 12A-12B) and TikTok (figures 12C-12D)

# Generative AI content on Instagram

We found significantly less content containing synthetic AI imagery in Instagram's search results compared to TikTok.

Given the limited results of synthetic AI imagery content in the search results on Instagram, we cannot derive statistically significant trends between synthetic AI imagery and the country, language, and topic queried for. The only cases of synthetic AI imagery content among the top 30 results, as ordered per the search page results on Instagram, were found for the following queries: for hashtags related to history (#historia), we found two cases in Poland and two cases in Spain.

For #history, we found two cases in Spain and two in Germany. For the hashtags related to the Pope, we found two cases in Poland (#papież). For the English hashtag #pope, we found two cases in Spain, one case in Poland, and one case in Germany. For #trump, we found one case in Spain and one case in Poland. For #Zelensky, we found three cases in Spain and in Poland, and two cases in Germany.

There are several explanations for the low number of synthetic AI imagery in Instagram's search results. Synthetic AI imagery may be more widespread on TikTok than on Instagram, as TikTok might be a more profitable platform for monetization of this type of content. TikTok, especially its For You Page (FYP), is optimized for one-off virality that often recommends content from content creators the user does not follow or who have not built a followers audience. Such affordance, alongside the fact that the dominant format of the platform, TikTok's short videos, might favor synthetic AI imagery. As such, it is likely that if this analysis focused solely on Instagram's Reels (a short-video format similar to TikTok videos 'tiktoks') rather than all Instagram content, the number of synthetic AI imagery would be higher.

It is also possible that Instagram has a different algorithm for surfacing non-personalized search results context compared to TikTok. That would suggest that TikTok and Instagram have taken different approaches to curating search results, which impact the presence of synthetic AI content. However, given that many memes and social media trends tend to first emerge on TikTok and then spread on other platforms, it might be that Instagram will face the same scale of the synthetic AI imagery problem as TikTok sees only later on.

# Generative AI content on TikTok

| Topic area | Hashtag (sum of results in English and relevant translation) | Average percentage of synthetic AI imagery per topic area and country | | | | | |
|---|---|---|---|---|---|---|---|
| | | Spain | | Germany | | Poland | |
| | | *16.06.* | *26.06.* | *16.06.* | *26.06.* | *16.06.* | *26.06.* |
| politics / current affairs | #trump #zelensky | 5.56% | 12.5% | 23.73% | 10.00% | 9.09% | 27.59% |
| entertainment | #history | 28.0% | 27.59% | 39.29% | 40.35% | 18.42% | 22.86% |
| wellbeing | #health | 10.00% | 5.41% | 16.95% | 13.33% | 13.16% | 2.63% |
| culture / current events | #pope | 4.76% | 6.25% | 23.53% | 21.05% | 35.71% | 37.84% |

Table 15. Average percentage of synthetic AI imagery in the content on TikTok divided by topic area.

We found synthetic AI imagery across all countries and topic areas, which suggests that it is a widespread and cross-national phenomenon. Most synthetic AI imagery was found in search results for the hashtag #history and its respective translations across the three countries examined. The highest number of synthetic AI imagery per single hashtag was found for #pope and its translation in Poland, over 53% in the second data collection on TikTok. The same hashtag, also with its translation in Spanish, resulted in zero cases of synthetic AI imagery in Spain. The lower number of synthetic AI cases in Spain might have been caused by the fact that the Spanish #papa resulted in many pieces of content related to 'father' and 'fatherhood,' which contaminated the data sample with content unrelated to the topic of the Pope.

We note that if all content appearing in the top 30 results on TikTok search pages is analyzed, posts containing synthetic AI imagery are younger. The average number of days since post publication is lower than for non-AI content, but this is likely related to the fact that GenAI content has not been around for that long, not because the content is favored by any algorithm (see Fig. 13).

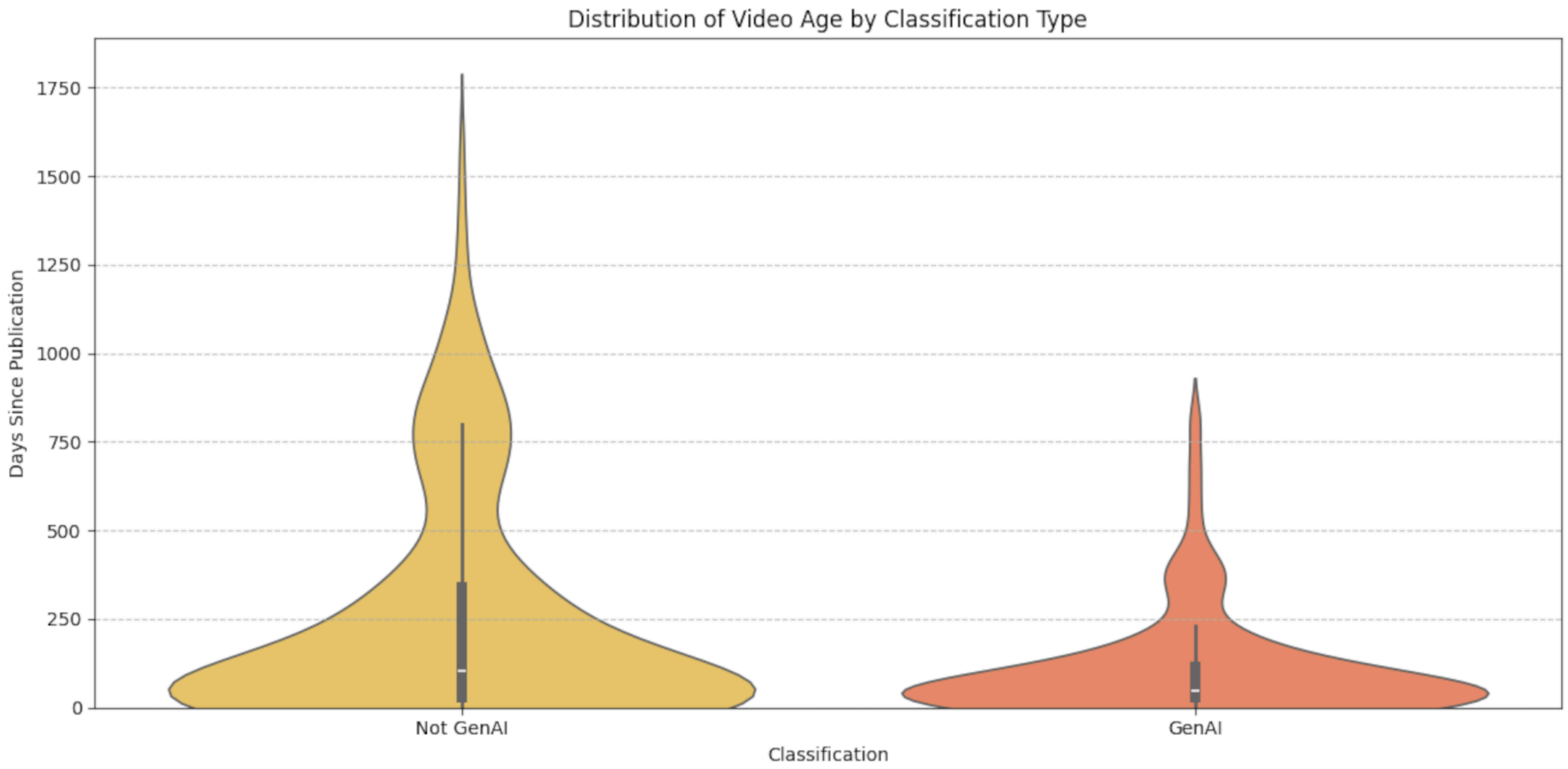

Figure 13. Distribution of video age in the search results on TikTok as a violin plot. The search results mostly return videos younger than 3 months, but the maximum age of GenAI-labeled videos is much higher.

Checking for correlations, we found that content containing synthetic AI imagery tends to be statistically significantly shared more often than other content on TikTok, approximately 1.15 times as often. This means users use the share function to send a link of a post to someone else, e.g. via messenger, more often for synthetic AI content than for non-AI content. This is true for both content containing synthetic AI imagery labeled as AI content and content not labeled and disclosed as such.

# ‘AI Label’ on TikTok and Instagram

The current labelling mechanism for AI content on Instagram and TikTok seems insufficient. We found that around half of all synthetic AI content is not labelled at all: on TikTok, 58 % of AI content in the first data collection and 41% of AI content in the second data collection were not labelled (see Table 9). On Instagram, only three of the thirteen posts containing synthetic AI imagery were labelled as such.

It is also worth noting that of all the labeled videos on TikTok, all but one video were labeled by the creators. We found only one instance with the label “AI-generated” instead of “Creator labeled as AI-generated,” indicating that the label was added by the platform. This is in contrast to the active stance TikTok took in their DSA risk assessment report, where TikTok emphasized having labeled over 4 million AI-generated videos within a year.

| | | Percentage of labelled unique content in 16.06.2025 data | | Percentage of labelled unique content in 26.06.2025 data | |
|---|---|---|---|---|---|
| | | TikTok labels | User disclaimer | TikTok labels | User disclosure |
| TikTok | Spain | 42.86% | 21.43% | 61.54% | 0 |
| | Germany | 15.52% | 22.41% | 22.45 | 22.45% |
| | Poland | 13.79% | 17.24% | 6.45 % | 16.13% |

Table 16. Percentage of unique pieces of content containing synthetic AI imagery and labeled as such on TikTok.

As mentioned before, the browser version of Instagram does not display AI labels at all. While TikTok’s AI labels are visible to users on both mobile and browser versions of TikTok, which is a positive development compared to Instagram’s limitation of label visibility on the web, TikTok’s alternative ‘disclousers’ of AI content, which are contained as posts’ hashtags or descriptions, are often hidden from the users’ view when encountering the post (see Table 2 in the section Labelling AI content on social media platforms). Similarly, Meta’s AI labels are often not visible to users unless they click to ‘expand’ the post description (see Table 3 in the section Labelling AI content on social media platforms). We found that on TikTok, content that only partially contains synthetic AI imagery, meaning it consists of a mix of both synthetic and regular imagery, is labelled in only 17% of cases we found, whereas content that is entirely made of synthetic AI imagery is labeled as such in 53% of cases based on our data.

We found no statistically significant difference between labelled and unlabelled synthetic AI content and engagement metrics on TikTok. Similarly, the positioning of content in TikTok's search results did not show signs of a strong correlation between content type, its labelling status, and higher or lower position.

# Taxonomy of Agentic AI Accounts

In our dataset, over 80% of the synthetic AI content on TikTok and almost 15% on Instagram was posted by Agentic AI Accounts.

What we term as 'Agentic AI Accounts' or 'AAA' in our datasets refers to accounts that post synthetic AI content by either partial or fully automated means in a pipeline which has been outlined in recent investigations into AI slop.[7] 'Agentic AI Accounts' are a new kind of automated social media account. These accounts leverage generative AI tools to automate content creation, facilitating rapid, repetitive testing of platform algorithms and audience interests to identify engaging content for posting. This technique aims to 'game' content recommendation and ordering algorithms through the sheer quantity of posted content to increase the likelihood of content going viral.

With respect to the content type and posting behavior, Agentic AI Accounts employ AI tools to either partially or fully automate their content creation and posting pipeline. We predict that with the introduction of AI agents, capable of carrying out complex tasks and interacting with environments independently, the creation of content can be fully automated. In our dataset, we considered accounts as 'agentic' if the most recent 10 posts consisted exclusively of synthetic AI imagery, even if in the past the account used to post non-AI content (a case which we encountered rarely).

On TikTok, out of 140 unique videos containing synthetic AI content, 121 videos (86,4%) were posted by Agentic AI Accounts, whereas only nine (6,4%) came from accounts that predominantly post regular content. On Instagram, out of 13 posts containing synthetic AI imagery, two posts (15,4%) came from Agentic AI Accounts, and eight posts (61,5%) came from accounts that regularly post both synthetic and regular content (given the limited sample of data on Instagram, this assessment requires further investigation).

Based on our data, we distinguish between three types of Agentic AI Accounts: Mono-Topic, Poly-Topic, and Hybrid (see Table 17). Mono-Topic AAA focus on one

[7] See 404 Media's reporting on "Where Facebook's AI Slop Comes From" and "Inside the Economy of AI Spammers Getting Rich By Exploiting Disasters and Misery"

format, usually including formal and subject matter characteristics. These accounts repeatedly post the same type of content (e.g., photorealistic synthetic footage of Donald Trump as a toddler) with slight changes in subject matter and formal qualities. Poly-Topic AAA attempt various formal and subject matter conventions, usually following or attempting to establish memetic-like trends. Such accounts may post synthetic AI imagery ranging from cartoon animals to public figures. Hybrid AAA use both synthetic and regular content (mostly stock images and found footage) to illustrate AI-generated and narrated stories. These accounts often follow clickbait and shocking content conventions (human misery, anomalies, mysteries), where even if some footage is regular (non-AI), the text, audio, and story are AI-generated.

| 1) Mono-Topic Agentic AI Accounts | | 2) Poly-Topic Agentic AI Accounts | | 3) Hybrid Agentic AI Accounts | |
|---|---|---|---|---|---|
| specialize in one convention, usually both in subject matter and formal qualities | | attempt various formal and subject matter conventions, often following memetic-like trends | | use both synthetic, stock, and found imagery with AI-generated audio voiceover | |
| 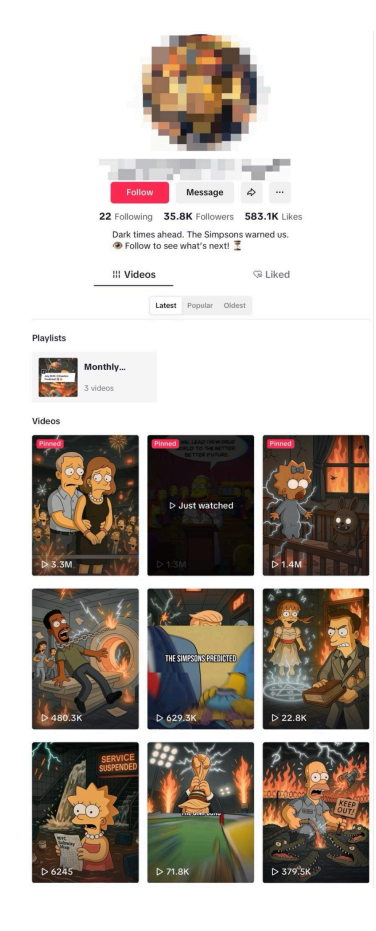 | 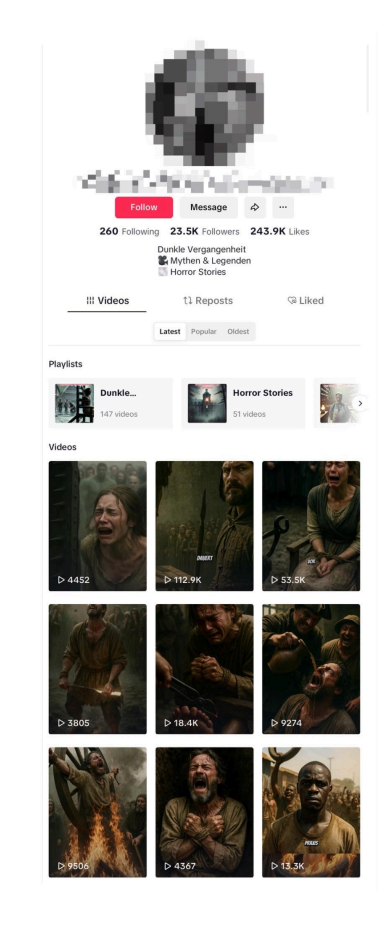 | 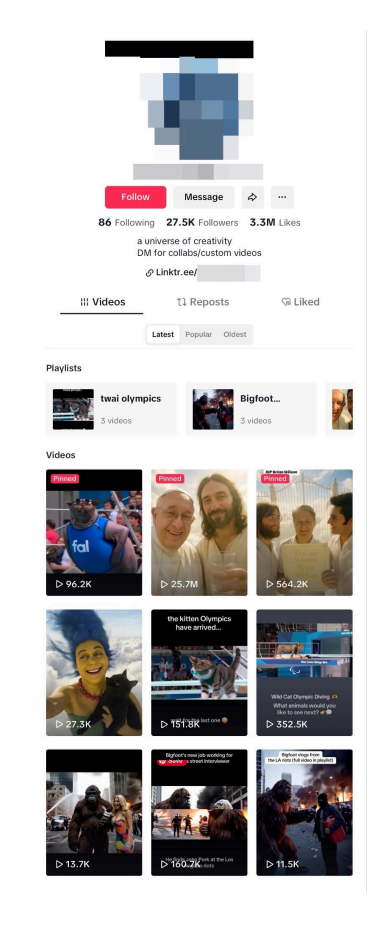 | 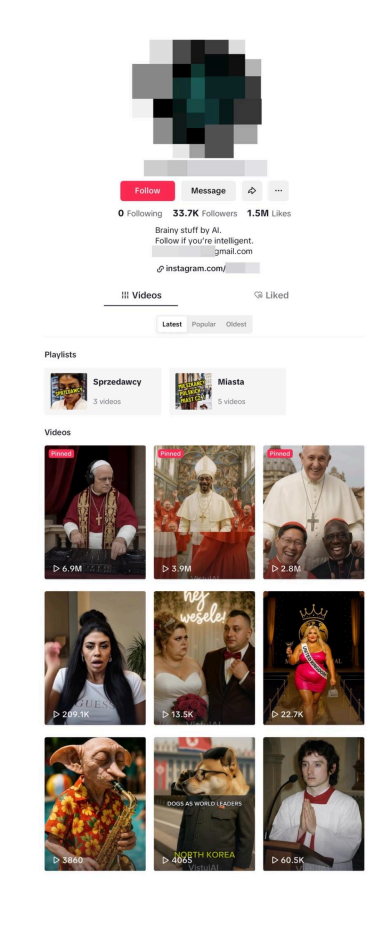 | 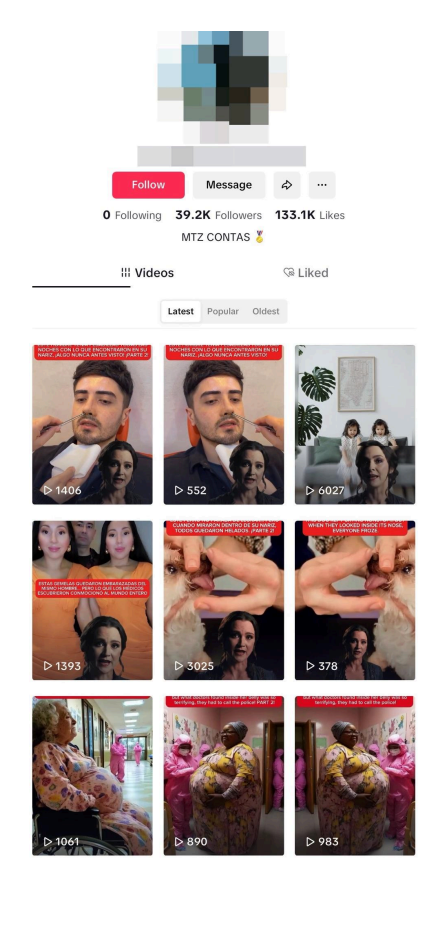 | 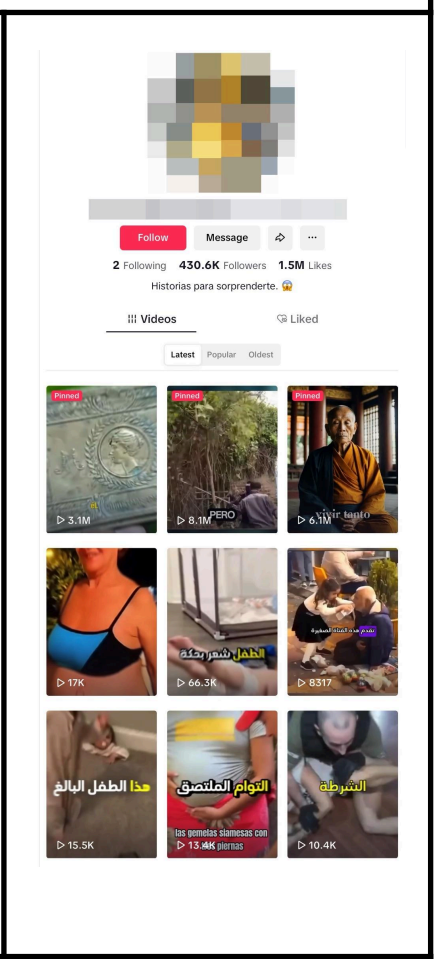 |
| fig. 16A [link] | fig. 16B [link] | fig. 16C [link] | fig. 16D [link] | fig. 16E [link] | fig. 16F [link] |

Table 17. Taxonomy of Agentic AI Accounts.

The primary content Agentic AI Accounts produce can be categorized under the 'AI slop' category. Indeed, AI slop content also gained the most engagement out of all synthetic AI content in our dataset; for example, four videos in the top 30 most viewed TikTok videos across our TikTok dataset (first data collection) constitute examples of AI slop (see Table 17).

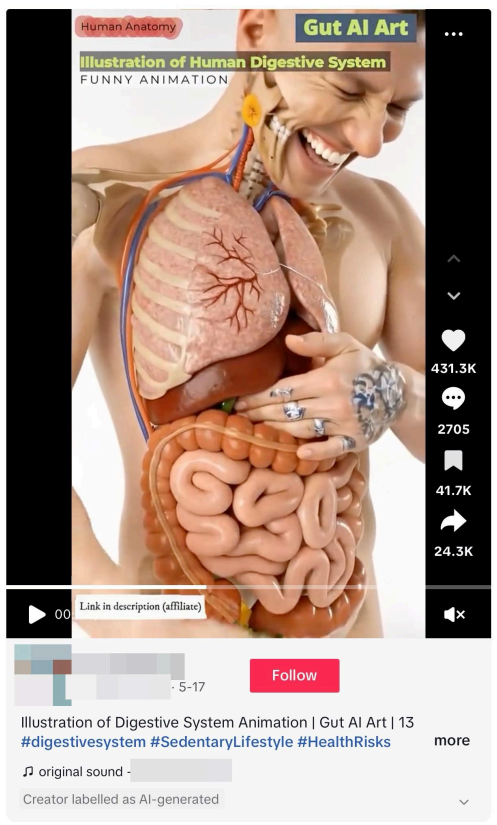

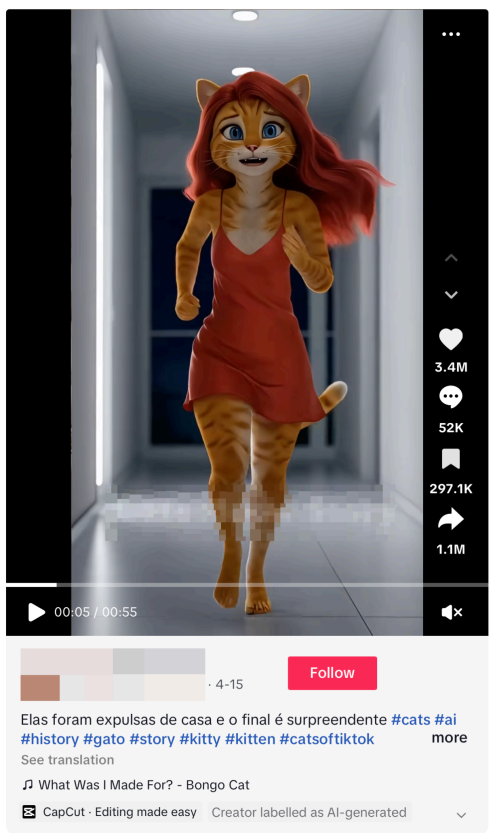

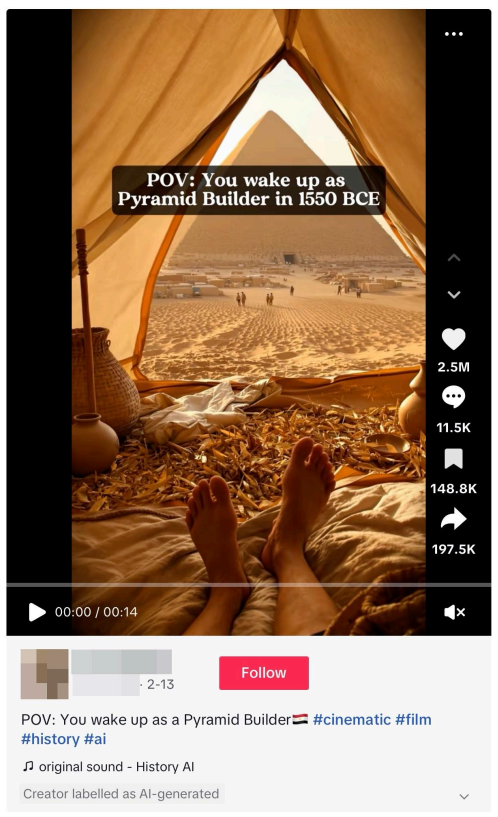

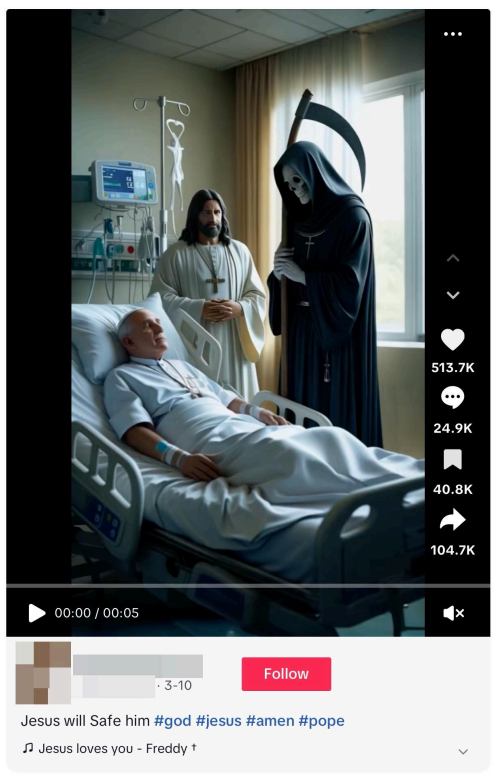

| | | | |
|---|---|---|---|
| fig. 17A [link] | fig. 17B [link] | fig. 17C [link] | fig. 17D [link] |
| views: 64,102,779 | views: 53,614,923 | views: 43,900,000 | views: 33,077,687 |
| 10th video of the 30 most viewed TikTok videos in the first data collection | 14th video of the 30 most viewed TikTok videos in the first data collection | 17th of the 30 most viewed TikTok videos in the first data collection | 25th video of the 30 most viewed TikTok videos in the first data collection |

Table 18. Most viewed posts containing synthetic AI content in the first TikTok dataset. All above posts were shared by specialized Agentic AI Accounts.

A particular subtype of specialized Agentic AI Accounts encountered in this investigation focuses on exploiting explicit content. Such accounts constitute a primary example of objectification of women's bodies (see Table 18) by turning to the production of synthetic imagery. These two cases can be related to the growing number of accounts specializing in fetishistic and sexual synthetic AI content and non-consensual synthetic porn content, for example, on Instagram. The first AAA case poses as a (fake) news media channel and posts solely highly sexualized synthetic clips of women delivering 'news' as media reporters dressed only in bikini costumes (figures 18A-18B). The second AAA case specializes in posting synthetic clips of sexualized women, alluding to historical and fantasy-like settings (figures 18C-18D).

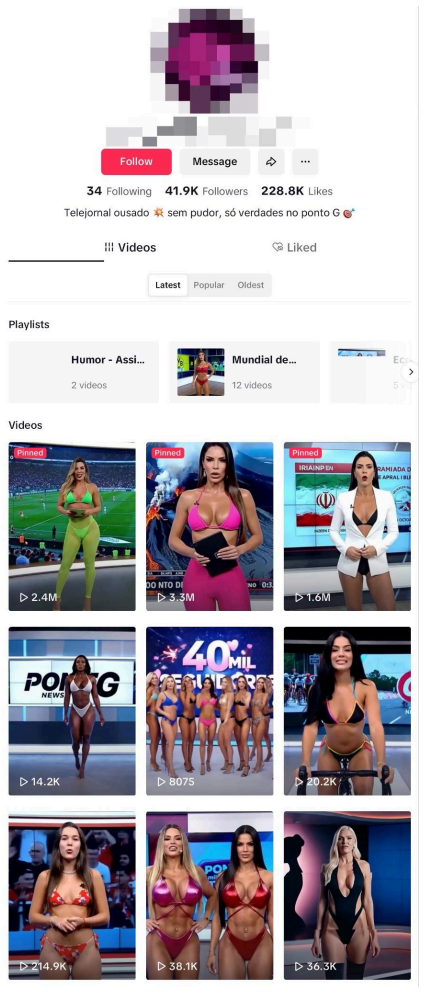

fig. 18A [link]

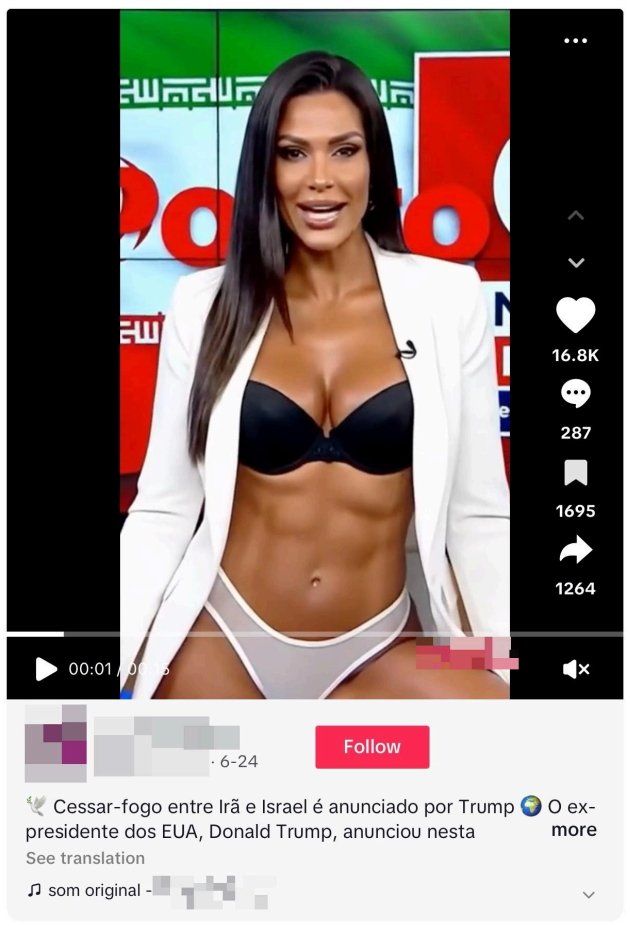

fig. 18B [link]

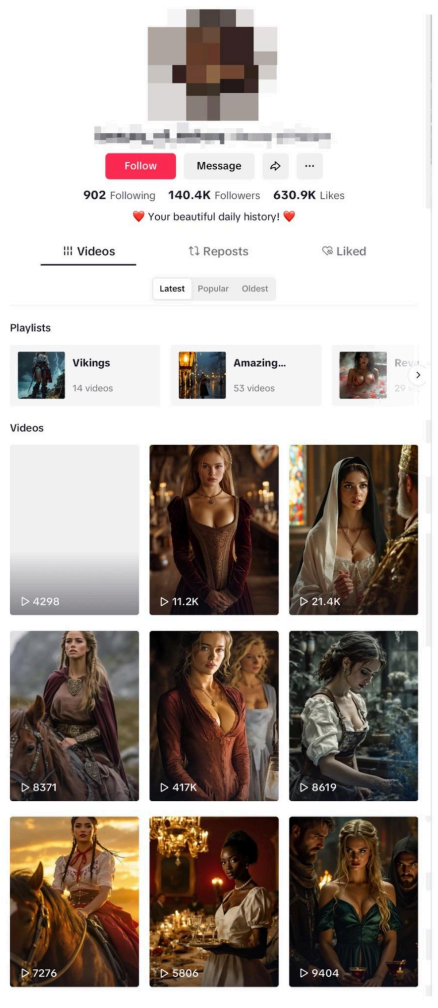

fig. 18C [link]

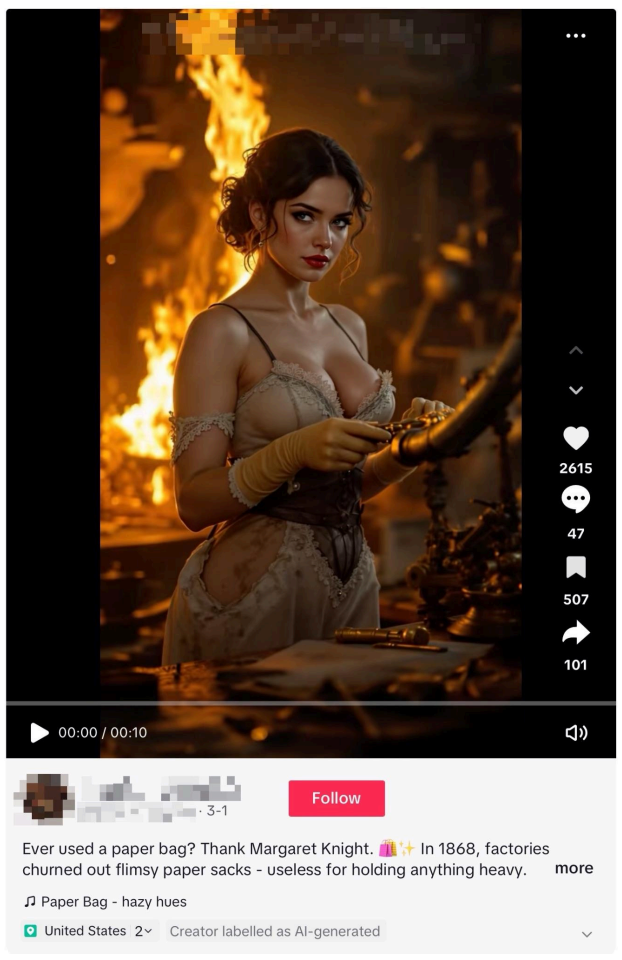

fig. 18D [link]

Table 19. Examples of two posts that appeared in TikTok search results (figure 18B and 18D), alongside screenshots of the respective Mono-Topic Agentic AI Accounts that exploit explicit content using synthetic female bodies.

Synthetic AI imagery appears not to be something that regular users (meaning, non-Agentic AI Accounts) are picking up often. In our dataset, only 6,42% TikTok accounts that shared a synthetic AI imagery post had a history of posting predominantly regular content. Similarly, only 6,42% of TikTok accounts that posted synthetic AI imagery had a history of posting both synthetic AI imagery and regular content on their feed. For Instagram, the percentages of accounts posting both synthetic AI imagery and regular content were higher: 7 accounts (53,84%) and 4 accounts (30,76%;) respectively; however, as the Instagram data sample was limited, further analysis is required to assess these posting patterns.

Users behind such Agentic AI Accounts take a systematic approach, producing and publishing content at scale, which could mean it is with the aim of achieving virality and subsequent monetization. We also noted cases of Agentic AI Accounts that posted what appeared to be sponsored content or possibly scams (see figures 19A-19C, table 20). The following three examples of accounts either use synthetic AI imagery to sell products images and footage of which are synthetic (figures 19A-19C), or to promote monetization guides for synthetic AI content on social media platforms (figure 19B, some posts in rows 1, 2, and 3 from the top). This is also consistent with the reports of AI slop content becoming a new form of monetizing content creation as well as selling nonexistent or deceptive products.

| 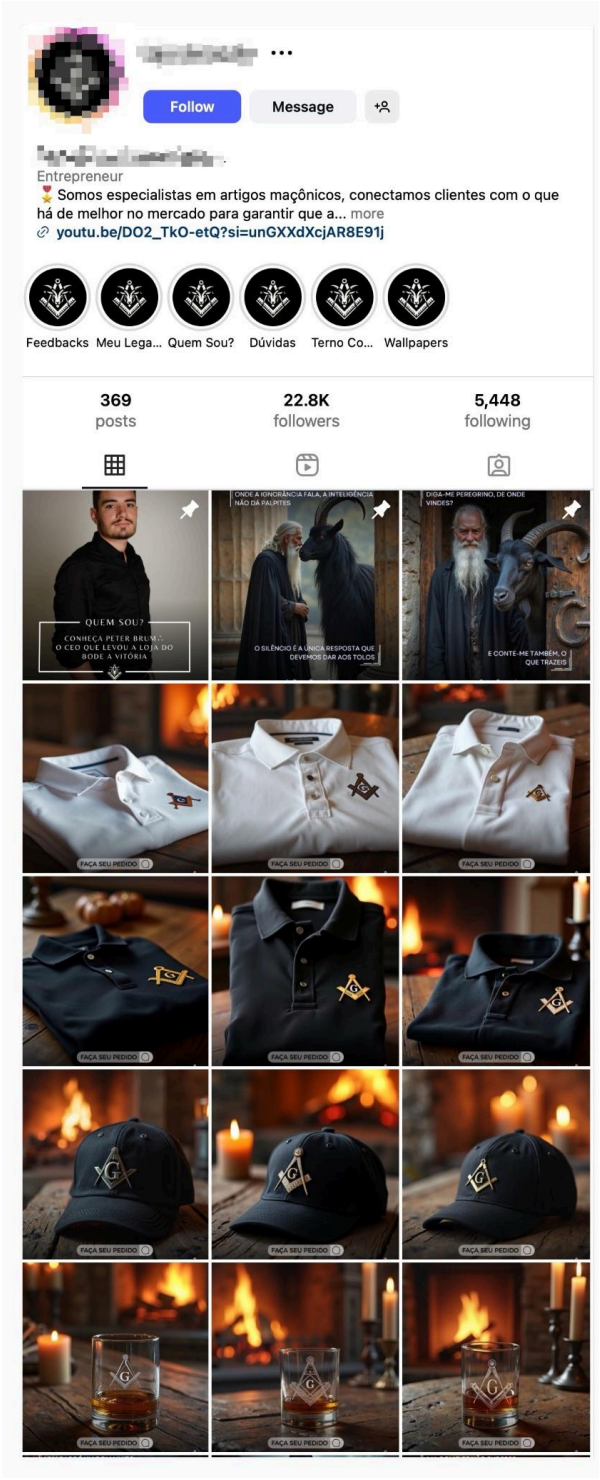 | 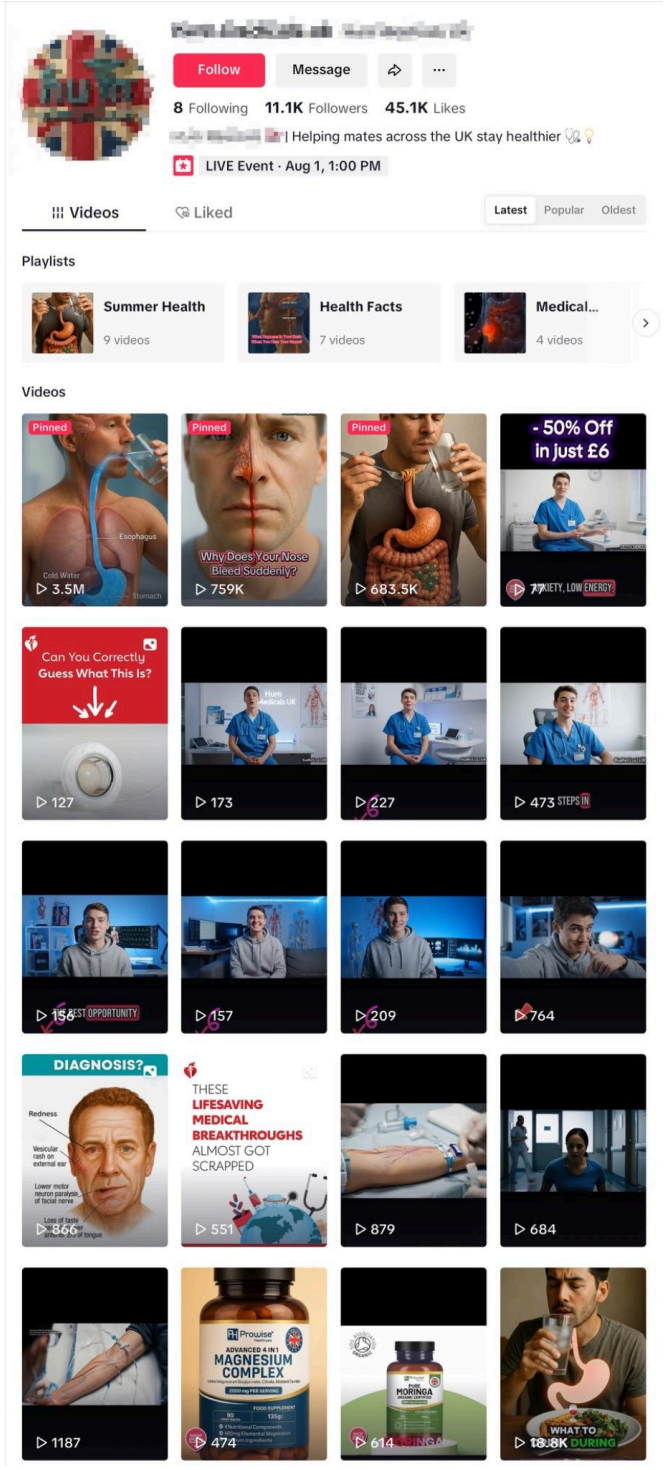 | 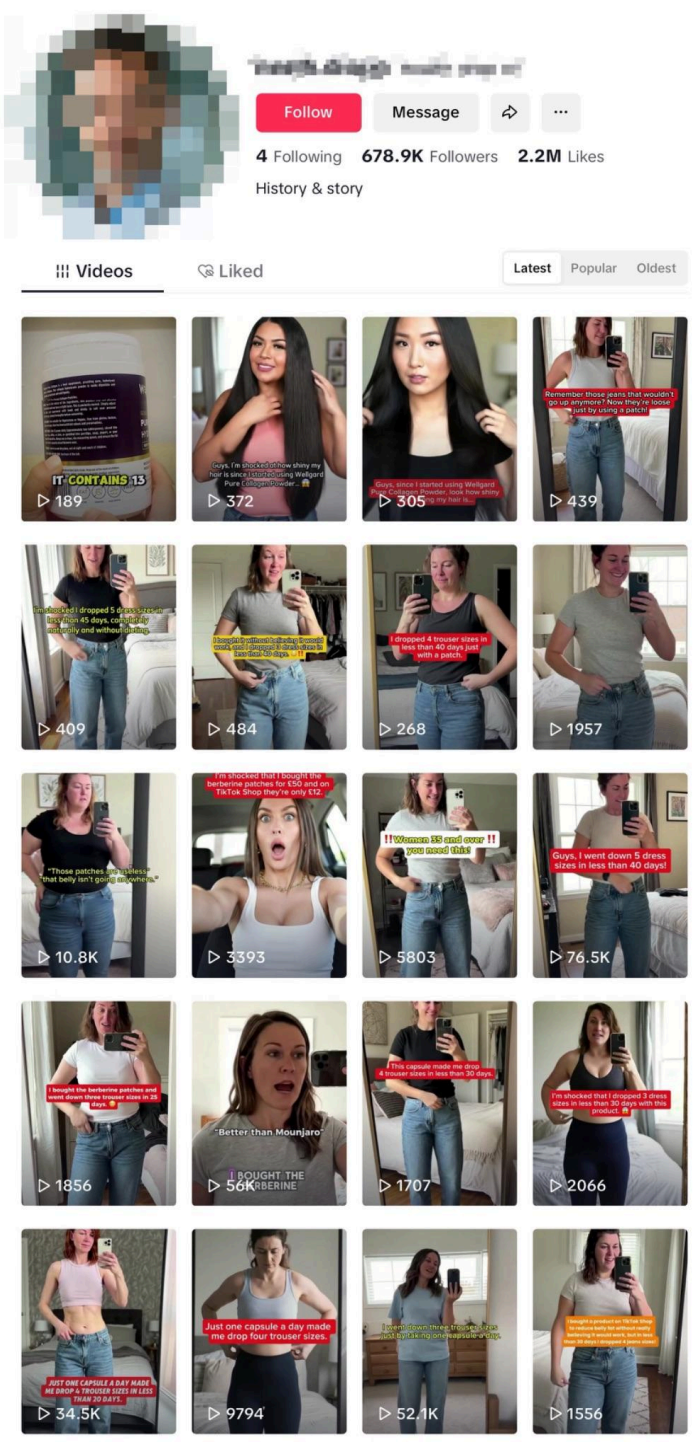 |
|---|---|---|
| fig. 19A [link] | fig. 19B [link] | fig. 19C [link] |

Table 20. Examples of accounts advertising AI-generated products.

A particular example of an Agentic AI Account and its monetization technique is also illustrated by the posting history of an account in Figure 20. The screenshot captures how, at first, the account in question posted solely variables of AI slop in a manner of a hybrid Agentic AI Account (image on the right, first five rows from the bottom). Then, the account switched to posting repetitive AI content, with each post being a synthetic advertisement for weight loss. We might assume that such a turn of strategy could be prompted by the fact that the account either used AI slop to first gain visibility and following and then turned to more outright monetized content, or that the account owner realized that the sponsored (or scam-like) content could be more profitable on TikTok.

We make available upon request the list of Agentic AI Accounts we encountered on TikTok and Instagram during our experiment, and all the individual videos we annotated.

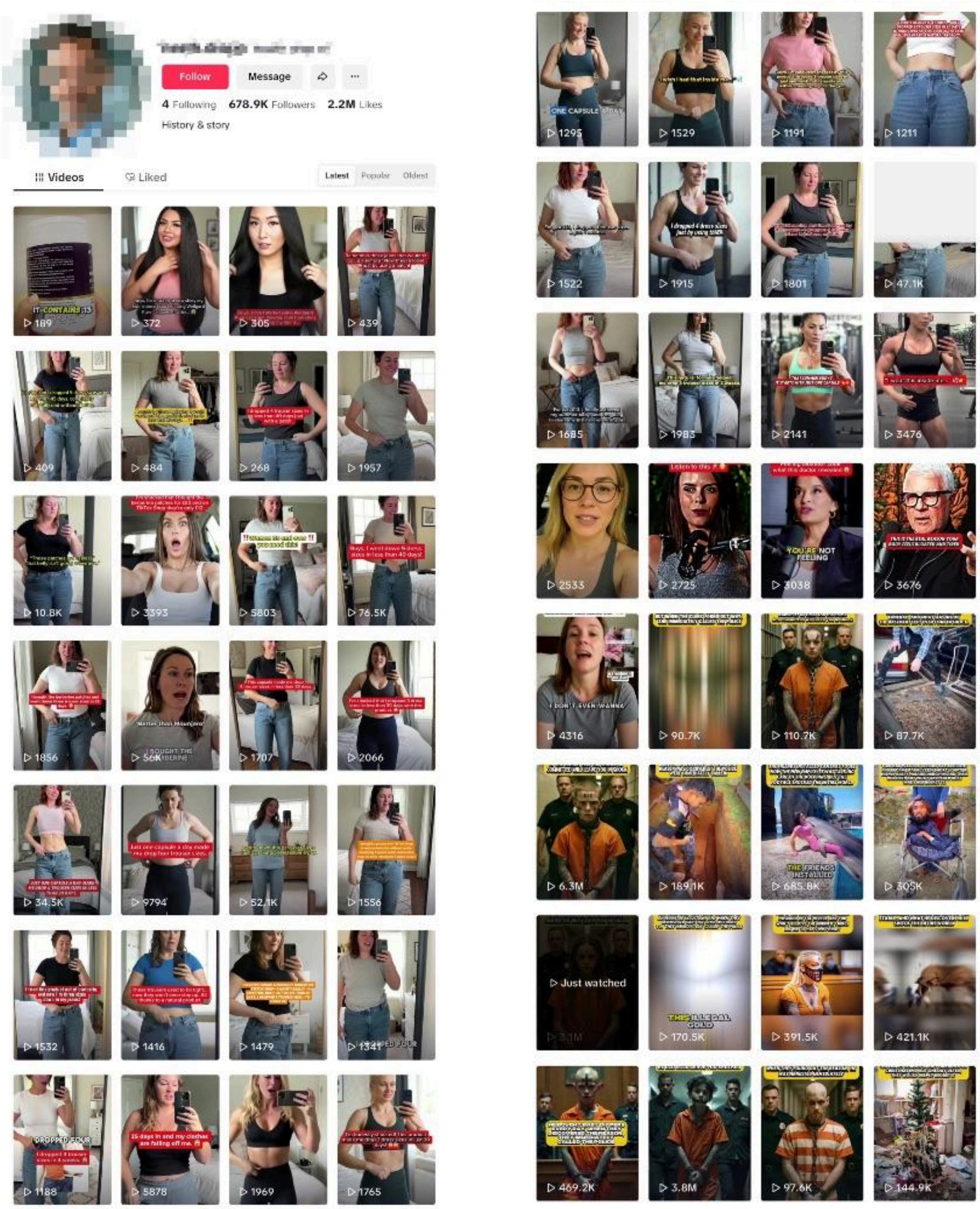

Figure 20. Example of an Agentic AI Account switching topics from crime (right) to weight loss (left). [link]

# Conclusion

Our investigation empirically demonstrates that the phenomenon of synthetic AI imagery in posts on social media platforms such as Instagram and TikTok is an existing and growing phenomenon: synthetic AI imagery content is equally distributed across various topic areas and across countries, taking up circa 25% of the top 30 results in search result pages on TikTok during our research. We found significantly more synthetic AI imagery in top search results on TikTok than on Instagram.

On both TikTok and Instagram, it appears that the majority of content made using synthetic AI imagery is not labeled appropriately neither by the content creators nor by the platform. Lack of labeling could be considered a breach of the applicable regulation when the content resembles existing persons, objects, places, entities, or events as outlined in the DSA. Although VLOPs such as TikTok and Instagram (part of Meta's report) included AI-generated content as a cross-cutting risk in their systemic risk assessment reports, it is surprising that the labeling is not done thoroughly, particularly in relation to prominent systemic risk areas outlined in Article 34 of the DSA.

According to their policies, platforms themselves seem to recognize the need to inform users when realistic content is AI-generated. However, we argue that creators and users should not be solely responsible for adding the appropriate AI labels. This can be particularly ineffective when platforms do not impose a standard and clear labeling obligation, but also when platforms allow unclear markers like #ai down in the post caption or description, or hide the labels in context menus. Commission Guidelines for providers of Very Large Online Platforms and Very Large Online Search Engines on the mitigation of systemic risks for electoral processes pursuant to Article 35(3) of the DSA clearly recommends that providers of VLOPs and VLOSEs apply efficient labels recognizable by users taking into account its graphics, position and timing drawing on the research on effectiveness of labels. Although VLOPs analyzed in our study outline their measures against AI-generated content in their last DSA risk assessment reports, they don't refer to any study or test aimed at assessing the efficiency of the deployed labeling mechanisms. In fact, despite their claims of having state-of-the-art detection mechanisms, we have found AI-generated content that is not being labeled, indicating that platforms fail to enforce their rules.

For our research, we relied on manual annotation in part because there are no reliable tools available for public interest researchers to classify synthetic content. We saw a rapid adoption of Google's Veo 3 model for video generation by content creators during the course of our study; in the meantime, Google had announced that they have implemented their watermarking technique synthID for content created with Veo3, yet the detection platform was still “waitlist only” two months after its announcement. Regardless, we have found videos that contain the visible “veo” watermark in the frame that were also not labeled.

The lack of proper labelling and inconsistent visibility of AI disclosures across TikTok and Instagram can be deceiving when the content has photorealistic formal qualities, and can be seen as plausible. We found several instances of unlabelled, photorealistic discourses of politicians, making statements that they never made in reality, which is a clearly prohibited practice across different regulations of the EU digital playbook. We believe that compliance with applicable regulations requires platforms to enforce their rules and respect their commitments on AI labels and to make labels visible by default across content types and all platform versions, including the web versions. This is particularly applicable to Instagram, where AI labels are often not visible by default and are lacking entirely on the web version of the platform (based on our research), a shortcoming we urge Instagram to address.

Most of the AI content we encountered originated from a new type of “Agentic AI Accounts” as outlined in the above taxonomy. These accounts utilize fully or partially generative AI tools and pipelines to generate, disseminate, and even autonomously adapt content to optimize for virality. The growing sophistication and affordability of those tools raise concerns about the future scale and impact of AI slop content. Beyond the impact on the content creator economy, their potential to escalate information operations and manipulation campaigns is a major concern outlined as the first pillar in the EU’s Democracy Shield. We therefore strongly recommend that platforms and regulators reflect on this specific type of Agentic AI Accounts, in order to flag and eventually moderate them properly.

The European regulatory frameworks offer clear pathways to allow users to distinguish synthetic content but its enforcement seems to be lagging behind. Although there are technical complexities involved such as watermarking and detecting AI-generated content, as platforms are increasingly becoming providers of genAI powered features and tools, it should be even easier for such platforms to deploy appropriate labeling mechanisms.

Finally, beyond labeling, the increasing presence and success of clickbait, deceptive content on platforms should be a wake-up call for the latter to reassess their recommender systems and business models for a healthy and trustworthy online information environment.

# Appendix

## Identifying Keywords

| | | | |
|---|---|---|---|
| artificial intelligence | AI | #artificialintelligence | #ai |
| künstliche Intelligenz | KI | #kunstlicheintelligenz | #ki |
| inteligencia artificial | IA | #inteligenciaartificial | #ia |
| aiart | | #aiart | |
| ai-generated | | | |
| aigenerated | | #aigenerated | |
| _ai | | | |

Table A. Relevant keywords for users' AI disclaimers in the content of the posts' descriptions, hashtags, and stickers

## Search Terms

| topic area | hashtag | location |
|---|---|---|
| | | Spain |
| | | Germany |
| | #trump | Poland |
| | | Spain |
| | | Germany |
| politics / current affairs | #zelensky | Poland |
| | | Spain |
| | | Germany |
| | #history | Poland |
| | | Spain |
| | #historia | Poland |
| entertainment | #geschichte | Germany |
| | | Spain |
| wellbeing | #health | Germany |
| | | Poland |
| | #zdrowie | Poland |
| | #gesundheit | Germany |
| | #salud | Spain |
| culture / current events | #pope | Spain |
| | | Germany |
| | | Poland |
| | #papież | Poland |
| | #papa | Spain |
| | #papst | Germany |

Table B. Query list of hashtags used for data collection, including relevant translations, shown alongside respective topic areas and countries.

# Prevalence of synthetic content on TikTok per country and topic

| topic area | hashtag | GenAI TikTok% | Location |
|---|---|---|---|
| politics / current affairs | #trump | na | Spain |
| | | 17.24% | Germany |
| | | 34.48% | Poland |
| | #zelensky | 7.14% | Spain |
| | | 13.95% | Germany |
| | | 0% | Poland |
| entertainment | #history | 39.53% | Spain |
| | | 20.00% | Germany |
| | | 27.78% | Poland |
| | #historia | 28.00% | Spain |
| | | 23.08% | Poland |
| | #geschichte | 31.58% | Germany |
| wellbeing | #health | 21.05% | Spain |
| | | 11.63% | Germany |
| | | 15.38% | Poland |
| | #zdrowie | 7.84% | Poland |
| | #gesundheit | 14.63% | Germany |
| | #salud | 2.22% | Spain |
| | #pope | 23.08% | Spain |
| | | 26.47% | Germany |
| | | 29.17% | Poland |
| | #papież | 42.42% | Poland |
| | #papa | 0.00% | Spain |
| culture / current events | #papst | 13.89% | Germany |

Table C. Prevalence of synthetic content on TikTok per country and topic

Codebook

SYNTHETIC AI IMAGERY ON TIKTOK AND INSTAGRAM

# CODING AI SLOP

AI FORENSICS

# Codebook: AI Slop

This codebook serves as a consistent coding scheme for detecting and labelling synthetic visual content. What follows is a brief description of each label alongside examples and remarks on particularly attention-demanding content types for the coder. This typology is non-exhaustive. The following coding categories are mutually exclusive and binary [yes/no].

1. GenAI - content consists of generative AI imagery, that is, synthetic visuals (moving and/or still images).
    1.1 photorealistic - content which is labelled as GenAI and consists of a representation of stylistic realism; it imitates the stylistic exactness of a photographic or film capture (unlike, e.g., a cartoon).
2. Partial GenAI - content includes both synthetic and non-synthetic imagery, for example, generative AI images intertwined with stock images.
    2.1 photorealistic - content which is labelled as Partial GenAI and consists of a representation of stylistic realism; it imitates the stylistic exactness of a photographic or film capture (unlike, e.g., a cartoon).
3. Unclear - no definitive conclusion on the nature of the content can be drawn from the information at hand.
4. Not GenAI - the content is definitely not made using generative AI tools.

This codebook accounts for a spectrum of synthetic content: moving images, still images, and deepfakes. The exact definitions, as well as visual analysis strategies to detect such content, can be found in the "AIF Guidebook: A Human Guide to Detecting Synthetic AI Imagery." In general, in assessing any piece of content that contains moving images, it is recommended for the coder to pay attention to transitions between frames, slow, pause, and replay suspicious frames.

We focus on visual content, not audio; therefore, audio deepfakes and AI-generated voiceovers are outside the scope of this codebook.

## 1. GenAI

Content consists solely of generative AI imagery, that is, synthetic visuals (moving and\or still images).

### 1.1 photorealistic

GenAI content that consists of a representation of stylistic realism; it imitates the stylistic exactness of a photographic or film capture (unlike, e.g., a cartoon) (see figures 1-3 and 4-5)

The examples may include moving or still plausibly photo-realistic images, yet display: visual artifacts (figure 1; figure 2); impossible or highly unlikely time, place, and/or circumstances in the composition (figure 1; figure 3); a mismatch of expression, setting, context, often combined with cutesified subject (animals, toddlers) (figure 2).

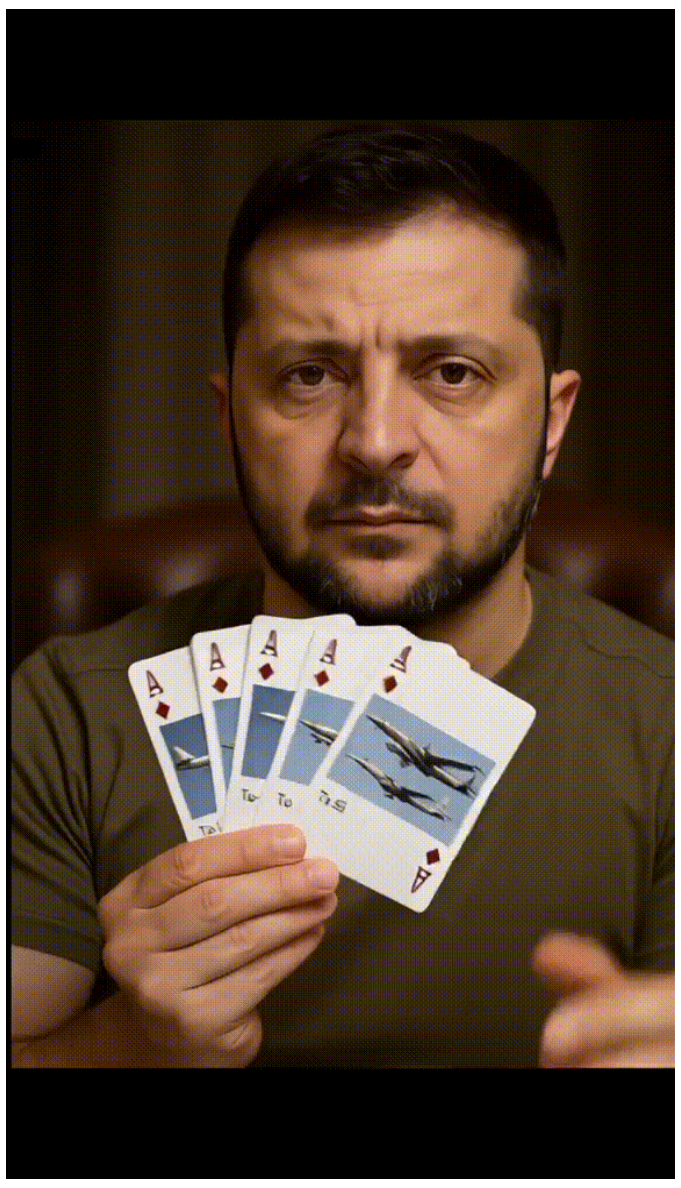

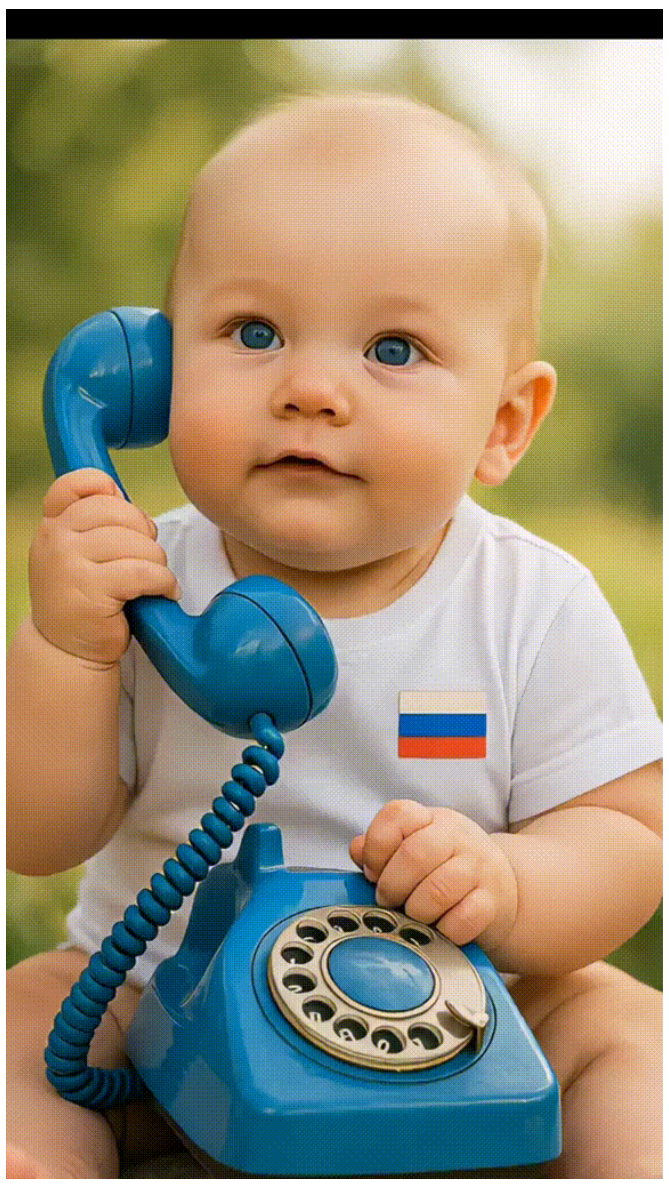

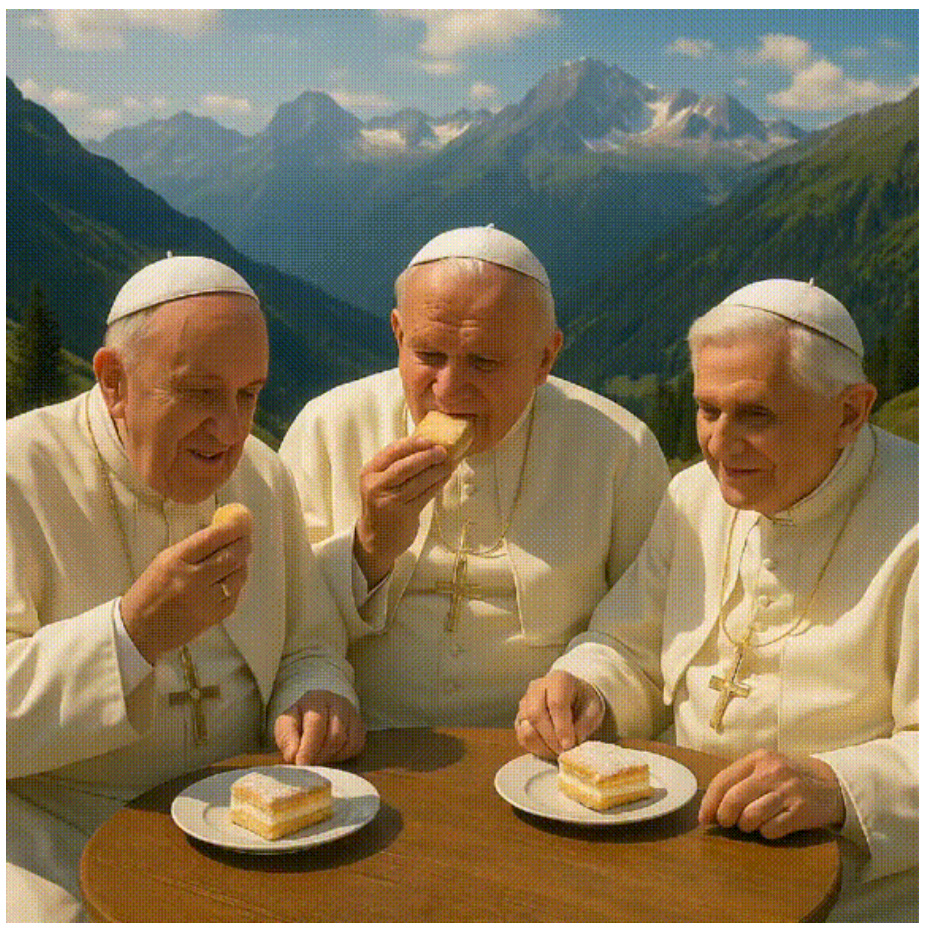

From left: *Figure 1*: A synthetic video of Volodymyr Zelensky playing cards, showing the deck with pictures of military planes to viewers. The cards merge with one another and disappear. The writing on the cards in the bottom corner is not in an existing alphabet. The letter “A” glitches.
*Figure 2*: The toddler’s face shows adult-like facial expressions and lip movements. The numbers on the phone are out of order and partially written in a gibberish alphabet.
*Figure 3*: Three Popes eating cake together, dressed in the same papal attire and of different ages in life, which could not physically take place.

The examples may also include representations of historic or historicized events, peoples, and places (figures 4-5), as well as depictions of current, historical, or plausible events “in the style of …” a popular media object (film, cartoon) or an art style (figure 6).

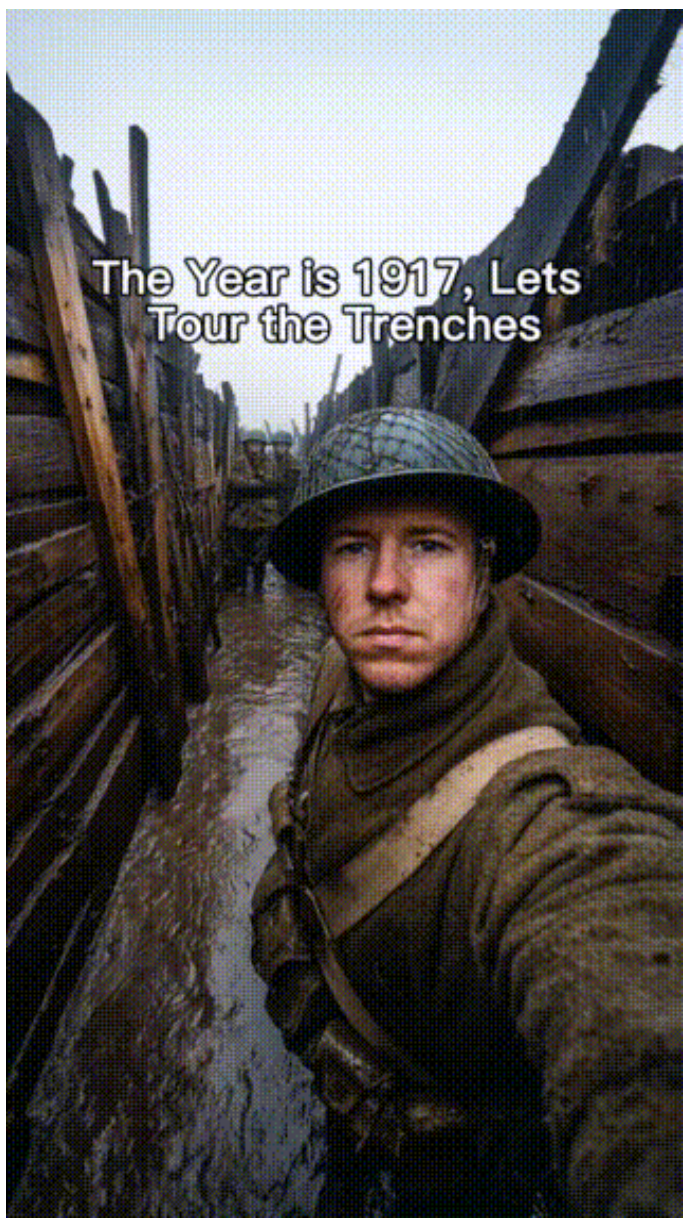

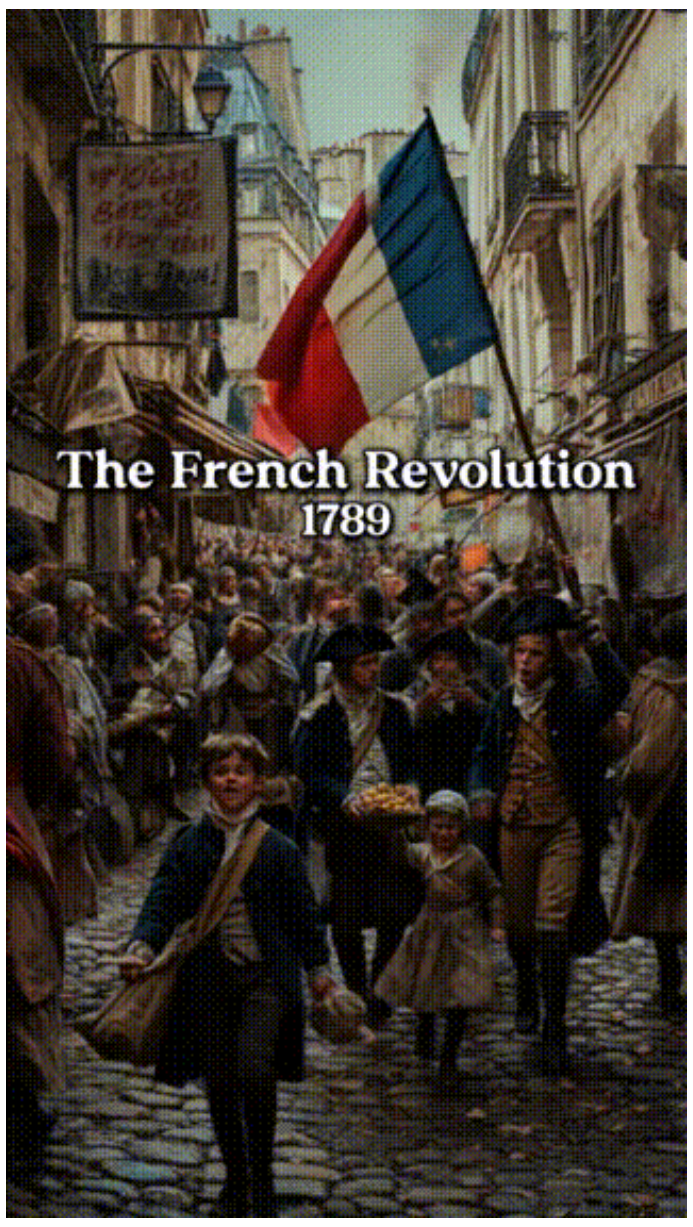

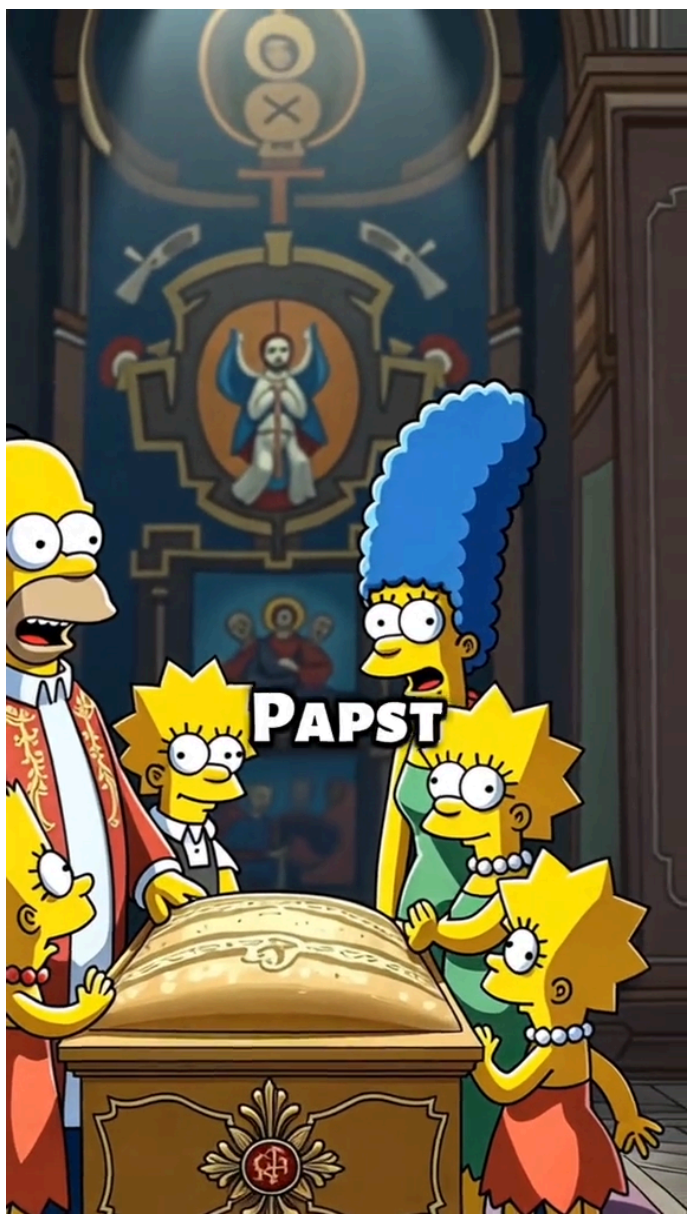

From left: *Figure 4*: A first-person view of a seemingly historical context. Aside from the aesthetics of synthetic imagery (oversaturation of colors, unrealistically exaggerated light and color, smooth, clay-like skin and attire), the content also contains several synthetic artifacts (objects appearing and disappearing from the hands of figures in the background) and physically unlikely events (unrealistic explosions).
*Figure 5*: Similarly to the previous example, the clip shows the same qualities for both synthetic aesthetics and artifacts.
*Figure 6*: Several images and semi-animated clips made in the style of a cartoon ‘The Simpsons.’ Aside from the unlikely chance of the sheer quantity of diverse images depicting the same concept (death of the Pope), the coherence of image details (particularly architectural backgrounds) does not follow the Simpsons’ style and depicts human figures.

## 2. Partial GenAI

Content that includes genAI imagery (still or moving images) only in some part.

### 2.2 photorealistic

Partial GenAI content that consists of a representation of stylistic realism; it imitates the stylistic exactness of a photographic or film capture (unlike, e.g., a cartoon) (see figures 7 and 11-13)

In other words, it is content that constitutes a mix of synthetic imagery with stock photos or other types of visual, non-synthetic content. This label is also to be used when a part of the content definitely is an example of generative AI, while other part(s) are of unclear origin. On the coding side, the possibility of Partial GenAI content requires the coder to watch (or pause through each frame) each piece of content, even if the content does not look like synthetic at first glance. To aid in detection, we divide Partial GenAI content into two types in which it may appear - the coding will not account for those types, as it is discussed below only to aid the coders in identifying content to be labelled as Partial GenAI.

### GENAI IN OPENING FRAMES

Generative AI content is used as an opening image/scene, followed by non-synthetic content or a mix of non-synthetic and synthetic content.

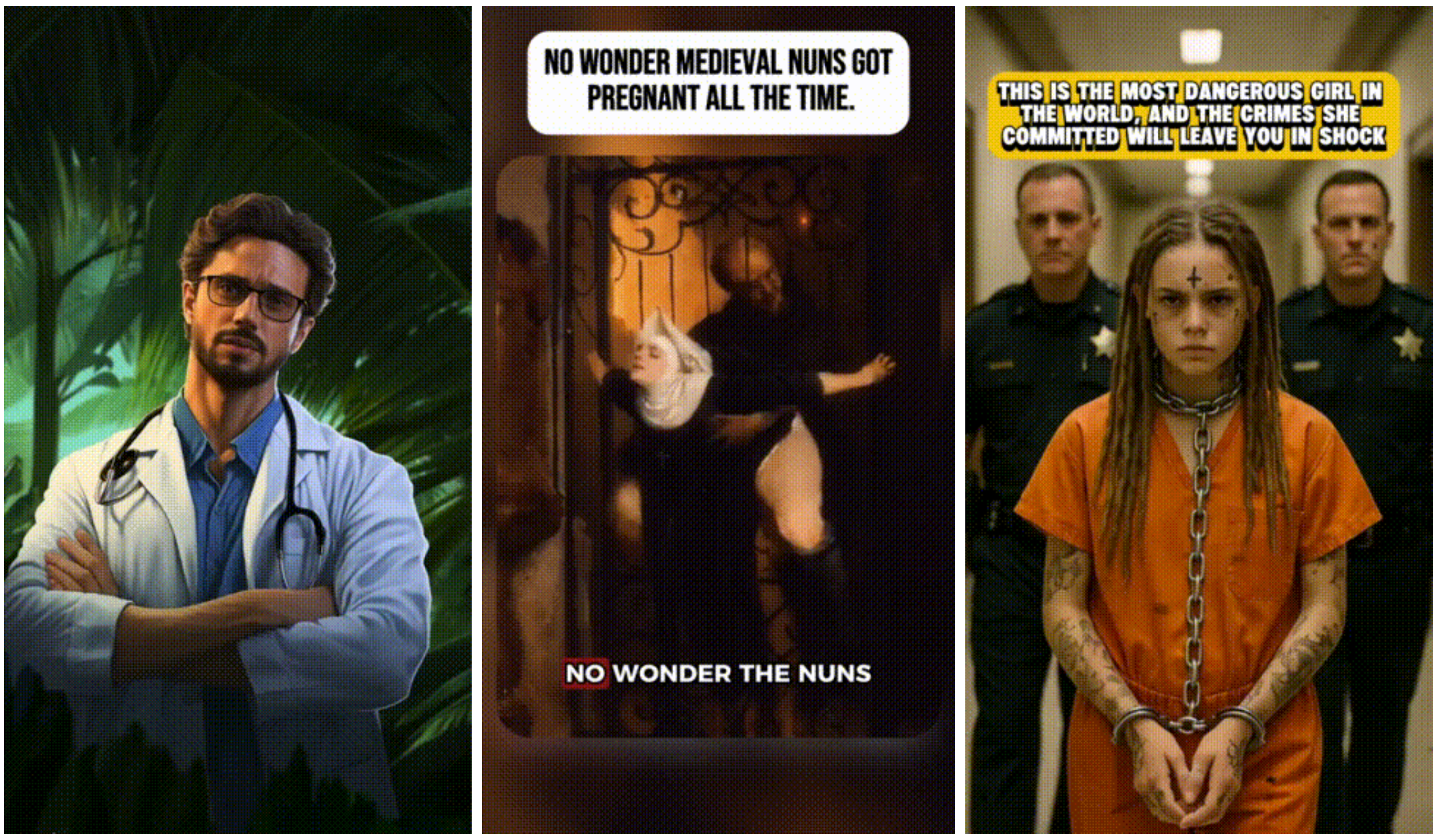

From left: *Figure 7*: A generative AI opening image of a doctor speaking (likely a deepfake from a still image due to exaggerated artificiality of the head movements). Highly visible are the aesthetics of genAI imagery, especially the unrealistic (cartoon or comics-like) attire, physically impossibly smooth details of skin, muscles, and hair, and exaggerated and unrealistic light and shadow. Generative AI images combined with possible organic footage and animation follow.

*Figure 8*: The opening moving image is a generative AI alteration of an existing painting, turned into a short clip. While other images that follow appear to be organic pictures/paintings, given that none of them contain evident genAI attributes, and to avoid double/checking their authenticity, this example is coded as Partial GenAI.

*Figure 9*: The opening clip bears the aesthetics of genAI imagery (harsh, cinematographic lighting with strong light-shadow contrasts, smooth details of skin and hair, luminescent clothing of clay-like texture) as well as highly unlikely circumstances (heavy chain around the neck of the young girl) and a senatorial text which might have referred to a movie or TV series clip if the footage did not contain the aesthetics of synthetic imagery. The clip is then followed by stock images, further emphasizing that the story is inauthentic and clickbait.

### GENAI IN INTERCHANGABLE FRAMES

Generative AI imagery is inserted in only some part of the content, not the opening frame.

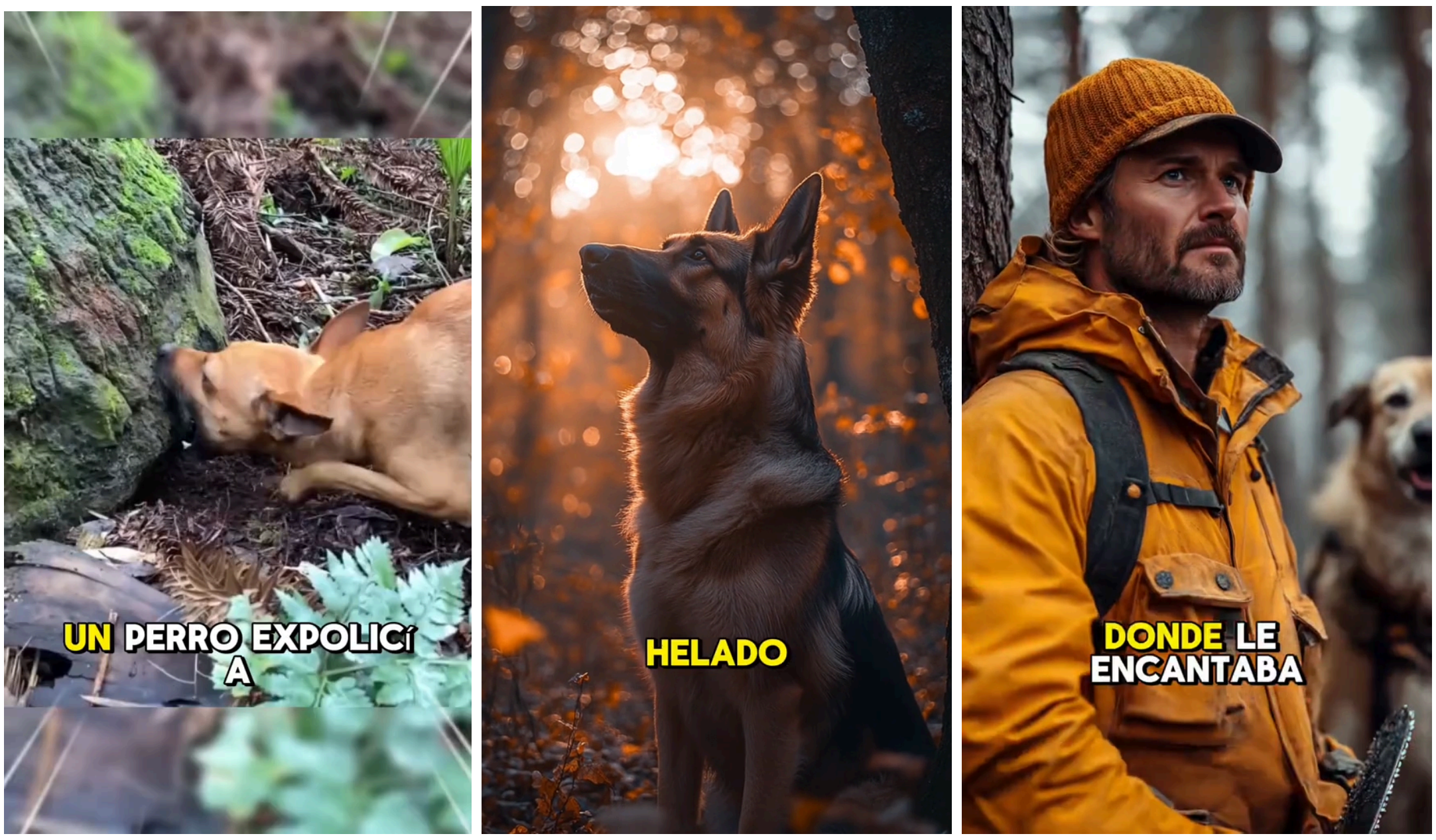

From left: *Figure 10*: While the opening clip shows an organic video footage of a dog digging in a forest, the following images bear the aesthetics of generative AI imagery (oversaturated colors, smooth, clay-like qualities of textures, and strong light-shadow contrast).

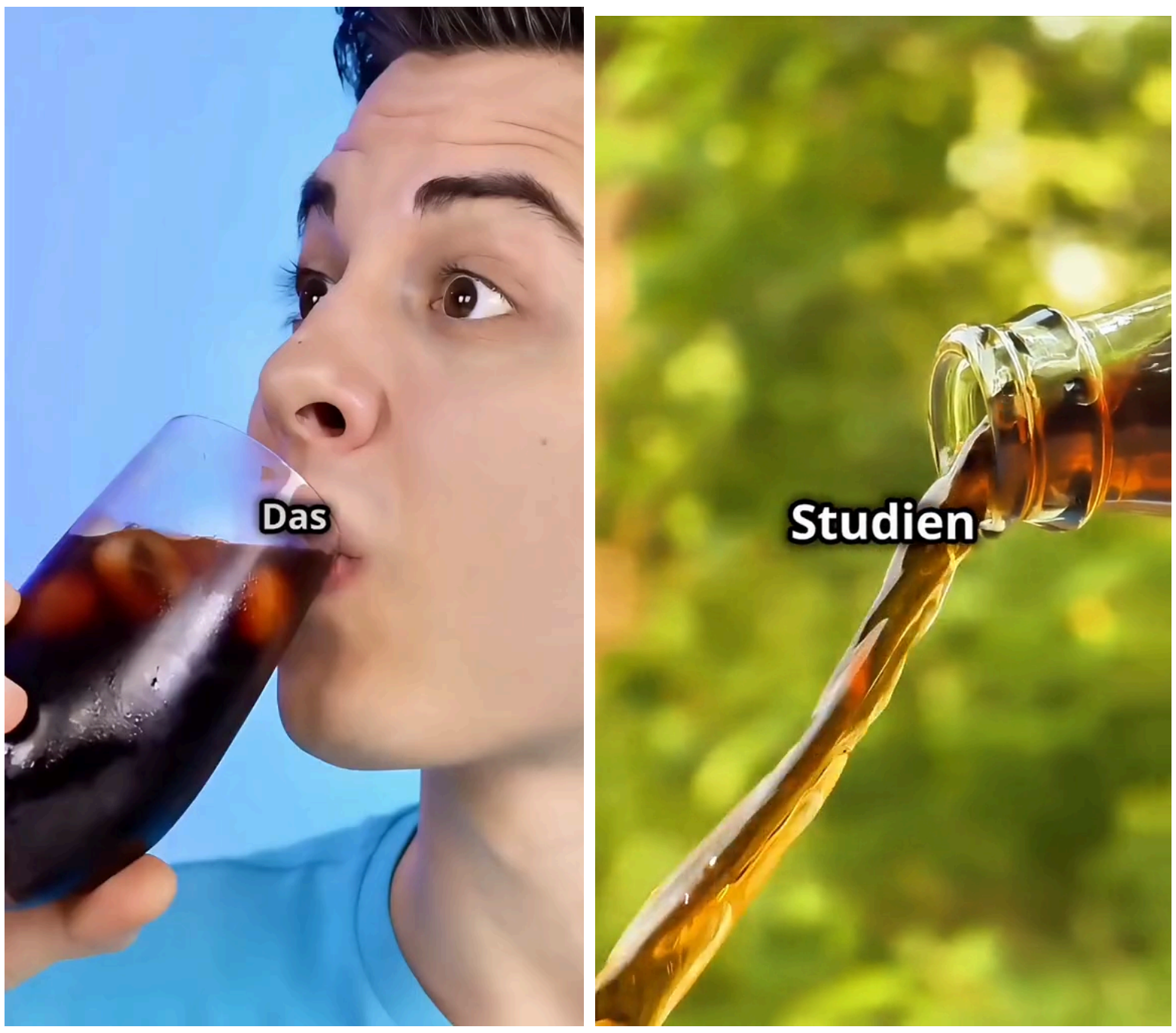

*Figure 11*: This piece of content, consisting of a series of short clips, contains the first genAI example more than halfway through, depicting a bottle of Coca-Cola where the liquid behaves unnaturally (flowing unlike this type of liquid would).

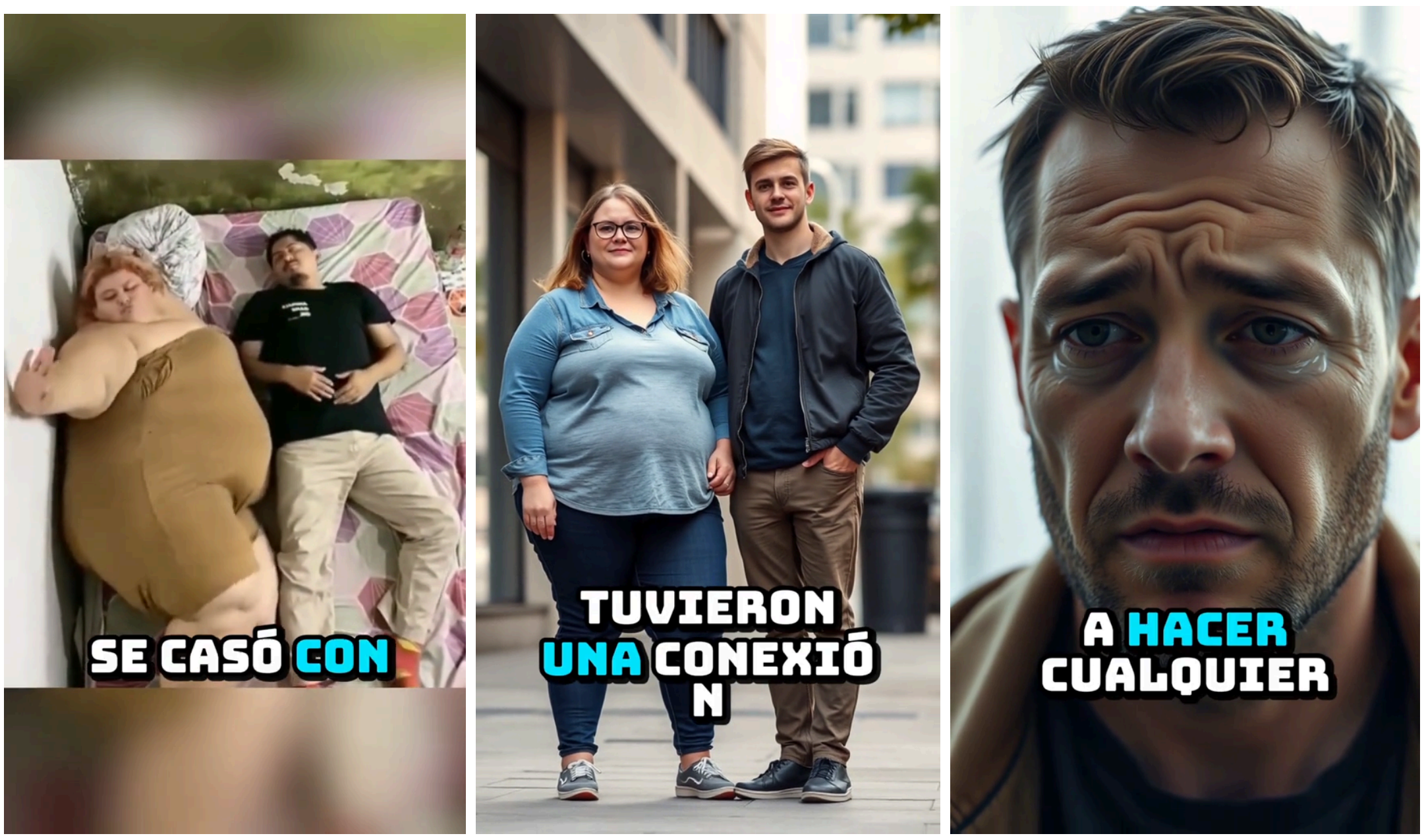

*Figure 12*: While the first clip seems to be an example of possibly found footage, the following images bear all aesthetic qualities of genAI content described in the examples above.

### 3. Unclear

Given that the coding is performed with a conservative approach, the "unclear" label shall be used only when the coder can point out evidence that makes one doubt whether a piece of content might be synthetic, yet the evidence is not strong enough to say that the piece of content indeed is genAI. These pieces of content will be later discussed among coders to find an agreement, while acknowledging that some pieces of content lack the details necessary to definitely label them as Generative AI. Otherwise, if no justification can be given to the doubt, the "no genAI" label should be utilized.

## DEALING WITH BORDERLINE CASES

If the coder is unsure whether a piece of content is genAI or not, the first step is to consider the TITLE, DESCRIPTION, HASHTAGS, and ACCOUNT NAME that shared the content. If there is any indication of genAI, it is highly likely that the content indeed was (at least partially) made with generative AI.

The second step is to take screenshots of suspicious frames or images and perform a reverse-image search, utilizing, for example, Google's reverse search. This strategy is particularly useful if the synthetic image or deepfake was very subtle or of poor quality. The incentive to conduct such a search may be that the video seems somewhat plausible, yet it could also be manipulated for virality or hate reasons, or it is considered "weird." It is a very subjective decision-making and context-dependent choice, requiring, to a degree, a coder to "follow the vibes." The two examples below (figure 13 and figure 14) are such cases.

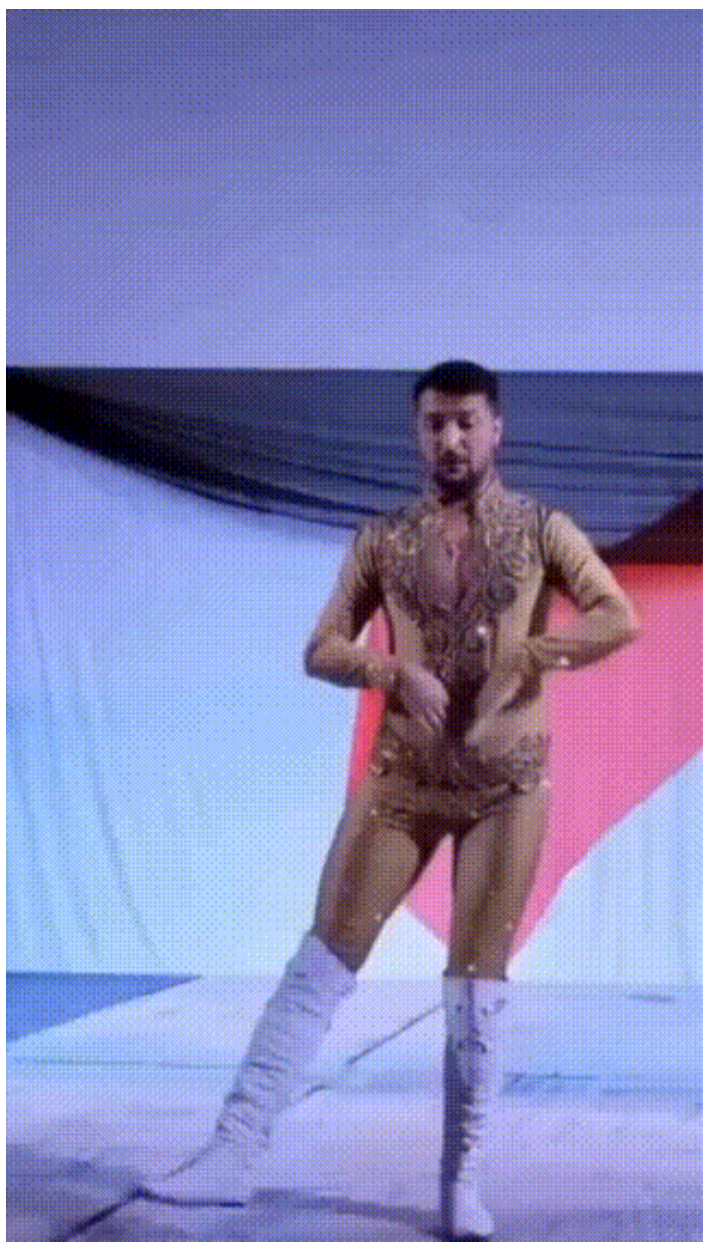
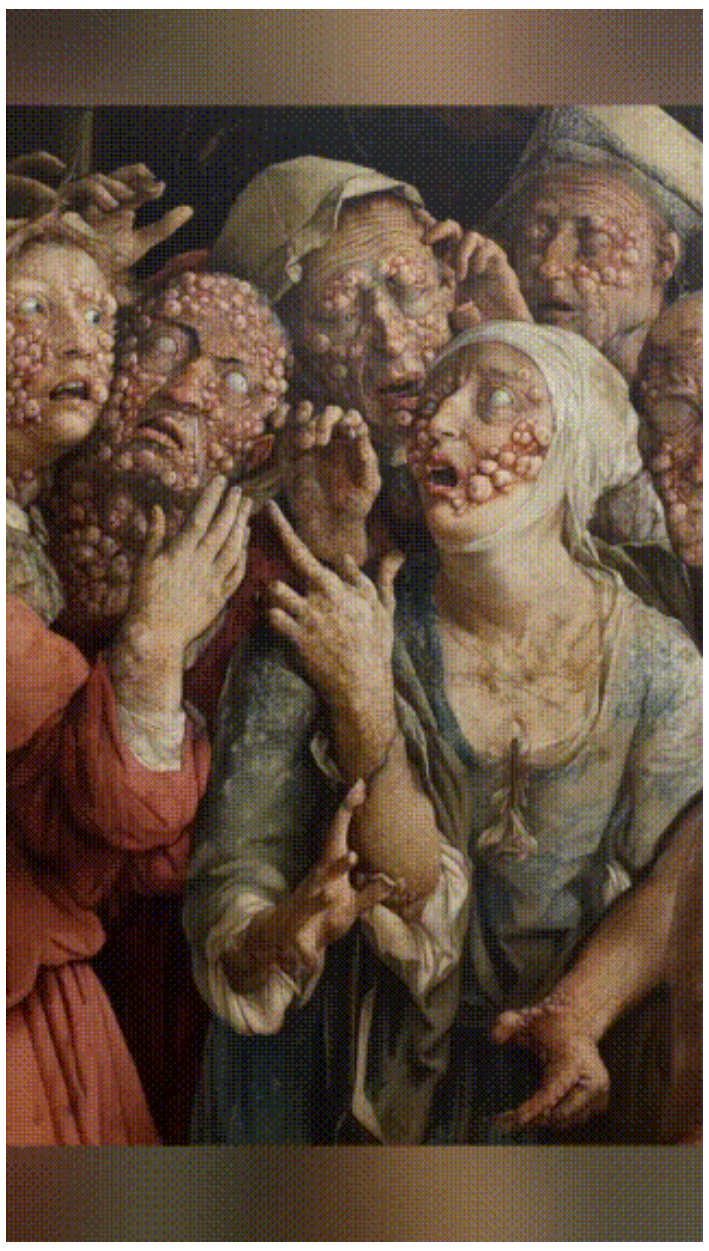
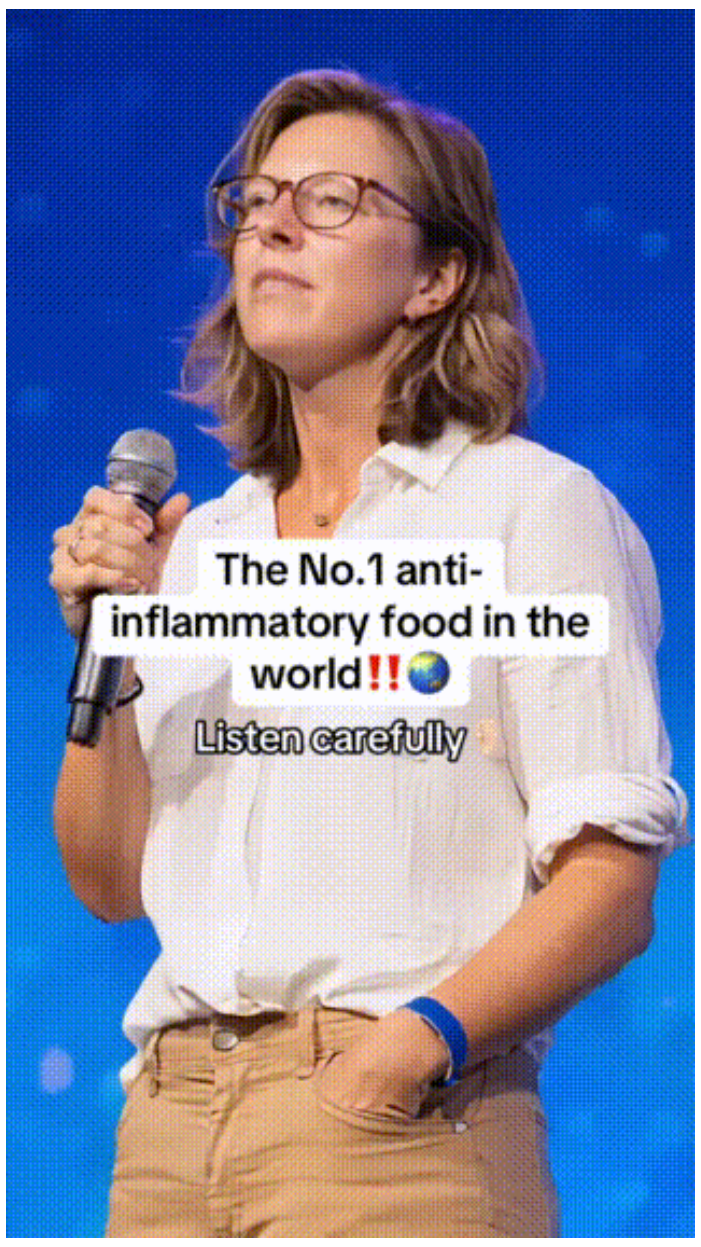

From left: *Figure 13*: while Zelensky was an actor known for comedic appearances before becoming the president of Ukraine, he is also particularly targeted with genAI content. While plausible and of low quality, the opening frame of this clip is suspicious, and there is a slight mismatch between the color of the face and neck. Following a reverse search of a screenshot of this opening clip, this footage was proven and debunked as a 2022 deepfake.
*Figure 14*: While the images across this post seem to reflect stylistic elements of actual paintings from previous centuries, the subjects seem unlikely, and the story, while plausible, does not ring a bell. Following a search for the referenced year and reverse image search (which concluded with some results of the same and similar images), this content was concluded not to have been made using GenAI.
*Figure 15*: While plausible as a recording of a woman speaking, the details give away a set of genAI artifacts, such as a disappearing earring and a shape-shifting ring.

## 4. Not GenAI

The content in question does not consist of generative AI imagery.